**This paper is a postscript-version (Authors Accepted Manuscript) of the paper accepted in Forest Ecology and Management**

# Tree species effects on topsoil carbon stock and concentration are mediated by tree species type, mycorrhizal association, and N-fixing ability at the global scale

Yan Peng[1*], Inger Kappel Schmidt[1], Haifeng Zheng[1], Petr Hedĕnec[1], Luciana Ruggiero Bachega[2], Kai Yue[3, 4], Fuzhong Wu[3], and Lars Vesterdal[1]

1. Department of Geosciences and Natural Resource Management, University of Copenhagen, Frederiksberg 1958, Denmark

2. Department of Environmental Science, Federal University of São Carlos (UFSCar), São Carlos, SP 13565-905, Brazil

3. State Key Laboratory for Subtropical Mountain Ecology, School of Geographical Sciences, Fujian Normal University, Fuzhou 350007, China

4. Center for Biodiversity Dynamics in a Changing World (BIOCHANGE) and Section for Ecoinformatics and Biodiversity, Department of Biology, Aarhus University, NyMunkegade 114, DK-8000, Aarhus C, Denmark

*Corresponding author: Yan Peng

Email address: yape@ign.ku.dk

Full address: Department of Geosciences and Natural Resource Management, University of Copenhagen, Rolighedsvej 23, DK-1958, Frederiksberg C, Denmark

## Abstract

Selection of appropriate tree species is an important forest management decision that may affect sequestration of carbon (C) in soil. However, information about tree species effects on soil C stocks at the global scale remains unclear. Here, we quantitatively synthesized 850 observations from field studies that were conducted in a common garden or monoculture plantations to assess how tree species type (broadleaf *vs.* conifer), mycorrhizal association (arbuscular mycorrhizal (AM) *vs.* ectomycorrhizal (ECM)), and N-fixing ability (N-fixing *vs.* non-N-fixing), directly and indirectly, affect topsoil (with a median depth of 10 cm) C concentration and stock, and how such effects were influenced by environmental factors such as geographical location and climate. We found that (1) tree species type, mycorrhizal association, and N-fixing ability were all important factors affecting soil C, with lower forest floor C stocks under broadleaved (44%), AM (39%), or N-fixing (28%) trees respectively, but higher mineral soil C concentration (11%, 22%, and 156 %) and stock (9%, 10%, and 6%) under broadleaved, AM, and N-fixing trees respectively; (2) tree species type, mycorrhizal association, and N-fixing ability affected forest floor C stock and mineral soil C concentration and stock directly or indirectly through impacting soil properties such as microbial biomass C and nitrogen; (3) tree species effects on mineral soil C concentration and stock were mediated by latitude, MAT, MAP, and forest stand age. These results reveal how tree species and their specific traits influence forest floor C stock and mineral soil C concentration and stock at a global scale. Insights into the underlying mechanisms of tree species effects found in our study would be useful to inform tree species selection in forest management or afforestation aiming to sequester more atmospheric C in soil for mitigation of climate change.

## 1. Introduction

In the face of global climate change and elevated atmospheric carbon dioxide ($CO_2$) concentrations, forest soils play an important role in global carbon (C) cycling and are potential terrestrial C sinks (Pan *et al.*, 2011). However, the degree to which forest soils can sequester C may vary significantly with forest management, and there is a lack of scientific consensus regarding the feasibility of management strategies aimed at promoting forest soil C sequestration (Vesterdal *et al.*, 2013; Jandl *et al.*, 2014). One important element of forest management strategies is the selection of tree species most conducive to C sequestration in biomass and soil (Mayer *et al.*, 2020). Several studies have assessed tree species effects on both forest floor and mineral soil C stocks (Vesterdal *et al.*, 2008; Mueller *et al.*, 2015; Cepáková *et al.*, 2016), however, our understanding of the quantitative response to tree species change and the drivers and underlying mechanisms of such effects remains limited (Mayer *et al.*, 2020). Therefore, more detailed and quantitative knowledge of tree species effects on soil C stocks is crucial for informing tree species selection strategies in the context of forest management or afforestation.

Tree species are known to affect soil C stocks through a variety of traits that are closely correlated with C sequestration and flux processes, such as the input of above- and belowground plant litter and the output of C mainly via heterotrophic respiration fluxes (Prescott and Vesterdal, 2013; Vesterdal *et al.*, 2013). Through differences in litter quality, tree species have the potential to impact the C concentrations, C stocks, and their distribution to forest floor and mineral soil (Wardle *et al.*, 2004; Bardgett and van der Putten, 2014). It is traditionally acknowledged that tree species with leaf litter traits driving slow rates of decomposition have been associated with accumulation of higher C stocks compared with tree species with fast litter decomposition rates (Berg, 2014; Lehmann and Kleber, 2015). This hypothesis has mainly been based on the observation of thick C-rich forest floors under coniferous tree species associated with ectomycorrhizae (ECM) (Keller and Phillips, 2019). However, a more recent hypothesis has suggested that tree species with foliar litter traits conducive to fast decomposition will lead to more pronounced microbial stabilization and transformation of plant litter C through greater production of microbial residues (Cotrufo *et al.*, 2013), ultimately supporting greater mineral soil C stocks. The latter tree species are mainly deciduous and in a temperate climate, some of these are associated with arbuscular mycorrhizae (AM). Deciduous AM tree species have also been suggested to enhance deeper incorporation of C by higher belowground rates of litter input, either via roots or bioturbation by soil fauna (Vesterdal *et al.*, 2013; Lin *et al.*, 2017; Craig *et al.*, 2018). A previous study also suggested that accumulation of forest floor C in conifer stands is mainly attributed to the adverse environmental conditions that retard decomposition, indicating the importance of environmental factors in regulating tree species effects on soil C stocks (Berger and Berger, 2012). Evidence of different nutrient economies in AM and ECM tree species (Phillips *et al.*, 2013) further suggests that soil C dynamics could differ in contrasting soil environments. Despite that these two hypotheses related to driving tree species traits seem rather contrasting because of their different mechanisms and opposite direction of effects, both suggest that tree species type (broadleaf vs. conifer) and their mycorrhizal association (AM vs. ECM) are important traits driving tree species effects on C stocks and their vertical distribution. It remains to be confirmed if one of these hypotheses has general validity or if they are valid within different contexts.

Although mycorrhizal association of tree species mainly relates to their phylogeny (Koele et al., 2012), the dominant mycorrhizal association of a forest community can also be influenced by climate and soil conditions (Barceló *et al.*, 2019). In general, AM trees increase in dominance in subtropical and lowland tropical forests where nutrient mineralization processes are fast, nevertheless ECM-associated tree species tend to dominate in cooler environments where the decomposition of organic matter occurs at a slower rate (Read and Perez-Moreno, 2003; Soudzilovskaia *et al.*, 2015), indicating that climate and location may indirectly affect tree species effects on soil C stocks. Climate may control soil C stock through its effect on forests microclimate (Berger and Berger, 2012), such as the more dry and cooler conditions under conifers as a result of greater interception of rain and less sunlight throughout the year. Besides, because geographical location is closely correlated with climate, soil properties such as soil respiration (Chen *et al.*, 2014), nutrient availability, and physicochemical characteristics vary remarkably at different locations and thus may control soil C stock (Yue *et al.*, 2016). Therefore, climate, location, and soil property may also influence the effects of tree species on soil C stocks, directly and/or indirectly. In addition, many AM tree species are also N-fixing, and the suggested effects on soil C may be related to this trait rather than the mycorrhizal association as such (Lin et al., 2017). Soil C stocks in the mineral soil are generally higher under N-fixing trees (Nave *et al.*, 2009; Mayer *et al.*, 2020), but the underlying mechanisms are not fully understood. Therefore, quantitative assessments of tree species effects and interactive drivers on soil C stocks are still scarce, especially at a global scale, which limits our understanding of targeted use of tree species to sequester soil C in existing forests and afforestation.

To examine tree species effects on soil C stocks, we comprehensively reviewed previously published articles and conducted a quantitative synthesis. Because different tree species in natural forest ecosystems usually follow certain gradients in soil physicochemical characteristics, climate, and forest successional stage that can also influence C stocks, comparing C stocks of natural forests would be a doubtful source of information on the true "tree species effects" (Vesterdal *et al.*, 2013). We thus compiled a database with 850 observations collected from 143 articles that reported forest floor C stock and mineral soil C concentration and stock in common garden and monoculture plantation studies. The objective of this study was to explore the quantitative influence of tree species on soil C concentrations, stocks, and the underlying drivers. We asked (1) whether and how biotic factors e.g. tree species type, mycorrhizal association, and N-fixing ability of different tree species influence soil C stocks; and (2) how abiotic factors such as climate, spatial location, and soil characteristics modulate tree species effects on soil C concentrations and stocks.

## 2. Materials and methods

### 2.1. Data collection and compiling

Peer-reviewed journal articles published before 20 December 2018 were searched using the *Web of Science*, and *Google Scholar* with the search terms of "(tree species OR plant species) AND (soil carbon OR carbon cycling) AND (common garden OR plantation)" in English and Chinese. We extracted studies from our search that met the following criteria: (1) studies were carried out in forest ecosystems (i.e., laboratory mesocosm studies were excluded); (2) at least one response variable of C (i.e., forest floor C stock, mineral soil C stock, or mineral soil C concentration) or litter production was reported; (3) tree species were identified by the Latin name; (4) only common garden or monoculture plantations with replicated plots within the same site were included in our database (i.e., mixed plantations and natural forests were excluded); (5) the means and sample sizes of the selected response variables were available or could be calculated from the related publications. If results from the same study sites and the same sampling year were reported in different articles, only one article was included in our database. After extraction, a total of 143 articles covering 850 observations matched these selection criteria and were thus included in our study (Fig. 1, Table S1, Appendix 1). The number of tree species across tree species type, mycorrhizal association, and N-fixing ability are shown in Table 1.

**Table 1** The numbers of tree species within species groups included in analysis of forest floor C stock, mineral soil C concentration, and mineral soil C stock.

| Soil layer | Tree species group | Species type | | Mycorrhizal association | | N-fixing ability | |
|---|---|---|---|---|---|---|---|
| | | Broadleaf | Conifer | AM | ECM | N-fixing | non-N-fixing |
| Forest floor C stock | Broadleaf | 30 | - | 17 | 13 | 5 | 25 |
| | Conifer | | 15 | 0 | 15 | 0 | 15 |
| | AM | | | 17 | - | 4 | 13 |
| | ECM | | | | 28 | 1 | 27 |
| | N-fixing | | | | | 5 | - |
| | non-N-fixing | | | | | | 40 |
| Mineral soil C concentration | Broadleaf | 127 | - | 73 | 54 | 14 | 113 |
| | Conifer | | 70 | 17 | 53 | 0 | 70 |
| | AM | | | 90 | - | 11 | 79 |
| | ECM | | | | 107 | 3 | 104 |
| | N-fixing | | | | | 14 | - |
| | non-N-fixing | | | | | | 183 |
| Mineral soil C stock | Broadleaf | 102 | - | 57 | 45 | 16 | 86 |
| | Conifer | | 53 | 11 | 42 | 0 | 53 |
| | AM | | | 68 | - | 12 | 56 |
| | ECM | | | | 87 | 4 | 83 |
| | N-fixing | | | | | 16 | - |
| | non-N-fixing | | | | | | 139 |

From each common garden or monoculture plantation at each study site, we extracted mean values of forest floor and mineral soil C stocks, mineral soil C concentration, and/or litter production for each tree species, if any of them was available. When only C concentrations were reported for mineral soil, we calculated C stocks according to Eqn 1:

$$C_{stock} = C_{concentration} \times \text{bulk density} \times \text{soil depth} \qquad (1)$$

if the information on soil bulk density and soil depth were available. For studies reporting several sampling depths, we only included the layer(s) from top of the mineral soil (i.e., 0 cm) to a certain depth (e.g., 0-5, 0-10, and 0-15cm, and maximum sampling depth), which was needed as a covariate in our analysis of tree species effects. Across all the data points of our constructed database, soil depths ranged from 1 to 100 cm, with a median of 10 cm. Therefore, we mainly addressed topsoil, but will hereafter use "soil" for conciseness. Meanwhile, if information about soil total nitrogen (N) concentration, nitrate ($NO_3^-$), ammonium ($NH_4^+$), plant-available phosphorus (PAP), C:N ratio, microbial biomass C (MBC), microbial biomass N (MBN), microbial biomass phosphorus (MBP), soil physical characteristics (i.e., the particle size distribution and bulk density), soil pH, and soil basal respiration were reported for a specific soil depth, we also recorded such information along with soil C data. To determine mycorrhizal association type (i.e., AM *vs.* ECM) of each tree species, we searched the *Web of Science* for studies that reported such information (Brundrett, 2009; Koele *et al.*, 2012). For tree species that have been reported to associate with both AM and ECM fungi (e.g., *Eucalyptus* and *Salix* spp.), we defined these trees as ECM because many ECM roots have very small amounts of AM fungi (Wagg *et al.*, 2008). In addition, we also classified by the N-fixing ability (i.e., N-fixing *vs.* non-N-fixing) and tree species type (i.e., broadleaf *vs.* conifer) of different tree species according to the literature, and recorded the stand ages if available. All original data were extracted from the text, tables, figures, and appendices of the publications. When data were presented graphically, the figures were digitized to extract the numerical values using the free software Engauge Digitizer version 5.1 (http://digitizer.sourceforge.net). In addition, mean annual temperature (MAT) and mean annual precipitation (MAP) were obtained directly from the selected articles, or extracted from the *WorldClim* version 2.0 (http://www.worldclim.org) using the information of latitude and longitude in case these data were not reported.

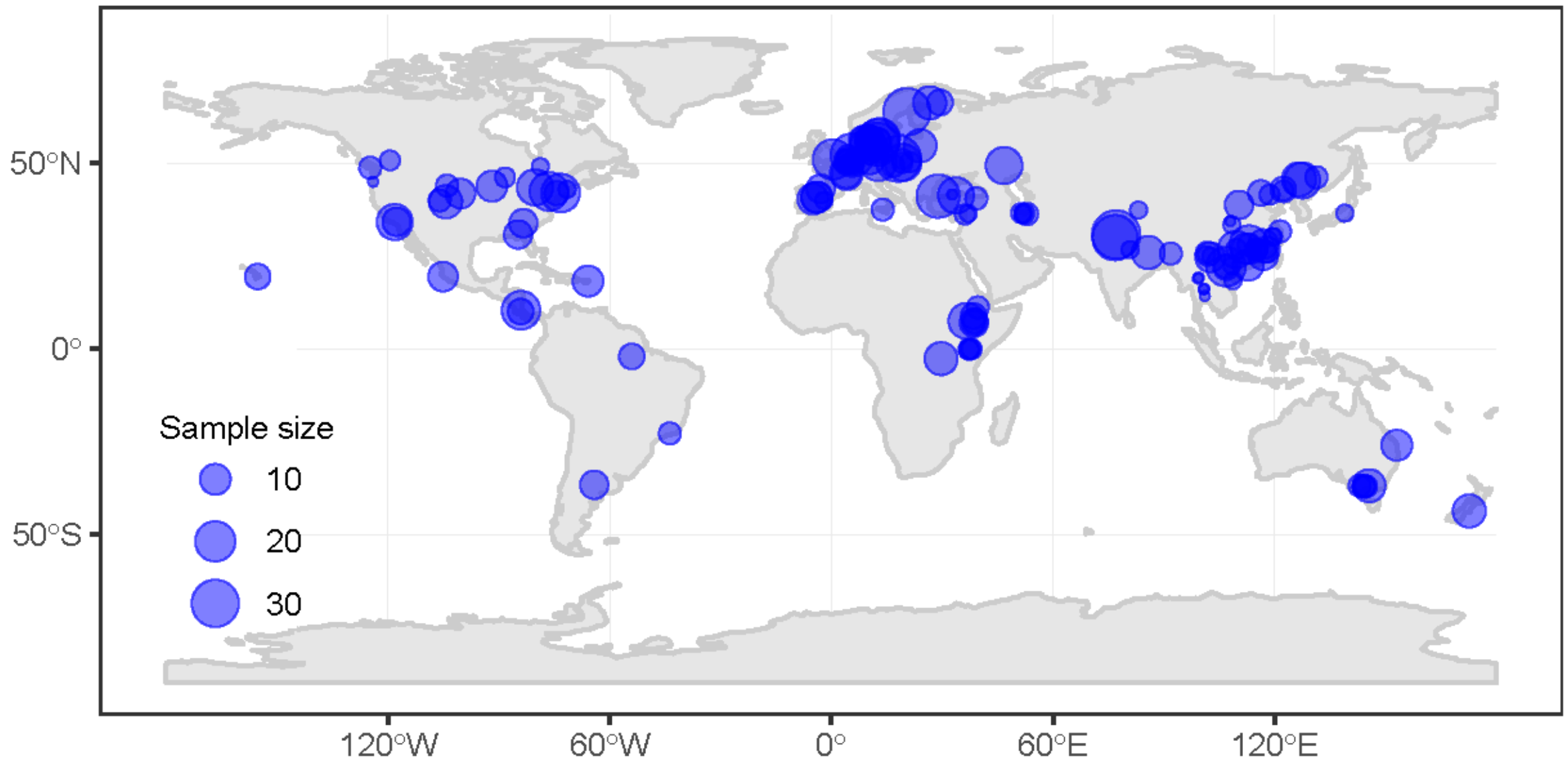

**Fig. 1** Map showing the location of the study sites from the compiled 143 articles used in the present study. The sample size (number of observations) from each site is represented by symbol size, and more detailed information are shown in Table S1 in Supporting Information.

## 2.2. Statistical analysis

For the entire dataset including paired (with both broadleaf and conifer, AM and ECM, and/or N-fixing and non-N-fixing) and non-paired plots, we first compared forest floor litter mass, forest floor and mineral soil C stocks, and mineral soil C concentration within each group (i.e., between AM and ECM trees, N-fixing and non-N-fixing trees, and broadleaved and coniferous trees). We used linear mixed models to estimate the mean values for each group by adding tree species type, mycorrhizal association, or N-fixing ability as fixed factors, while soil depth and "study" (the identity of each primary study) as random-effect factors. The difference within each group was assessed using a two-tailed Wilcoxon rank-sum test to account for small and uneven group size and/or non-normal error distribution. All these analyses were performed using R version 3.6.2 (R Core Team, 2019).

We then used structural equation models (SEMs) to further assess the relative importance of climate (i.e., MAT and MAP), spatial location (i.e., latitude, longitude, and altitude), tree species type (broadleaf and conifer), mycorrhizal association (i.e., AM and ECM), N-fixing ability, and soil properties (N stock (for forest floor) or N concentration (for mineral soil), C:N ratio, and pH) on soil C stocks and concentrations. We proposed an *a priori* model of hypothesized relationships among predictor and response variables within a path diagram (Fig. 2) and tested the fit of the *a priori* model against each sub-grouped dataset. This *a priori* model also tested whether tree species effects on soil C were direct and/or indirect through changes in other soil properties. The same model was hypothesized for forest floor C stock, mineral soil C stock, and mineral soil C concentration. The overall impacts of "climate" and "location" on C stocks or concentration were modeled as exogenous categorical effects using dummy variables (Grace, 2006). For each predictor variable in the SEM, we conducted principal component analyses (PCA) to extract a reduced number of variables that captured most of the variance to avoid redundancy among correlated variables (Grace, 2006; Yue *et al.*, 2018) and only used the first PCA axes with eigenvalues > 1 in the SEM. Also, a reduced number of predictor variables by using the first PCA axes can facilitate a good model fit (Grace, 2006). Tree species type, mycorrhizal association, and N-fixing ability were coded as binary variables for which "1" indicated the broadleaved, AM, or N-fixing trees, and "0" the coniferous, ECM, or non-N-fixing trees. The factors of study and soil depth were also treated as random-effect factors within each model to account for the potential non-independence among data from the same study and the influences of sampling depth, respectively. We used Fisher's C to test the overall goodness-of-fit of each model. The SEM analyses were performed in R using the *piecewiseSEM* package (Lefcheck, 2016). Because of the limited co-occurring data for $NH_4^+$, $NO_3^-$, PAP, MBC, MBN, MBP, and basal respiration along with soil C stock or concentration, we did not include these variables in the SEM analysis. Instead, we first assessed how tree species type, mycorrhizal type, and N-fixing ability affected these properties, by considering each of them as fixed factor and study and soil depth as random factors, and then tested the correlations between these soil properties and soil C stock and concentration based on Pearson correlation coefficients. Each soil property variable was assessed separately.

To further assess the effects of tree species type, mycorrhizal association, and N-fixing ability on soil C stocks and concentrations based on rigorous common garden or paired plot designs controlling for site effects, we conducted meta-analyses using data from paired plots within a common garden (e.g., one AM species and two ECM species in a common garden, would mean two effect sizes) in the case the initial soil properties were the same. For each paired study, the effect sizes of tree species type, mycorrhizal association, and N-fixing ability were calculated as the normalized effects using the natural log response ratio (lnRR) as:

$$\mathrm{lnRR}_{\text{species type}} = \ln(\text{broadleaf/conifer}) \quad (2)$$

$$\mathrm{lnRR}_{\text{mycorrhizal association}} = \ln(\text{AM/ECM}) \quad (3)$$

$$\mathrm{lnRR}_{\text{N-fixing ability}} = \ln(\text{N-fixing/non-N-fixing}) \quad (4)$$

To conduct our meta-analyses, we initially ran intercept-only models to calculate the overall effect sizes ($\mathrm{lnRR}_{++}$). These intercept-only models fitted lnRR as the response variable and included "study" as a random factor given that calculating effect sizes of all possible pairs from a single common garden may be non-independent. Meta-regression models, which included fixed effects, were then run to explore the effects of climate zone (i.e., boreal, temperate, and subtropical/tropical), MAT, MAP, latitude, altitude, stand age, and soil depth on $\ln RR_{++}$ by fitting these variables as fixed factors. For aiding the interpretation of results, $\ln RR_{++}$ and its corresponding 95% confidence intervals (CIs) were transformed back to percentage change as $(e^{lnRR} - 1) \times 100\%$. If the 95% CIs of $\ln RR_{++}$ did not overlap zero, the effects were considered to be significant at $\alpha = 0.05$ for tree species type, mycorrhizal association, or N-fixing ability.

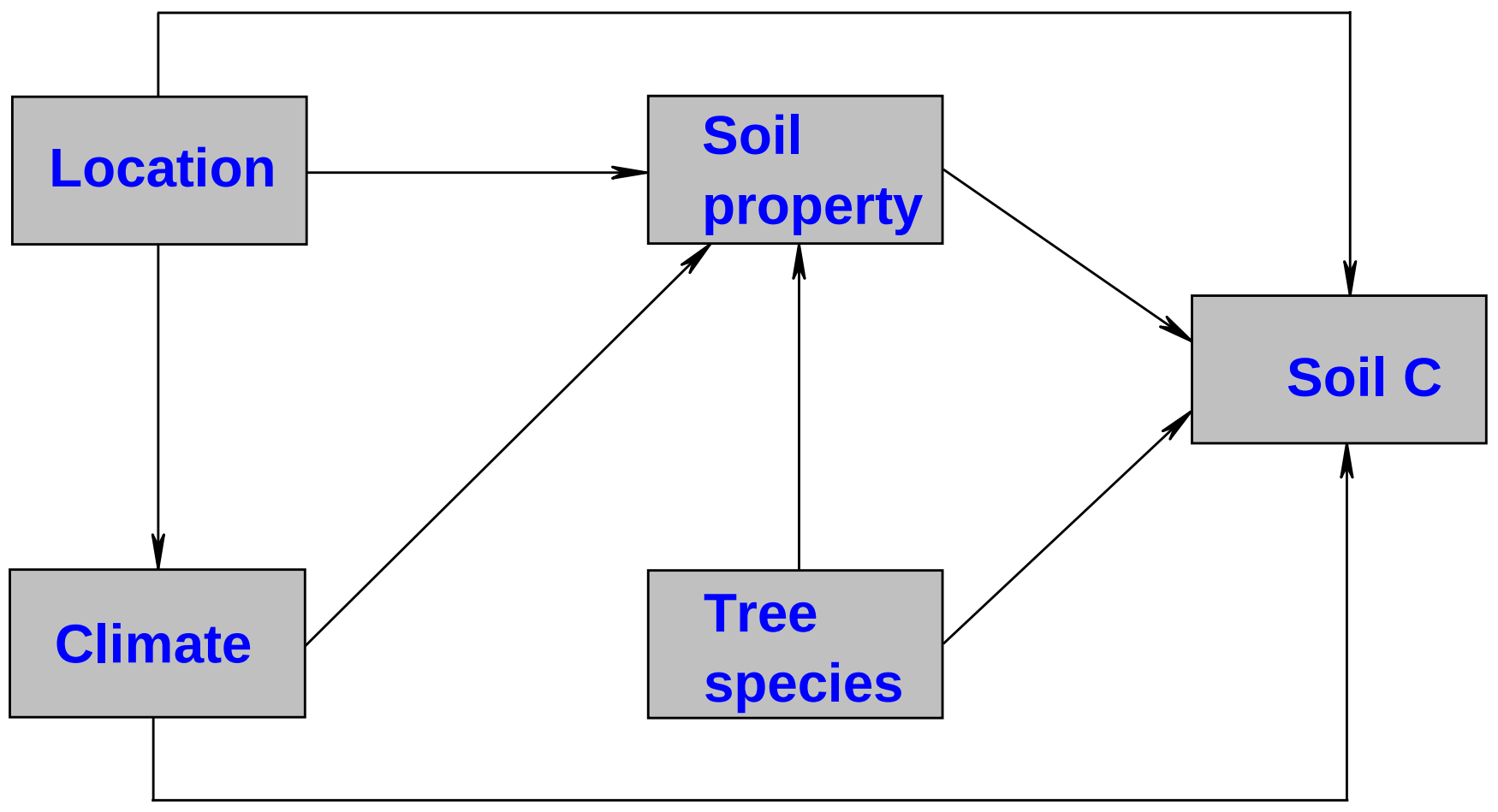

**Fig. 2** The proposed *a priori* structural equation model (SEM) of the causal relationship among tree species (explained by species type, mycorrhizal association, N-fixing ability), soil properties (explained by N concentration, C:N ratio, and pH), location (explained by latitude, longitude, and altitude), climate (explained by MAT and MAP) and soil C stocks or concentrations. The single-headed arrows indicate a hypothesized causal effect of one variable on another. All the predictor variables in the model represent the first principal component analysis (PCA) axes with eigenvalues > 1.

## 3. Results

Globally, litter production, forest floor C stock, and mineral soil C concentration and stock all differed significantly according to tree species type, mycorrhizal association, and N-fixing ability (Fig. 3). Within the three tree species categories, broadleaves had a higher litter production than conifers, but ECM and non-N-fixing trees produced more litter than AM and N-fixing trees, respectively (Fig. 3a). Forest floor C stocks were higher under conifers despite lower litter production, whereas ECM and non-N-fixing trees were significantly highest in forest floor C in line with the higher rates of production (Fig. 3b). In contrast, mineral soil C concentrations and stocks were significantly higher under broadleaved, AM, and N-fixing trees, respectively (Fig. 3c-d).

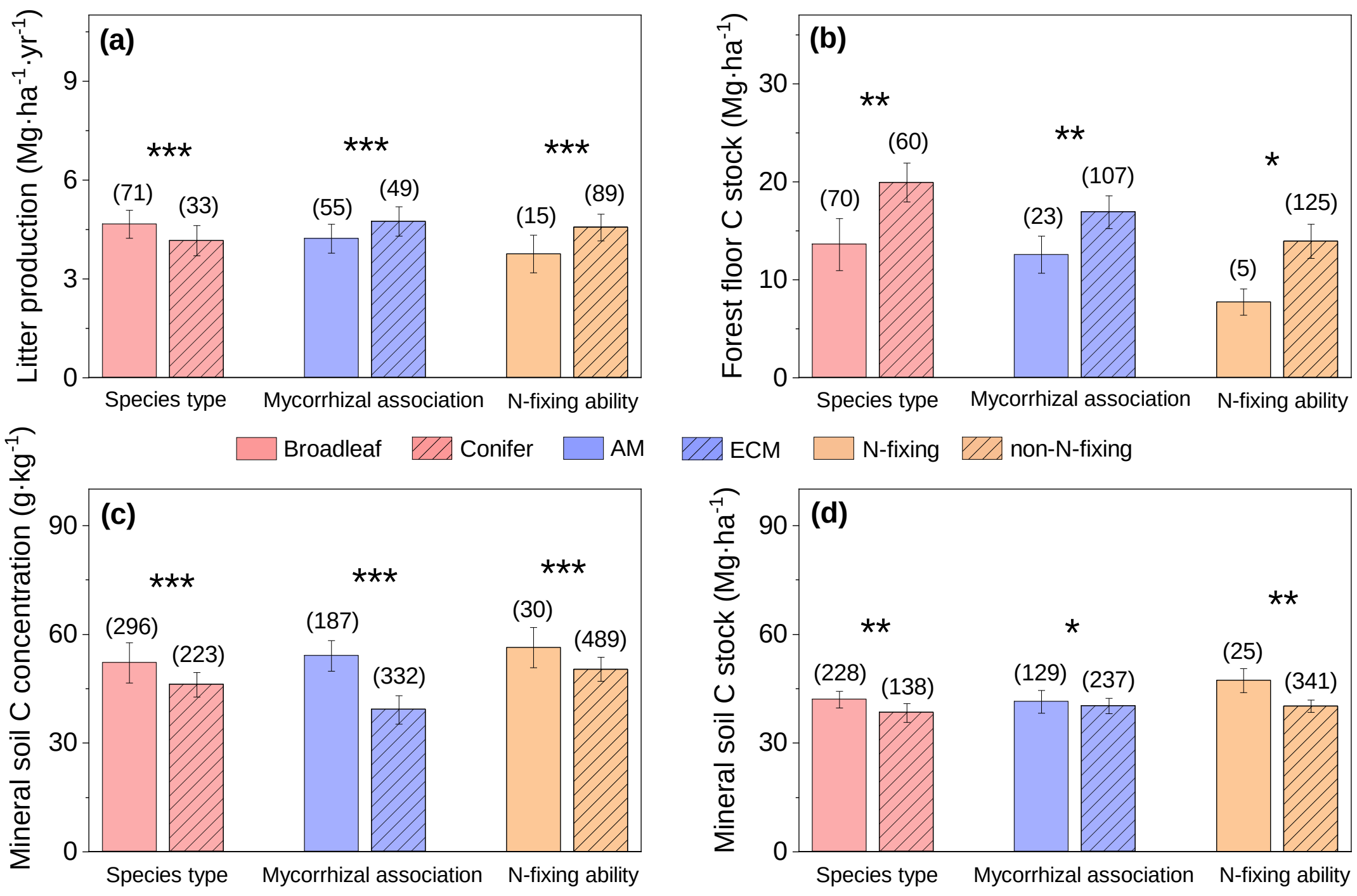

**Fig. 3** Means (± 1 SE) of litter production, forest floor carbon (C) stock, and mineral soil C concentration and stock under different tree species type (broadleaf *vs.* conifer), mycorrhizal association (AM *vs.* ECM), and N-fixing ability (N-fixing *vs.* non-N-fixing) across all data points as estimated from linear mixed models, in which the identity of primary study and soil depth (for mineral soil only) were treated as random

factors. Statistically significant differences are shown with asterisks ($^{*}p< 0.05$, $^{**}p< 0.01$, $^{***}p< 0.001$), and the number in parentheses are the number of data points.

Tree species type, mycorrhizal association, and N-fixing ability all significantly affected soil properties such as N concentration, MBC, and basal respiration (Table 2). By simultaneously assessing the effects of soil properties in combination with climate, location, and tree species using SEM analysis, we found that climate and tree species, the variances of which were mainly explained by MAP and tree species type, respectively, showed significant direct effects on forest floor C stock (Fig. 4). Specifically, the first PCA axis of climate had a negative effect on forest floor C stock and negatively correlated with MAP, indicating a positive effect of MAP on forest floor C stock (Fig. 4a). The first PCA axis of tree species, which was negatively correlated with tree species type and mycorrhizal association, showed a positive effect on forest floor C stock, suggesting that C stock was higher under coniferous or ECM trees because broadleaved and AM trees were coded as "1" and coniferous and ECM trees as "0" in the SEM analysis. In contrast, no significant effect of tree species or other variables on soil mineral C concentration or stock was found, and the marginal $R^2$ of SEMs for C concentration and stock were only 0.03 and 0.08, respectively (Fig. 4b, c). Location and soil properties did not affect forest floor C stock or mineral soil C concentration and stock. The indirect effects of tree species were examined by first testing the differences in soil variables between the groups of tree species (Table 2) and next testing their correlations with mineral soil C concentration and stock (Table 3). Mycorrhizal association and N-fixing ability strongly affected MBC, with higher values under AM or non-N-fixing trees (Table 2, 3), and MBC was again significantly positively correlated with soil C concentration and stock.

Tree species type significantly affected MBN, with a higher value under broadleaved trees (Table 2, 3), and MBN was again tightly positively correlated with soil C concentration.

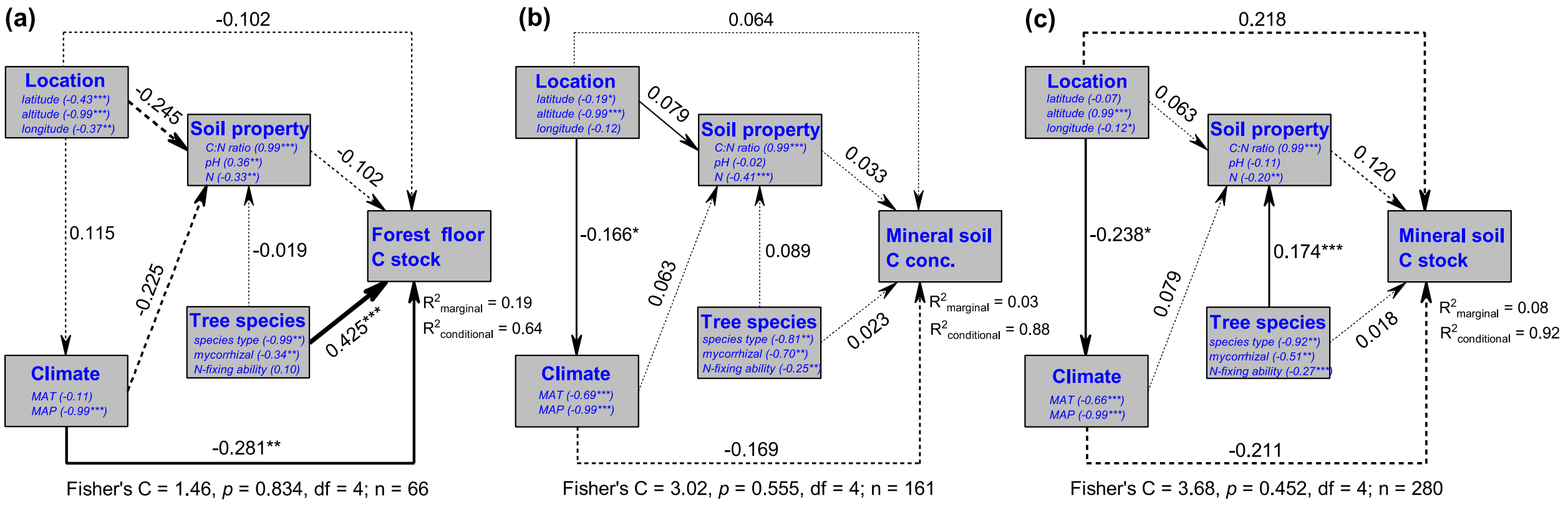

**Fig. 4** Structural equation models (SEMs) for forest floor C stock (a), mineral soil C concentration (b) and mineral soil C stock (c) describing the influence of climate (mean annual temperature (MAT) and mean annual precipitation (MAP)), location (latitude, longitude, and altitude), tree species (tree species type, mycorrhizal association, N-fixing ability), and soil properties (N stock for forest floor, N concentration for mineral soil, C:N ratio and pH) on forest floor carbon (C) stocks and mineral soil C concentration stock for the entire dataset. Positive and negative coefficients indicate positive and negative effects of the first principal component analysis (PCA) axes. Values in brackets are the correlation coefficients between the first PCA axes and each variable. Significant effects are indicated by solid lines and non-signifante effects by dashed lines. Arrow widths are proportional to path coefficient. The goodness-of-fit tests are shown at the bottom of each path diagram. The values of Fisher's C are very small and all *p*-values >0.05, indicating very good fits of all the three models. $^{*}p < 0.05$, $^{**}p < 0.01$, $^{***}p < 0.001$.

**Table 2** Effects of tree species type, mycorrhizal association, and N-fixing ability on mineral soil chemical and biotic properties as assessed using linear mixed models in which each of the tested variables was treated as fixed effect and study identity and soil depth were treated as random effects. Values are means ± 1SE, and *p*-values of tree species effects are also shown.

| Soil property | Species type | | | Mycorrhizal association | | | N-fixing ability | | |
|---|---|---|---|---|---|---|---|---|---|
| | Broadleaf | Conifer | *p*-value | AM | ECM | *p*-value | N-fixing | non-N-fixing | *p*-value |
| N (g $kg^{-1}$) | 3.8 ± 0.63 | 3.4 ± 0.64 | **0.037** | 3.9 ± 0.64 | 3.5 ± 0.63 | **0.021** | 4.0 ± 0.70 | 3.6 ± 0.63 | **0.026** |
| C:N | 15.7 ± 1.4 | 19.5 ± 1.5 | **0.009** | 15.2 ± 1.6 | 18.5 ± 1.3 | **0.001** | 14.9 ± 3.8 | 17.5 ± 1.3 | **0.017** |
| pH | 5.4 ± 0.14 | 5.2 ± 0.14 | 0.417 | 5.5 ± 0.15 | 5.2 ± 0.15 | **0.008** | 5.3 ± 0.17 | 5.3 ± 0.14 | 0.422 |
| $NH_4^+$ (mg $kg^{-1}$) | 214.8 ± 140.1 | 575.9 ± 155.4 | 0.262 | 309.1 ± 189.7 | 411.2 ± 225.9 | 0.431 | 327.5 ± 124.8 | 377.5 ± 164.0 | 0.453 |
| $NO_3^-$ (mg $kg^{-1}$) | 14.7 ± 5.7 | 13.2 ± 5.7 | 0.084 | 17.2 ± 5.6 | 11.9 ± 5.5 | **0.024** | 17.2 ± 7.8 | 14.0 ± 5.6 | 0.065 |
| PAP (mg $kg^{-1}$) | 312.1 ± 146.1 | 463.8 ± 258.8 | 0.411 | 293.9 ± 208.2 | 447.8 ± 265.2 | 0.401 | 246.3 ± 152.1 | 391.4 ± 191.7 | 0.404 |
| MBC (mg $kg^{-1}$) | 599.8 ± 211.7 | 577.7 ± 215.2 | 0.054 | 684.8 ± 215.2 | 535.9 ± 208.1 | **0.006** | 560.7 ± 250.8 | 594.8 ± 209.4 | **0.040** |
| MBN (mg $kg^{-1}$) | 85.7 ± 31.8 | 57.2 ± 32.9 | **0.002** | 76.4 ± 33.9 | 73.5 ± 33.7 | 0.089 | 70.8 ± 38.9 | 75.2 ± 33.2 | 0.053 |
| MBP (mg $kg^{-1}$) | 26.4 ± 13.4 | 15.4 ± 13.5 | 0.202 | 19.8 ± 13.6 | 24.9 ± 13.1 | 0.375 | 23.3 ± 13.9 | 23.4 ± 12.9 | 0.483 |
| Basal respiration (mg $CO_2$ $g^{-1}$ $day^{-1}$) | 27.6 ± 12.1 | 29.4 ± 12.1 | **0.027** | 24.2 ± 12.5 | 29.7 ± 12.3 | **0.012** | 27.1 ± 12.4 | 28.8 ± 12.1 | **0.030** |

PAP: plant available phosphorus; MBC: microbial biomass carbon; MBN: microbial biomass nitrogen; MBP: microbial biomass phosphorus; Data in bold indicate statistical significance.

**Table 3** Pearson correlations (ρ) between soil properties and mineral soil C concentration and C stock under different tree species.

| Soil properties | C concentration | | C stock | |
|---|---|---|---|---|
| | ρ | *p*-value | ρ | *p*-value |
| $NH_4^+$ | 0.125 | 0.601 | 0.141 | 0.236 |
| $NO_3^-$ | 0.252 | 0.283 | 0.118 | 0.381 |
| PAP | -0.011 | 0.923 | 0.061 | 0.453 |
| MBC | 0.444 | **0.001** | 0.270 | **0.002** |
| MBN | 0.253 | 0.194 | 0.282 | **0.029** |
| MBP | 0.949 | **< 0.001** | 0.711 | **0.021** |
| Basal respiration | 0.217 | 0.357 | 0.093 | 0.536 |

PAP: plant available phosphorus; MBC: microbial biomass carbon; MBN: microbial biomass nitrogen; MBP: microbial biomass phosphorus; *n*: number of data point; Data in bold indicate statistical significance.

By conducting a meta-analysis using data from paired plots, we found similar patterns as in results generated from the total dataset (shown in Fig. 3). Forest floor C stock was 43.9% lower under broadleaved trees than under coniferous trees at the global level (Fig. 5a). In contrast, mineral soil concentration and stock were significantly higher under broadleaf than conifer, with averages 12% and 9% higher, respectively, but no difference was found within climate zones. As to mycorrhizal association, forest floor C stock was 39% lower under AM than under ECM trees at the global scale (Fig. 5b). However, mineral soil C stock under AM trees was significantly higher than under ECM trees, with an average of 10% globally. A similar trend was also found for mineral soil C concentrations, which were 22% higher under AM trees at the global scale, and 22% and 23% higher under AM trees compared with ECM trees in temperate and in subtropical/tropical regions, respectively (Fig. 5b). Nitrogen-fixing trees tended to have lower forest floor C stocks at the global scale ($p$ =0.056) and in temperate zones ($p$ = 0.058), but not in tropical zones (Fig. 5c), but both mineral soil C concentration and stock were significantly higher under N-fixing trees than under non-N-fixing trees (Fig. 5c). Meta-regressions revealed that the effects of tree species type and mycorrhizal association on forest floor C stock were negatively affected by latitude and forest stand age, but positively with effects of N-fixing ability (Table 4). Tree species type effects on mineral soil C concentration were positively related to MAT and MAP and negatively to stand age (Table 4). N-fixing ability effects on mineral soil C concentration and stock were positively related to latitude soil depth and stand age. These results suggested greater effects of species type and mycorrhizal association on soil C in younger (lower stand age) forests located in lower latitude and warmer climate, but greater effects of N-fixing ability on soil C in higher latitude mature (higher stand age) forests.

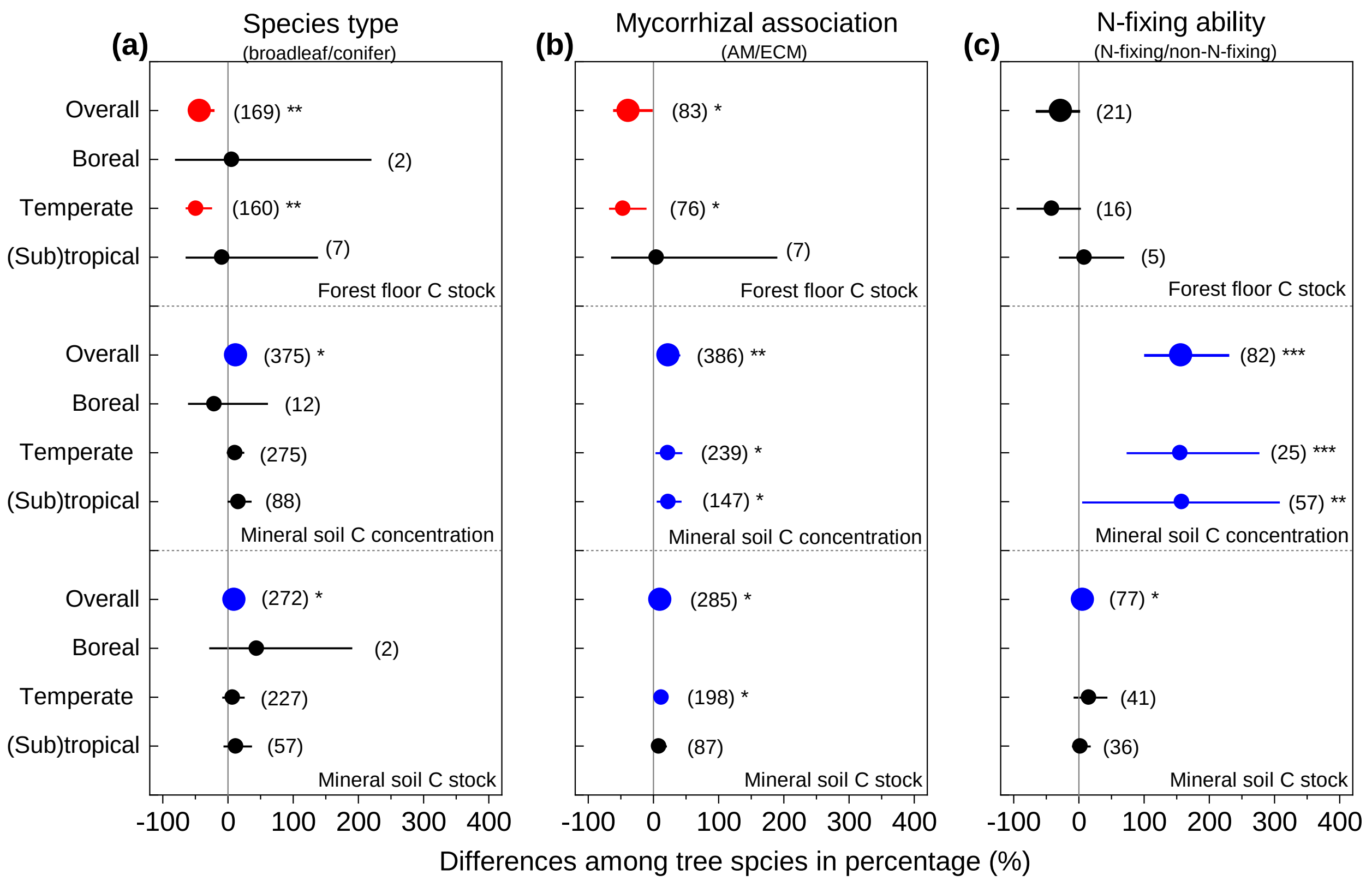

**Fig. 5** Effects of (a) tree species type (broadleaf *vs.* conifer), (b) mycorrhizal association (AM *vs.* ECM), and (c) N-fixing ability (N-fixing *vs.* non-N-fixing) of tree species on C stocks of forest floor and mineral soil and C concentration of mineral soil at the global scale and within the different climate zones (boreal, temperate, and subtropical/tropical zones). Results are expressed as the percentage differences (%) between tree species type, mycorrhizal type, or N-fixing ability and based on paired plots. Values indicate the means with 95% confidence intervals (CIs) and the numbers of data points are shown. Dots in blue and red indicate significant difference, and dot size represents effect sizes across all variables or within subgroups. $^{*}p < 0.05$, $^{**}p < 0.01$, $^{***}p < 0.001$.

**Table 4** The effects of latitude, altitude, mean annual temperature (MAT), mean annual precipitation (MAP), forest stand age, and soil depth on the natural log response ratios (lnRR) of tree species type (broadleaf *vs.* conifer), mycorrhizal association (AM *vs.* ECM), and N-fixing ability (N-fixing *vs.* non-N-fixing) generated from linear mixed models by adding each variable as fixed effect and study as a random effect. Soil depth was treated as a random effect when addressing latitude, altitude, MAT, MAP, and stand age for mineral soil, but as fixed effect when addressing itself. Direction of the estimates (in parentheses) and *p*-values are shown, and values in bold indicate statistical significance.

| | Tree species type | | | Mycorrhizal association | | | N-fixing ability | | |
|---|---|---|---|---|---|---|---|---|---|
| | Forest floor C stock | Mineral C concentration | Mineral C stock | Forest floor C stock | Mineral C concentration | Mineral C stock | Forest floor C stock | Mineral C concentration | Mineral C stock |
| Latitude | (-) **0.006** | (+) 0.237 | (-) 0.733 | (-) **0.016** | (+) 0.249 | (-) 0.209 | (+) **0.004** | (+) **0.035** | (+) 0.448 |
| Altitude | (-) 0.624 | (+) 0.090 | (+) 0.631 | (-) 0.750 | (+) 0.217 | (+) 0.271 | (+) 0.476 | (+) 0.787 | (-) 0.514 |
| MAT | (-) 0.115 | (+) **0.009** | (+) 0.469 | (-) 0.387 | (+) 0.310 | (-) 0.211 | (+) 0.413 | (-) 0.466 | (-) 0.224 |
| MAP | (-) 0.364 | (+) **0.025** | (+) 0.433 | (-) 0.697 | (+) 0.295 | (-) 0.107 | (+) 0.667 | (-) 0.538 | (-) 0.228 |
| Stand age | (-) **0.009** | (-) 0.844 | (-) 0.893 | (-) **0.048** | (-) 0.695 | (-) 0.229 | (+) **0.007** | (+) 0.170 | (+) 0.542 |
| Soil depth | - | (-) 0.076 | (+) 0.869 | - | (-) 0.192 | (-) 0.429 | - | (+) 0.309 | (+) **0.006** |

## 4. Discussion

Forest floor C stock was significantly higher in coniferous forests, but mineral soil C concentration and stock were higher in broadleaved forests. Generally, litter quality of broadleaved trees is higher than that of coniferous trees, resulting in a higher decomposition rate of broadleaf litter than conifer litter (Aerts, 1997; Augusto *et al.*, 2015). For example, compared with conifer litter, broadleaf litter generally has lower C:N ratio, lignin:N ratio, and concentration of phenolics that can lead to rapid decomposition rate and faster and more efficient accumulation of mineral-associated organic matter (Cotrufo *et al.*, 2013), resulting in a higher input of C from forest floor to mineral soil. Therefore, although litter production in broadleaf forests was larger than in conifer forests (Fig. 3a), forest floor C stock was lower whereas mineral soil C concentration and stock were higher in broadleaf forests compared with conifer forests. The difference in forest floor C stock between broadleaved and coniferous trees may also be attributed to the adverse microclimatic conditions, such as soil moisture, that control litter decomposition (Berger and Berger, 2012). However, the lack of data on such microclimatic conditions in our study limited the assessment of its impact in regulating tree species effects on soil C. Our findings were partly in line with a previous meta-analysis suggesting that forest floor C stock was 38% higher under conifers, while mineral soil C stocks were similar between broadleaves and conifers (Boča *et al.*, 2014). The inconsistency regarding mineral soil C may be attributed to different data and analysis methods between our study and that of Boča *et al.*, (2014), in which they had less data points, and the issue of non-independence of collected data points as well as soil depth were not considered when calculating the overall mean effects.

Mycorrhizal association has recently been recognized as an important factor regulating forest ecosystem functions such as C and N cycling (Phillips *et al.*, 2013; Lin *et al.*, 2017). Our results showed that AM trees had significantly lower forest floor C stocks compared with ECM forests, which may partly be attributed to the lower litter production and higher litter decomposition rate compared with ECM forests (Keller and Phillips, 2019). Previous studies found that foliar litter from ECM trees usually had lower litter quality (i.e., higher C:N ratio and lignin:N ratios and lower base cation concentrations) than that from AM trees, which contributes to a slower litter decomposition rate (Hobbie *et al.*, 2006; Lin *et al.*, 2017). Previous research has indicated that ECM and AM fungi can have opposite effects on litter decomposition, which can contribute to a higher forest floor C stock under ECM trees than AM trees. Specifically, ECM fungi can indirectly decrease litter decomposition rate by competing with saprotrophic microorganisms for soil water and N (Gadgil and Gadgil, 1971), while AM fungi can promote litter decomposition via their positive effects on the activity of soil bacterial communities (Nuccio *et al.*, 2013). Overall, the higher forest floor C stock in ECM forests supports the traditional hypothesis that tree species with low litter quality driving slow decomposition rates have been associated with accumulation of higher C stocks compared with tree species with fast rates of litter decomposition (Berg, 2014; Lehmann and Kleber, 2015).

In contrast to the forest floor pattern, our results showed that mineral soil C concentration and stock were significantly higher under AM trees than ECM trees. Generally, litter from AM trees has higher quality, suggesting a higher content of labile C that would be easier leached into mineral soil (Lin *et al.*, 2017). Our result supports the proposed hypothesis that tree species with foliar litter traits conducive to fast decomposition, such as in AM tree species, will lead to more pronounced microbial stabilization and transformation of plant litter C through greater production of microbial residues (Cotrufo *et al.*, 2013). This hypothesized *Microbial Efficiency-Matrix Stabilization* framework suggested that labile plant constituents, which are utilized more efficiently by microbes, are the dominant source of microbial products, and these microbial products of decomposition would become the main precursors of stable soil organic C by promoting aggregation and via strong chemical bonding to the mineral soil matrix (Cotrufo *et al.*, 2013; Tamura and Tharayil, 2014). The higher mineral soil C concentration under AM trees compared with ECM trees may also be attributed to a greater proportion of forest floor C incorporation into mineral soil contributed by soil fauna (Schelfhout *et al.*, 2017; Walmsley *et al.*, 2019) because litter C incorporated into soil aggregates by active soil fauna are more protected from decomposition (Frouz *et al.*, 2013; Frouz, 2018). This potential mechanism was supported by studies that reported leaf litter quality was positively correlated with soil fauna abundance (Hobbie *et al.*, 2006; Frouz *et al.*, 2013), and this was particularly the case for AM species such as *Acer pseudoplatanus* and *Fraxinus excelsior* (Schelfhout *et al.*, 2017).

The N-fixing ability of different tree species also had significant effects on forest floor and mineral soil C. The results of the linear mixed model using the total dataset and the meta-analysis using paired data both revealed that forest floor C stock was higher under non-N-fixing trees (although marginally significant ($p$ = 0.056) in the meta-analysis) while mineral soil C concentration and stock were higher in N-fixing trees. The non-significant results may be attributed to the small sample sizes that limited the statistical power of our analyses (Yue *et al.*, 2017), but similar forest floor C stocks have also been reported in other studies (Bachega *et al.*, 2016). Previous studies found that N-fixing trees are likely to promote mineral soil C accumulation by both retaining old C and accreting new C from plant litter (Resh *et al.*, 2002), indicating higher soil C concentrations and stocks in N-fixing forests. On the other hand, N-fixing is often more likely to form symbiosis with AM fungi while non-N-fixing trees mainly associated with ECM fungi (Pawlowska *et al.*, 1997), indicating that the possible effect of N-fixing species could be confounded with effects attributed to AM association. However, among the 102 AM tree species included in our dataset, only 14 were N-fixing trees, suggesting only limited possible confounding of mycorrhizal association with N-fixing ability effects on soil C.

It is noteworthy that the effects of tree species on mineral soil C concentration and stock were non-significant in our SEM analyses (Fig. 4b, c). Three potential mechanisms may explain the inconsistent findings between SEM analyses and meta-analysis: (1) tree species effects on soil C were not confounded with other factors such as microbial biomass and soil respiration in the meta-analysis, but this may be an issue in SEM because MBC and soil respiration, which could not be included in the model because of limited data points, could be confounded with tree species effects; (2) the marginal $R^2$ in these SEM were only 0.03 and 0.08, indicating very low interpretability of these models; and (3) the heterogeneity among the unpaired data used in the SEM limited the detection of significant results. Nevertheless, the significant effects of tree species on soil C as assessed by linear mixed models using total dataset and by meta-analyses using pairwise datasets were quite similar, indicating that our results are robust across sites on a global scale.

In the meta-regressions, location, climate, or soil properties were found to be important moderators of the effects of tree species type, mycorrhizal association, and N-fixing ability on forest floor C stock and mineral soil C concentration and stock. Because spatial location, climate, and soil properties are directly and indirectly linked to plant physiology and ecology, such as plant functional traits, litter quality, and litter decomposition (Wright *et al.*, 2005; García-Palacios *et al.*, 2013), it is not surprising that they significantly moderated tree species effects on soil C. For example, ECM forests usually have much lower soil inorganic N concentrations compared with AM forests (Lin *et al.*, 2017), which may be attributed to the ability of ECM fungi to directly mobilize N from soil organic matter via extracellular enzyme production (Mao *et al.*, 2019). Ectomycorrhizal trees are therefore less dependent on saprotrophic microbes for N uptake (Lindahl and Tunlid, 2015), while AM fungi mainly take up inorganic N produced by saprotrophic microorganisms (Phillips *et al.*, 2013; Lehmann and Kleber, 2015). Thus, the effects of tree species on soil C stocks can be exerted indirectly through the mediation of soil nutrient economies. The greater effects of tree species type and mycorrhizal association on soil C in lower latitude and warmer climate may be attributed to the fact that element cycling processes are faster, and effects may occur faster and be sooner detectable than in cooler environments where the inputs and decomposition of organic matter occur at a slower rate (Zhang *et al.*, 2008). In addition to these tested moderator variables, soil texture may also be an important variable moderating tree species effects on soil C, as previous studies found that clay-rich soils have a greater capacity to accumulate soil C than soils with lower clay content (Laganiere *et al.*, 2010). However, because of limited soil texture data, we could not assess the moderating influence of soil texture on tree species effects on soil C in this study.

Microbial biomass C and N, which were significantly correlated with soil C concentration and stock, were significantly affected by tree species (Table 3). Soil microbes play an important role in soil structure and function such as soil organic matter decomposition and biogeochemical cycling (Douterelo *et al.*, 2010). A recent study found that soil C stocks were positively correlated with MBC and MBN (Zhao *et al.*, 2016), and the effects of microbes on soil C dynamics were mainly through controlling C fluxes such as soil organic matter degradation and soil$CO_2$ emission (Iqbal *et al.*, 2010). In addition, soil fauna has an important role in mediating tree species effects on soil C (Frouz *et al.*, 2013). Among soil fauna, earthworms are one of the most important organisms in forest ecosystems because they can incorporate organic materials into the soil and also affect the activities of other soil organisms (Lee, 1985). Moreover, earthworm communities were found to be significantly affected by tree species (Schelfhout *et al.*, 2017), suggesting that tree species affect soil C stocks indirectly through their influence on earthworm communities. However, owing to the scarcity of such data, we were unable to assess how soil fauna communities under different tree species may modulate soil C dynamics at the global scale.

## 5. Conclusions

Our results from data syntheses and meta-analyses provided further evidence that tree species with specific functional traits have significant effects on forest floor C stock and mineral soil C concentration and stock at the global scale. Broadleaved, AM, and N-fixing trees had lower forest floor C stocks but higher mineral soil C concentrations and stocks than coniferous, ECM, and non-N-fixing trees, respectively. Tree species type, mycorrhizal association, and N-fixing ability affected forest floor C stock and mineral soil C concentration and stock directly or indirectly via impacting soil properties such as MBC and MBN. In addition, tree species effects on soil C concentration and stock were mediated by latitude, MAT, MAP, or forest stand age, with higher effects of species type and mycorrhizal association on soil C in lower latitude, warmer climate, and young forests, but higher effects of N-fixing ability in higher latitude and mature forests. These results show how tree species groups differ in their influence on soil C concentration and stock at the global scale, and the insights into the underlying mechanisms of tree species effects found in our study contribute to informing tree species selection for forest management or afforestation. Future studies should focus more on the potential indirect tree species-mediated effects via soil fauna and belowground litter input on soil C dynamics.

## Acknowledgments

We would like to thank the two anonymous reviewers for providing insightful comments and useful suggestions that significantly improved the quality of our study. We are grateful to all the researchers whose published data were used in this study. Yan Peng acknowledges China Scholarship Council for supporting a Ph.D. program grant (201606910045). Kai Yue was financially supported by the National Natural Science Foundation of China (31922052 and 31800373). Petr Heděnec was supported by the Marie Curie European Fellowship (747824-AFOREST-H2020-MSCA-IF-2016/H2020-MSCA-IF-2016). Luciana Ruggiero Bachega was supported by The São Paulo Research Foundation, Brazil (2015/14785-5, and 2017/26019-0).

# SUPPORTING INFORMATION FOR

# Tree species effects on soil carbon stock and concentration are mediated by tree species type, mycorrhizal association, and N-fixing ability

**Table S1** Raw data used in this study that were extracted from the 143 primary studies

| Reference | Latitude | Longitude | Altitude | MAT | MAP | Species type | Mycorrhizal association | N-fixing ability | layer | Litter | Cs | Ns | CN | Cc | Nc | NH4 | NO3 | PAP | MBC | MBN | MBP | respiration | pH |
|---|---|---|---|---|---|---|---|---|---|---|---|---|---|---|---|---|---|---|---|---|---|---|---|
| Son & Gower 1992 | 43.867 | -91.85 | 240 | 7 | 819 | boradleaf | ECM | non-N-fixing | forest floor | NA | NA | 0.079 | NA | NA | NA | NA | NA | NA | NA | NA | NA | NA | NA |
| Son & Gower 1992 | 43.867 | -91.85 | 240 | 7 | 819 | conifer | ECM | non-N-fixing | forest floor | NA | NA | 0.26 | NA | NA | NA | NA | NA | NA | NA | NA | NA | NA | NA |
| Son & Gower 1992 | 43.867 | -91.85 | 240 | 7 | 819 | conifer | ECM | non-N-fixing | forest floor | NA | NA | 0.239 | NA | NA | NA | NA | NA | NA | NA | NA | NA | NA | NA |
| Son & Gower 1992 | 43.867 | -91.85 | 240 | 7 | 819 | conifer | ECM | non-N-fixing | forest floor | NA | NA | 0.306 | NA | NA | NA | NA | NA | NA | NA | NA | NA | NA | NA |
| Son & Gower 1992 | 43.867 | -91.85 | 240 | 7 | 819 | conifer | ECM | non-N-fixing | forest floor | NA | NA | 0.225 | NA | NA | NA | NA | NA | NA | NA | NA | NA | NA | NA |
| Son & Gower 1992 | 43.867 | -91.85 | 240 | 7 | 819 | boradleaf | ECM | non-N-fixing | mineral | NA | 29.38 | 3.19 | 9.2 | 8.3 | 0.9 | NA | NA | 20.6 | NA | NA | NA | NA | 5.4 |
| Son & Gower 1992 | 43.867 | -91.85 | 240 | 7 | 819 | conifer | ECM | non-N-fixing | mineral | NA | 31.48 | 3.29 | 9.6 | 8.6 | 0.9 | NA | NA | 35.7 | NA | NA | NA | NA | 5.3 |
| Son & Gower 1992 | 43.867 | -91.85 | 240 | 7 | 819 | conifer | ECM | non-N-fixing | mineral | NA | 28.18 | 2.93 | 9.6 | 7.7 | 0.8 | NA | NA | 41.4 | NA | NA | NA | NA | 5.4 |
| Son & Gower 1992 | 43.867 | -91.85 | 240 | 7 | 819 | conifer | ECM | non-N-fixing | mineral | NA | 36.97 | 3.66 | 10.1 | 10.1 | 1 | NA | NA | 28.2 | NA | NA | NA | NA | 5.5 |
| Son & Gower 1992 | 43.867 | -91.85 | 240 | 7 | 819 | conifer | ECM | non-N-fixing | mineral | NA | 43.73 | 3.73 | 11.7 | 12.9 | 1.1 | NA | NA | 21.2 | NA | NA | NA | NA | 5.2 |
| Kooch et al. 2018 | 36.617 | 51.433 | 215 | 15.7 | 1035 | boradleaf | ECM | non-N-fixing | mineral | NA | 55.97 | 5.48 | 10.3 | 40 | 3.9 | 30.67 | 29.13 | 22.61 | 667.37 | 55.18 | 68.18 | 0.42 | 6.16 |
| Kooch et al. 2018 | 36.617 | 51.433 | 215 | 15.7 | 1035 | boradleaf | ECM | non-N-fixing | mineral | NA | 49.13 | 2.62 | 19.1 | 45.9 | 2.4 | 24.37 | 20.39 | 13.48 | 546 | 43.55 | 54.12 | 0.41 | 6.48 |
| Kooch et al. 2018 | 36.617 | 51.433 | 215 | 15.7 | 1035 | conifer | AM | non-N-fixing | mineral | NA | 59.17 | 3.38 | 17.2 | 41.2 | 2.4 | 22.85 | 18.18 | 15.73 | 501.68 | 43.24 | 44.75 | 0.4 | 6.45 |
| Rachid et al. 2013 | -22.767 | -43.683 | 13 | 23.4 | 1250 | boradleaf | ECM | non-N-fixing | mineral | NA | NA | NA | 12.85 | 4.49 | 0.35 | 1.4815 | 0.0287 | NA | NA | NA | NA | NA | 5.45 |
| Rachid et al. 2013 | -22.767 | -43.683 | 13 | 23.4 | 1250 | boradleaf | ECM | N-fixing | mineral | NA | NA | NA | 12.92 | 4.15 | 0.32 | 2.0201 | 0.459 | NA | NA | NA | NA | NA | 5.38 |
| Rachid et al. 2013 | -22.767 | -43.683 | 13 | 23.4 | 1250 | boradleaf | ECM | non-N-fixing | mineral | NA | NA | NA | 12.72 | 6.43 | 0.51 | 1.6816 | 0.3443 | NA | NA | NA | NA | NA | 5.73 |
| Rachid et al. 2013 | -22.767 | -43.683 | 13 | 23.4 | 1250 | boradleaf | ECM | N-fixing | mineral | NA | NA | NA | 12.92 | 7.38 | 0.58 | 1.2688 | 2.7684 | NA | NA | NA | NA | NA | 5.88 |
| Xu et al. 2006 | 36.6 | 139 | 1014 | 9.8 | 1536 | conifer | AM | non-N-fixing | mineral | NA | NA | NA | 14 | 118.5 | 8.4 | 10.2 | 11.5 | NA | 8005.61 | NA | NA | 31.9902 | 5.1 |
| Xu et al. 2006 | 36.6 | 139 | 1014 | 9.8 | 1536 | conifer | ECM | non-N-fixing | mineral | NA | NA | NA | 18.1 | 169.8 | 9.4 | 13.9 | 4.6 | NA | 3487.67 | NA | NA | 65.4322 | 4.8 |
| Chodak et al. 2015 | 66.433 | 29.45 | 274 | -0.8 | 554 | conifer | ECM | non-N-fixing | forest floor | NA | NA | NA | 46.7 | 345 | 7.5 | NA | NA | NA | 3575 | NA | NA | NA | 3.6 |
| Chodak et al. 2015 | 66.433 | 29.45 | 274 | -0.8 | 554 | conifer | ECM | non-N-fixing | forest floor | NA | NA | NA | 44.6 | 362 | 8.6 | NA | NA | NA | 3518 | NA | NA | NA | 3.7 |
| Chodak et al. 2015 | 66.433 | 29.45 | 274 | -0.8 | 554 | boradleaf | ECM | non-N-fixing | forest floor | NA | NA | NA | 26.3 | 360 | 14.1 | NA | NA | NA | 3391 | NA | NA | NA | 4.6 |
| Chodak et al. 2015 | 66.433 | 29.45 | 274 | -0.8 | 554 | conifer | ECM | non-N-fixing | mineral | NA | NA | NA | 23.5 | 12.5 | 0.48 | NA | NA | NA | 237 | NA | NA | NA | 4.3 |
| Chodak et al. 2015 | 66.433 | 29.45 | 274 | -0.8 | 554 | conifer | ECM | non-N-fixing | mineral | NA | NA | NA | 22.6 | 22.3 | 0.82 | NA | NA | NA | 199 | NA | NA | NA | 4.4 |
| Chodak et al. 2015 | 66.433 | 29.45 | 274 | -0.8 | 554 | boradleaf | ECM | non-N-fixing | mineral | NA | NA | NA | 26.2 | 31.9 | 1.21 | NA | NA | NA | 208 | NA | NA | NA | 4.8 |
| Vesterdal et al. 2008 | 55.134 | 12.036 | 11 | 7.5 | 703 | boradleaf | AM | non-N-fixing | forest floor | 2.7 | 1.48 | 0.027 | 53.9 | NA | NA | NA | NA | NA | NA | NA | NA | NA | NA |
| Vesterdal et al. 2008 | 55.134 | 12.036 | 11 | 7.5 | 703 | boradleaf | ECM | non-N-fixing | forest floor | 3.08 | 4.7 | 0.15 | 31.2 | NA | NA | NA | NA | NA | NA | NA | NA | NA | NA |
| Vesterdal et al. 2008 | 55.134 | 12.036 | 11 | 7.5 | 703 | boradleaf | ECM | non-N-fixing | forest floor | 3.12 | 1.473 | 0.041 | 34.93 | NA | NA | NA | NA | NA | NA | NA | NA | NA | NA |
| Vesterdal et al. 2008 | 55.134 | 12.036 | 11 | 7.5 | 703 | boradleaf | ECM | non-N-fixing | forest floor | 3.6 | 1.472 | 0.04327 | 34.01 | NA | NA | NA | NA | NA | NA | NA | NA | NA | NA |
| Vesterdal et al. 2008 | 55.134 | 12.036 | 11 | 7.5 | 703 | boradleaf | AM | non-N-fixing | forest floor | 3.54 | 3.192 | 0.12306 | 25.94 | NA | NA | NA | NA | NA | NA | NA | NA | NA | NA |
| Vesterdal et al. 2008 | 55.134 | 12.036 | 11 | 7.5 | 703 | conifer | ECM | non-N-fixing | forest floor | 4.2 | 15.12 | 0.56352 | 26.83 | NA | NA | NA | NA | NA | NA | NA | NA | NA | NA |
| Vesterdal et al. 2008 | 55.134 | 12.036 | 11 | 7.5 | 703 | boradleaf | AM | non-N-fixing | mineral | NA | 14.1 | 1.1 | 12.8 | 36.7 | 2.87 | NA | NA | NA | NA | NA | NA | NA | NA |
| Vesterdal et al. 2008 | 55.134 | 12.036 | 11 | 7.5 | 703 | boradleaf | ECM | non-N-fixing | mineral | NA | 15.3 | 1.04 | 14.9 | 36.3 | 2.44 | NA | NA | NA | NA | NA | NA | NA | NA |
| Vesterdal et al. 2008 | 55.134 | 12.036 | 11 | 7.5 | 703 | boradleaf | ECM | non-N-fixing | mineral | NA | 14.9 | 1.04 | 14.5 | 39.6 | 2.73 | NA | NA | NA | NA | NA | NA | NA | NA |
| Vesterdal et al. 2008 | 55.134 | 12.036 | 11 | 7.5 | 703 | boradleaf | ECM | non-N-fixing | mineral | NA | 14.2 | 1.05 | 13.6 | 33.7 | 2.48 | NA | NA | NA | NA | NA | NA | NA | NA |
| Vesterdal et al. 2008 | 55.134 | 12.036 | 11 | 7.5 | 703 | boradleaf | AM | non-N-fixing | mineral | NA | 15.5 | 1.04 | 15 | 43 | 2.86 | NA | NA | NA | NA | NA | NA | NA | NA |
| Vesterdal et al. 2008 | 55.134 | 12.036 | 11 | 7.5 | 703 | conifer | ECM | non-N-fixing | mineral | NA | 17 | 1.01 | 16.9 | 43.7 | 2.58 | NA | NA | NA | NA | NA | NA | NA | NA |
| Baum et al. 2009 | 51.083 | 10.45 | 439 | 7.7 | 870 | boradleaf | ECM | non-N-fixing | mineral | NA | NA | NA | 12 | 48 | 4 | NA | NA | NA | NA | NA | NA | NA | 5.1 |
| Olsson et al. 2012 | 56.667 | 13.05 | 63 | 6.9 | 952 | boradleaf | ECM | non-N-fixing | forest floor | NA | 9.671 | 0.46576 | 20.8 | NA | NA | NA | NA | NA | NA | NA | NA | NA | 4.58 |
| Olsson et al. 2012 | 56.667 | 13.05 | 63 | 6.9 | 952 | boradleaf | ECM | non-N-fixing | mineral | NA | 20.7 | 1.09589 | 18.9 | NA | NA | NA | NA | NA | NA | NA | NA | NA | 4.41 |
| Olsson et al. 2012 | 56.667 | 13.05 | 63 | 6.9 | 952 | conifer | ECM | non-N-fixing | forest floor | NA | 29.67 | 1.20548 | 24.6 | NA | NA | NA | NA | NA | NA | NA | NA | NA | 4.13 |
| Olsson et al. 2012 | 56.667 | 13.05 | 63 | 6.9 | 952 | conifer | ECM | non-N-fixing | mineral | NA | 24.85 | 1.20548 | 20.6 | NA | NA | NA | NA | NA | NA | NA | NA | NA | 4.12 |
| Olsson et al. 2012 | 56.667 | 13.05 | 63 | 6.9 | 952 | conifer | ECM | non-N-fixing | forest floor | NA | 59.33 | 2.16438 | 27.4 | NA | NA | NA | NA | NA | NA | NA | NA | NA | 3.7 |
| Olsson et al. 2012 | 56.667 | 13.05 | 63 | 6.9 | 952 | conifer | ECM | non-N-fixing | mineral | NA | 30.37 | 1.26027 | 24.1 | NA | NA | NA | NA | NA | NA | NA | NA | NA | 4.01 |
| Olsson et al. 2012 | 66.333 | 26.667 | 63 | 0.1 | 580 | boradleaf | ECM | non-N-fixing | forest floor | NA | 15.19 | 0.46576 | 32.6 | NA | NA | NA | NA | NA | NA | NA | NA | NA | 4.21 |
| Olsson et al. 2012 | 66.333 | 26.667 | 63 | 0.1 | 580 | boradleaf | ECM | non-N-fixing | mineral | NA | 5.543 | 0.21918 | 25.3 | NA | NA | NA | NA | NA | NA | NA | NA | NA | 4.77 |
| Olsson et al. 2012 | 66.333 | 26.667 | 63 | 0.1 | 580 | conifer | ECM | non-N-fixing | forest floor | NA | 17.26 | 0.46576 | 37.1 | NA | NA | NA | NA | NA | NA | NA | NA | NA | 4.07 |
| Olsson et al. 2012 | 66.333 | 26.667 | 63 | 0.1 | 580 | conifer | ECM | non-N-fixing | mineral | NA | 5.539 | 0.16438 | 33.7 | NA | NA | NA | NA | NA | NA | NA | NA | NA | 4.51 |
| Olsson et al. 2012 | 66.333 | 26.667 | 63 | 0.1 | 580 | conifer | ECM | non-N-fixing | forest floor | NA | 19.33 | 0.54794 | 35.3 | NA | NA | NA | NA | NA | NA | NA | NA | NA | 4 |
| Olsson et al. 2012 | 66.333 | 26.667 | 63 | 0.1 | 580 | conifer | ECM | non-N-fixing | mineral | NA | 4.168 | 0.13699 | 30.4 | NA | NA | NA | NA | NA | NA | NA | NA | NA | 4.59 |
| Smolander & Kitunen 2002 | 66.333 | 26.667 | 63 | 0.1 | 580 | boradleaf | ECM | non-N-fixing | forest floor | NA | NA | NA | 30 | 600 | 20 | 194.674 | NA | NA | 14.7718 | 1.5627 | NA | NA | 2.8 |
| Smolander & Kitunen 2002 | 66.333 | 26.667 | 63 | 0.1 | 580 | conifer | ECM | non-N-fixing | forest floor | NA | NA | NA | 32 | 559 | 17 | 58.4476 | NA | NA | 11.1852 | 1.2283 | NA | NA | 2.7 |
| Smolander & Kitunen 2002 | 66.333 | 26.667 | 63 | 0.1 | 580 | conifer | ECM | non-N-fixing | forest floor | NA | NA | NA | 37 | 570 | 15 | 6.70235 | NA | NA | 9.84338 | 0.9839 | NA | NA | 2.3 |
| Selmants et al. 2005 | 45.05 | -124 | 69 | 15 | 2400 | boradleaf | ECM | non-N-fixing | mineral | NA | NA | NA | 16.4 | 160 | 9.74 | NA | NA | NA | 2172 | 419 | NA | NA | 3.9 |
| Chatterjee et al. 2008 | 44 | -104 | 2060 | 4 | 509 | conifer | ECM | non-N-fixing | mineral | NA | 18.2 | 0.73 | 24.9 | NA | NA | NA | NA | NA | NA | NA | NA | NA | NA |

| | | | | | | | | | | | | | | | | | | | | | |
|---|---|---|---|---|---|---|---|---|---|---|---|---|---|---|---|---|---|---|---|---|---|
| Chatterjee et al. 2008 | 44 | -104 | 2060 | 4 | 509 | conifer | ECM | non-N-fixing | mineral | NA | 20.9 | 1.21 | 17.3 | NA | NA | NA | NA | NA | NA | NA | NA | NA | NA |
| Chatterjee et al. 2008 | 40 | -106 | 2060 | 2 | 501 | conifer | ECM | non-N-fixing | mineral | NA | 18.3 | 0.66 | 27.7 | NA | NA | NA | NA | NA | NA | NA | NA | NA | NA |
| Chatterjee et al. 2008 | 40 | -106 | 2060 | 2 | 501 | conifer | ECM | non-N-fixing | mineral | NA | 22.2 | 0.85 | 26.1 | NA | NA | NA | NA | NA | NA | NA | NA | NA | NA |
| Malchair & Carnol 2009 | 50.017 | 4.4 | 333 | 8.8 | 881 | boradleaf | ECM | non-N-fixing | mineral | NA | NA | NA | 26.3 | 120.98 | 4.6 | 190 | NA | NA | 2374.19 | 166.67 | NA | 1.47241 | 4.1 |
| Malchair & Carnol 2009 | 50.017 | 4.4 | 333 | 8.8 | 881 | boradleaf | ECM | non-N-fixing | mineral | NA | NA | NA | 23.2 | 129.92 | 5.6 | 500 | NA | NA | 2219.35 | 222.22 | NA | 1.68004 | 4 |
| Malchair & Carnol 2009 | 50.567 | 6.016 | 350 | 8.8 | 1043 | boradleaf | ECM | non-N-fixing | mineral | NA | NA | NA | 24.6 | 228.78 | 9.3 | 900 | NA | NA | 2529.03 | 225.69 | NA | 2.0257 | 3.9 |
| Malchair & Carnol 2009 | 50.567 | 6.016 | 350 | 8.8 | 1043 | boradleaf | ECM | non-N-fixing | mineral | NA | NA | NA | 24 | 230.4 | 9.6 | 2100 | NA | NA | 3174.19 | 350.69 | NA | 2.60432 | 4.1 |
| Malchair & Carnol 2009 | 50.567 | 6.016 | 350 | 8.8 | 1043 | conifer | ECM | non-N-fixing | mineral | NA | NA | NA | 29.2 | 230.68 | 7.9 | 0 | NA | NA | 2812.9 | 208.33 | NA | 2.2788 | 3.8 |
| Malchair & Carnol 2009 | 50.3 | 5.967 | 400 | 8 | 1072 | boradleaf | ECM | non-N-fixing | mineral | NA | NA | NA | 22 | 129.8 | 5.9 | 100 | NA | NA | 1987.1 | 180.56 | NA | 1.88333 | 4 |
| Malchair & Carnol 2009 | 50.3 | 5.967 | 400 | 8 | 1072 | boradleaf | ECM | non-N-fixing | mineral | NA | NA | NA | 23.4 | 154.44 | 6.6 | 3000 | NA | NA | 2141.94 | 187.5 | NA | 1.92874 | 4 |
| Malchair & Carnol 2009 | 50.3 | 5.967 | 400 | 8 | 1072 | conifer | ECM | non-N-fixing | mineral | NA | NA | NA | 30.2 | 199.32 | 6.6 | 11400 | NA | NA | 1935.48 | 166.67 | NA | 2.18335 | 3.9 |
| Malchair & Carnol 2009 | 50.3 | 5.967 | 400 | 8 | 1072 | conifer | ECM | non-N-fixing | mineral | NA | NA | NA | 27.7 | 182.82 | 6.6 | 0 | NA | NA | 1806.45 | 159.72 | NA | 1.9274 | 3.8 |
| Malchair & Carnol 2009 | 49.817 | 5.713 | 503 | 7.9 | 1042 | boradleaf | ECM | non-N-fixing | mineral | NA | NA | NA | 22.4 | 161.28 | 7.2 | 0 | NA | NA | 1858.06 | 163.19 | NA | 1.20775 | 3.9 |
| Malchair & Carnol 2009 | 49.817 | 5.713 | 503 | 7.9 | 1042 | boradleaf | ECM | non-N-fixing | mineral | NA | NA | NA | 24.5 | 198.45 | 8.1 | 700 | NA | NA | 1832.26 | 180.56 | NA | 1.55478 | 3.7 |
| Malchair & Carnol 2009 | 49.817 | 5.713 | 503 | 7.9 | 1042 | conifer | ECM | non-N-fixing | mineral | NA | NA | NA | 29.4 | 267.54 | 9.1 | 1600 | NA | NA | 2993.55 | 208.33 | NA | 2.89823 | 3.7 |
| Malchair & Carnol 2009 | 49.817 | 5.713 | 503 | 7.9 | 1042 | conifer | ECM | non-N-fixing | mineral | NA | NA | NA | 284 | 198.8 | 0.7 | 2400 | NA | NA | 2296.77 | 121.53 | NA | 1.22867 | 3.7 |
| Huang et al. 2011 | -43.65 | 172.7 | 50 | 11.8 | 724 | conifer | ECM | non-N-fixing | mineral | NA | 18.5 | 1.7 | 1.1 | 42 | 39 | NA | NA | NA | NA | NA | NA | NA | 5.2 |
| Huang et al. 2011 | -43.65 | 172.7 | 50 | 11.8 | 724 | conifer | ECM | non-N-fixing | mineral | NA | 21.5 | 1.7 | 1.3 | 50 | 39 | NA | NA | NA | NA | NA | NA | NA | 5.2 |
| Huang et al. 2011 | -43.65 | 172.7 | 50 | 11.8 | 724 | boradleaf | ECM | non-N-fixing | mineral | NA | 19.1 | 1.8 | 1.1 | 44 | 41 | NA | NA | NA | NA | NA | NA | NA | 5.2 |
| Huang et al. 2011 | -43.65 | 172.7 | 50 | 11.8 | 724 | boradleaf | ECM | non-N-fixing | mineral | NA | 23.3 | 1.9 | 1.3 | 54 | 43 | NA | NA | NA | NA | NA | NA | NA | 5.4 |
| Huang et al. 2011 | -43.65 | 172.7 | 50 | 11.8 | 724 | conifer | AM | non-N-fixing | mineral | NA | 19.1 | 1.7 | 1.2 | 46 | 40 | NA | NA | NA | NA | NA | NA | NA | 5.3 |
| Huang et al. 2011 | -43.65 | 172.7 | 50 | 11.8 | 724 | conifer | AM | non-N-fixing | mineral | NA | 21.9 | 1.7 | 1.3 | 52 | 41 | NA | NA | NA | NA | NA | NA | NA | 5.5 |
| Huang et al. 2011 | -43.65 | 172.7 | 50 | 11.8 | 724 | conifer | ECM | non-N-fixing | mineral | NA | 70.1 | 6.4 | 11 | NA | NA | NA | NA | NA | NA | NA | NA | NA | 5.2 |
| Huang et al. 2011 | -43.65 | 172.7 | 50 | 11.8 | 724 | conifer | ECM | non-N-fixing | mineral | NA | 74.3 | 6.4 | 11.6 | NA | NA | NA | NA | NA | NA | NA | NA | NA | 5.2 |
| Huang et al. 2011 | -43.65 | 172.7 | 50 | 11.8 | 724 | boradleaf | ECM | non-N-fixing | mineral | NA | 71.6 | 6.6 | 10.8 | NA | NA | NA | NA | NA | NA | NA | NA | NA | 5.2 |
| Huang et al. 2011 | -43.65 | 172.7 | 50 | 11.8 | 724 | boradleaf | ECM | non-N-fixing | mineral | NA | 77.6 | 6.8 | 11.4 | NA | NA | NA | NA | NA | NA | NA | NA | NA | 5.4 |
| Huang et al. 2011 | -43.65 | 172.7 | 50 | 11.8 | 724 | conifer | AM | non-N-fixing | mineral | NA | 71.6 | 6.5 | 11 | NA | NA | NA | NA | NA | NA | NA | NA | NA | 5.3 |
| Huang et al. 2011 | -43.65 | 172.7 | 50 | 11.8 | 724 | conifer | AM | non-N-fixing | mineral | NA | 70.6 | 6.2 | 11.4 | NA | NA | NA | NA | NA | NA | NA | NA | NA | 5.5 |
| Huang et al. 2013 | 26.8 | 117.967 | 267 | 19.1 | 1673 | boradleaf | AM | non-N-fixing | mineral | 9.5 | 18.57 | 1.03 | 18 | 39.5 | 2.2 | NA | NA | NA | NA | NA | NA | NA | 4.3 |
| Huang et al. 2013 | 26.8 | 117.967 | 267 | 19.1 | 1673 | conifer | AM | non-N-fixing | mineral | 4.3 | 14.5 | 0.91 | 15.9 | 30.2 | 1.9 | NA | NA | NA | NA | NA | NA | NA | 4.6 |
| Huang et al. 2014 | 22.167 | 106.833 | 360 | 21.3 | 1344 | boradleaf | ECM | non-N-fixing | mineral | NA | NA | NA | 2.3 | 16.9 | 7.2 | 2.03 | 0.39 | NA | 348.05 | NA | NA | NA | NA |
| Fu et al. 2015 | 26.733 | 115.05 | 139 | 17.9 | 1489 | conifer | ECM | non-N-fixing | mineral | 4.94 | NA | NA | 12.6 | 14.833 | 1.179 | NA | NA | NA | NA | NA | NA | NA | 4.47 |
| Fu et al. 2015 | 26.733 | 115.05 | 139 | 17.9 | 1489 | conifer | ECM | non-N-fixing | mineral | 3.25 | NA | NA | 11.6 | 12.791 | 1.105 | NA | NA | NA | NA | NA | NA | NA | 4.49 |
| Fu et al. 2015 | 26.733 | 115.05 | 139 | 17.9 | 1489 | conifer | AM | non-N-fixing | mineral | 2.86 | NA | NA | 14.2 | 21.145 | 1.486 | NA | NA | NA | NA | NA | NA | NA | 4.07 |
| Kang et al. 2018 | 31.683 | 121.467 | 6 | 15.5 | 1039 | boradleaf | AM | non-N-fixing | mineral | NA | NA | NA | 1.2 | 23.8 | 20 | 1.5 | 11.7 | NA | NA | NA | NA | NA | 7.78 |
| Kang et al. 2018 | 31.683 | 121.467 | 6 | 15.5 | 1039 | conifer | AM | non-N-fixing | mineral | NA | NA | NA | 1.4 | 13.8 | 10 | 1.89 | 6.7 | NA | NA | NA | NA | NA | 7.8 |
| Kang et al. 2018 | 31.683 | 121.467 | 6 | 15.5 | 1039 | conifer | AM | non-N-fixing | mineral | NA | NA | NA | 1.3 | 21.4 | 16 | 2.73 | 13.7 | NA | NA | NA | NA | NA | 7.77 |
| Kang et al. 2018 | 31.683 | 121.467 | 6 | 15.5 | 1039 | conifer | AM | non-N-fixing | mineral | NA | NA | NA | 1.2 | 20.7 | 17 | 3.48 | 15.3 | NA | NA | NA | NA | NA | 7.81 |
| Jozefowska et al. 2017 | 50.336 | 21.345 | 155 | 7 | 650 | conifer | ECM | non-N-fixing | mineral | NA | NA | NA | 13.4 | 13.4 | 1 | NA | NA | NA | NA | NA | NA | NA | 6.2 |
| Jozefowska et al. 2017 | 50.336 | 21.345 | 155 | 7 | 650 | boradleaf | ECM | non-N-fixing | mineral | NA | NA | NA | 18.7 | 13.1 | 0.7 | NA | NA | NA | NA | NA | NA | NA | 7.2 |
| Jozefowska et al. 2017 | 50.336 | 21.345 | 155 | 7 | 650 | boradleaf | ECM | non-N-fixing | mineral | NA | NA | NA | 10.4 | 19.7 | 1.9 | NA | NA | NA | NA | NA | NA | NA | 6.8 |
| Jozefowska et al. 2017 | 50.144 | 19.251 | 238 | 8 | 700 | conifer | ECM | non-N-fixing | mineral | NA | NA | NA | 35 | 7 | 0.2 | NA | NA | NA | NA | NA | NA | NA | 6 |
| Jozefowska et al. 2017 | 50.144 | 19.251 | 238 | 8 | 700 | boradleaf | ECM | non-N-fixing | mineral | NA | NA | NA | 40.5 | 8.1 | 0.2 | NA | NA | NA | NA | NA | NA | NA | 6.5 |
| Jozefowska et al. 2017 | 50.144 | 19.251 | 238 | 8 | 700 | boradleaf | ECM | non-N-fixing | mineral | NA | NA | NA | 68 | 6.8 | 0.1 | NA | NA | NA | NA | NA | NA | NA | 6.1 |
| Jozefowska et al. 2017 | 51.132 | 19.257 | 240 | 7.6 | 580 | conifer | ECM | non-N-fixing | mineral | NA | NA | NA | 37 | 7.4 | 0.2 | NA | NA | NA | NA | NA | NA | NA | 6.5 |
| Jozefowska et al. 2017 | 51.132 | 19.257 | 240 | 7.6 | 580 | boradleaf | ECM | non-N-fixing | mineral | NA | NA | NA | 49 | 4.9 | 0.1 | NA | NA | NA | NA | NA | NA | NA | 6.4 |
| Jozefowska et al. 2017 | 51.132 | 19.257 | 240 | 7.6 | 580 | boradleaf | ECM | non-N-fixing | mineral | NA | NA | NA | 20.9 | 20.9 | 1 | NA | NA | NA | NA | NA | NA | NA | 5.9 |
| Hagen-Thorn et al. 2004 | 54.75 | 24.067 | 90 | 6.3 | 622.7 | boradleaf | AM | non-N-fixing | mineral | NA | 28.62 | 1.854 | 15.5 | 26.2 | 1.691 | NA | NA | 18.5 | NA | NA | NA | NA | 5.1 |
| Hagen-Thorn et al. 2004 | 54.75 | 24.067 | 90 | 6.3 | 622.7 | boradleaf | ECM | non-N-fixing | mineral | NA | 27.82 | 1.605 | 17.3 | 25.9 | 1.497 | NA | NA | 27.5 | NA | NA | NA | NA | 4.7 |
| Hagen-Thorn et al. 2004 | 54.75 | 24.067 | 90 | 6.3 | 622.7 | boradleaf | ECM | non-N-fixing | mineral | NA | 24.79 | 1.452 | 17.3 | 21.7 | 1.253 | NA | NA | 23.8 | NA | NA | NA | NA | 5 |
| Hagen-Thorn et al. 2004 | 54.75 | 24.067 | 90 | 6.3 | 622.7 | boradleaf | ECM | non-N-fixing | mineral | NA | 27.38 | 1.69 | 16.2 | 23.2 | 1.43 | NA | NA | 20.4 | NA | NA | NA | NA | 5.5 |
| Hagen-Thorn et al. 2004 | 54.75 | 24.067 | 90 | 6.3 | 622.7 | boradleaf | ECM | non-N-fixing | mineral | NA | 26.44 | 1.543 | 16.4 | 25.1 | 1.533 | NA | NA | 23.8 | NA | NA | NA | NA | 5.1 |
| Hagen-Thorn et al. 2004 | 54.75 | 24.067 | 90 | 6.3 | 622.7 | conifer | ECM | non-N-fixing | mineral | NA | 31.39 | 1.543 | 20.2 | 24.8 | 1.23 | NA | NA | 41.2 | NA | NA | NA | NA | 4.3 |
| Schulp et al. 2008 | 52.25 | 5.683 | 64 | 9.4 | 860 | boradleaf | ECM | non-N-fixing | forest floor | NA | 48.5 | NA | NA | 245 | NA | NA | NA | NA | NA | NA | NA | NA | NA |
| Schulp et al. 2008 | 52.25 | 5.683 | 64 | 9.4 | 860 | boradleaf | ECM | non-N-fixing | forest floor | NA | 11.06 | NA | NA | 149 | NA | NA | NA | NA | NA | NA | NA | NA | NA |
| Schulp et al. 2008 | 52.25 | 5.683 | 64 | 9.4 | 860 | boradleaf | ECM | non-N-fixing | forest floor | NA | 24.62 | NA | NA | 193 | NA | NA | NA | NA | NA | NA | NA | NA | NA |
| Schulp et al. 2008 | 52.25 | 5.683 | 64 | 9.4 | 860 | conifer | ECM | non-N-fixing | forest floor | NA | 29.64 | NA | NA | 326 | NA | NA | NA | NA | NA | NA | NA | NA | NA |
| Schulp et al. 2008 | 52.25 | 5.683 | 64 | 9.4 | 860 | conifer | ECM | non-N-fixing | forest floor | NA | 26.33 | NA | NA | 240 | NA | NA | NA | NA | NA | NA | NA | NA | NA |
| Schulp et al. 2008 | 52.25 | 5.683 | 64 | 9.4 | 860 | conifer | ECM | non-N-fixing | forest floor | NA | 26.65 | NA | NA | 240 | NA | NA | NA | NA | NA | NA | NA | NA | NA |
| Schulp et al. 2008 | 52.25 | 5.683 | 64 | 9.4 | 860 | boradleaf | ECM | non-N-fixing | mineral | NA | 43.89 | NA | NA | 34 | NA | NA | NA | NA | NA | NA | NA | NA | NA |
| Schulp et al. 2008 | 52.25 | 5.683 | 64 | 9.4 | 860 | boradleaf | ECM | non-N-fixing | mineral | NA | 39.51 | NA | NA | 37 | NA | NA | NA | NA | NA | NA | NA | NA | NA |
| Schulp et al. 2008 | 52.25 | 5.683 | 64 | 9.4 | 860 | boradleaf | ECM | non-N-fixing | mineral | NA | 59.69 | NA | NA | 43 | NA | NA | NA | NA | NA | NA | NA | NA | NA |
| Schulp et al. 2008 | 52.25 | 5.683 | 64 | 9.4 | 860 | conifer | ECM | non-N-fixing | mineral | NA | 66.42 | NA | NA | 52 | NA | NA | NA | NA | NA | NA | NA | NA | NA |
| Schulp et al. 2008 | 52.25 | 5.683 | 64 | 9.4 | 860 | conifer | ECM | non-N-fixing | mineral | NA | 45.36 | NA | NA | 36 | NA | NA | NA | NA | NA | NA | NA | NA | NA |
| Schulp et al. 2008 | 52.25 | 5.683 | 64 | 9.4 | 860 | conifer | ECM | non-N-fixing | mineral | NA | 47.4 | NA | NA | 35 | NA | NA | NA | NA | NA | NA | NA | NA | NA |
| Schulp et al. 2008 | 52.25 | 5.683 | 64 | 9.4 | 860 | boradleaf | ECM | non-N-fixing | mineral | NA | 68.27 | NA | NA | NA | NA | NA | NA | NA | NA | NA | NA | NA | NA |
| Schulp et al. 2008 | 52.25 | 5.683 | 64 | 9.4 | 860 | boradleaf | ECM | non-N-fixing | mineral | NA | 53.28 | NA | NA | NA | NA | NA | NA | NA | NA | NA | NA | NA | NA |
| Schulp et al. 2008 | 52.25 | 5.683 | 64 | 9.4 | 860 | boradleaf | ECM | non-N-fixing | mineral | NA | 81.61 | NA | NA | NA | NA | NA | NA | NA | NA | NA | NA | NA | NA |
| Schulp et al. 2008 | 52.25 | 5.683 | 64 | 9.4 | 860 | conifer | ECM | non-N-fixing | mineral | NA | 97.08 | NA | NA | NA | NA | NA | NA | NA | NA | NA | NA | NA | NA |
| Schulp et al. 2008 | 52.25 | 5.683 | 64 | 9.4 | 860 | conifer | ECM | non-N-fixing | mineral | NA | 62.51 | NA | NA | NA | NA | NA | NA | NA | NA | NA | NA | NA | NA |
| Schulp et al. 2008 | 52.25 | 5.683 | 64 | 9.4 | 860 | conifer | ECM | non-N-fixing | mineral | NA | 69.74 | NA | NA | NA | NA | NA | NA | NA | NA | NA | NA | NA | NA |
| Diaz-Pines et al. 2011 | 40.4 | -4.967 | 1380 | 10.2 | 1068 | boradleaf | ECM | non-N-fixing | mineral | NA | 26.13 | 0.84 | 18.1 | 55.6 | 3.1 | NA | NA | NA | NA | NA | NA | NA | 6.3 |
| Diaz-Pines et al. 2011 | 40.4 | -4.967 | 1380 | 9.9 | 1071 | conifer | ECM | non-N-fixing | mineral | NA | 48.81 | 1.33 | 23 | 184.2 | 7.8 | NA | NA | NA | NA | NA | NA | NA | 5.4 |
| Diaz-Pines et al. 2011 | 40.9 | -3.867 | 1129 | 9.7 | 889 | boradleaf | ECM | non-N-fixing | mineral | NA | 38.14 | 0.89 | 17.7 | 104.5 | 5.8 | NA | NA | NA | NA | NA | NA | NA | 5.7 |
| Diaz-Pines et al. 2011 | 40.9 | -3.867 | 1129 | 9.1 | 948 | conifer | ECM | non-N-fixing | mineral | NA | 54.23 | 1.48 | 24.2 | 159.5 | 6.4 | NA | NA | NA | NA | NA | NA | NA | 5.4 |
| Diaz-Pines et al. 2011 | 40.867 | -4.017 | 1239 | 8.6 | 855 | boradleaf | ECM | non-N-fixing | mineral | NA | 22.6 | 1.08 | 13.1 | 39.3 | 3 | NA | NA | NA | NA | NA | NA | NA | 6.1 |
| Diaz-Pines et al. 2011 | 40.867 | -4.017 | 1239 | 8.9 | 806 | conifer | ECM | non-N-fixing | mineral | NA | 43.49 | 1.12 | 24.5 | 127.9 | 5 | NA | NA | NA | NA | NA | NA | NA | 5.2 |
| Hansson et al. 2011 | 56.667 | 13.05 | 60 | 6.9 | 952 | boradleaf | ECM | non-N-fixing | forest floor | 1.2 | 32.63 | 1.36758 | 23.7634 | NA | NA | NA | NA | NA | NA | NA | NA | NA | NA |
| Hansson et al. 2011 | 56.667 | 13.05 | 60 | 6.9 | 952 | conifer | ECM | non-N-fixing | forest floor | 2.3 | 16.65 | 0.83736 | 20 | NA | NA | NA | NA | NA | NA | NA | NA | NA | NA |
| Hansson et al. 2011 | 56.667 | 13.05 | 60 | 6.9 | 952 | conifer | ECM | non-N-fixing | forest floor | 2 | 5.942 | 0.36528 | 15.3763 | NA | NA | NA | NA | NA | NA | NA | NA | NA | NA |
| Hansson et al. 2011 | 56.667 | 13.05 | 60 | 6.9 | 952 | boradleaf | ECM | non-N-fixing | mineral | NA | 15.94 | 0.88942 | 17.6344 | NA | NA | NA | NA | NA | NA | NA | NA | NA | NA |
| Hansson et al. 2011 | 56.667 | 13.05 | 60 | 6.9 | 952 | conifer | ECM | non-N-fixing | mineral | NA | 13.01 | 0.72991 | 18.0645 | NA | NA | NA | NA | NA | NA | NA | NA | NA | NA |
| Hansson et al. 2011 | 56.667 | 13.05 | 60 | 6.9 | 952 | conifer | ECM | non-N-fixing | mineral | NA | 14.1 | 0.86875 | 16.6667 | NA | NA | NA | NA | NA | NA | NA | NA | NA | NA |
| Wang et al. 2013 | 22.167 | 106.833 | 350 | 21 | 1400 | conifer | ECM | non-N-fixing | mineral | NA | 46.94 | 2.58246 | NA | NA | NA | NA | NA | NA | NA | NA | NA | NA | NA |
| Wang et al. 2013 | 22.167 | 106.833 | 350 | 21 | 1400 | boradleaf | ECM | non-N-fixing | mineral | NA | 49.49 | 3.27018 | NA | NA | NA | NA | NA | NA | NA | NA | NA | NA | NA |
| He et al. 2013 | 22.067 | 106.85 | 267 | 21.8 | 1259 | boradleaf | ECM | non-N-fixing | mineral | NA | 124.6 | NA | NA | 39.72 | NA | NA | NA | NA | NA | NA | NA | NA | NA |
| He et al. 2013 | 22.067 | 106.85 | 267 | 21.8 | 1259 | conifer | ECM | non-N-fixing | mineral | NA | 73.23 | NA | NA | 24.93 | NA | NA | NA | NA | NA | NA | NA | NA | NA |
| Frouz et al. 2013 | 50.233 | 12.683 | 473 | 6.8 | 650 | boradleaf | AM | N-fixing | mineral | NA | 36.9 | NA | NA | 71 | NA | NA | NA | 15.3 | 402 | NA | NA | 79.2 | 6.6 |
| Frouz et al. 2013 | 50.233 | 12.683 | 473 | 6.8 | 650 | boradleaf | ECM | non-N-fixing | mineral | NA | 38 | NA | NA | 95 | NA | NA | NA | 19 | 562 | NA | NA | 76.8 | 6.5 |
| Frouz et al. 2013 | 50.233 | 12.683 | 473 | 6.8 | 650 | boradleaf | ECM | non-N-fixing | mineral | NA | 16.5 | NA | NA | 59 | NA | NA | NA | 11 | 240 | NA | NA | 88.8 | 6.9 |
| Frouz et al. 2013 | 50.233 | 12.683 | 473 | 6.8 | 650 | conifer | ECM | non-N-fixing | mineral | NA | 23.1 | NA | NA | 65 | NA | NA | NA | 14.8 | 237 | NA | NA | 86.4 | 6.9 |
| Wen et al. 2014 | 28.1 | 113.033 | 50 | 17.2 | 1422 | conifer | ECM | non-N-fixing | mineral | NA | 13.67 | 1.5 | 9.18 | 13.95 | 1.53 | NA | NA | NA | 270.103 | 30.103 | NA | NA | 4.4 |
| Wen et al. 2014 | 28.1 | 113.033 | 50 | 17.2 | 1422 | boradleaf | AM | non-N-fixing | mineral | NA | 16.15 | 1.47 | 11.08 | 17 | 1.55 | NA | NA | NA | 340.206 | 65.567 | NA | NA | 4.37 |
| Wen et al. 2014 | 28.1 | 113.033 | 50 | 17.2 | 1422 | conifer | ECM | non-N-fixing | mineral | NA | 21.02 | 1.42 | 15.1 | 13.92 | 0.94 | NA | NA | NA | 140.206 | 23.918 | NA | NA | 4.14 |
| Wen et al. 2014 | 28.1 | 113.033 | 50 | 17.2 | 1422 | boradleaf | AM | non-N-fixing | mineral | NA | 24.38 | 1.94 | 13.25 | 17.93 | 1.43 | NA | NA | NA | 204.124 | 40.825 | NA | NA | 4.51 |
| Wen et al. 2014 | 28.1 | 113.033 | 50 | 17.2 | 1422 | conifer | ECM | non-N-fixing | mineral | NA | 18.29 | 1.48 | 12.41 | 12.36 | 1 | NA | NA | NA | 113.402 | 18.969 | NA | NA | 4.21 |
| Wen et al. 2014 | 28.1 | 113.033 | 50 | 17.2 | 1422 | boradleaf | AM | non-N-fixing | mineral | NA | 26.34 | 1.66 | 17.19 | 19.37 | 1.22 | NA | NA | NA | 195.876 | 39.588 | NA | NA | 4.27 |
| Song et al. 2017 | 41.55 | 118.617 | 886 | 6.5 | 462 | conifer | ECM | non-N-fixing | mineral | NA | 54 | 6.3 | 8.7 | 12 | 1.4 | NA | NA | NA | NA | NA | NA | NA | NA |
| Song et al. 2017 | 41.55 | 118.617 | 886 | 6.5 | 462 | boradleaf | ECM | non-N-fixing | mineral | NA | 52.27 | 5.62 | 8.88 | 12.1 | 1.3 | NA | NA | NA | NA | NA | NA | NA | NA |
| Song et al. 2017 | 41.55 | 118.617 | 886 | 6.5 | 462 | boradleaf | AM | non-N-fixing | mineral | NA | 45.99 | 6.13 | 7.99 | 10.5 | 1.4 | NA | NA | NA | NA | NA | NA | NA | NA |
| Wang et al. 2017 | 45.717 | 126.617 | 145 | 3.6 | 600 | boradleaf | ECM | non-N-fixing | mineral | NA | 35.6 | 1.2 | 30.2 | 13.3 | 0.44 | NA | NA | 5.84 | NA | NA | NA | NA | 6.27 |
| Wang et al. 2017 | 45.717 | 126.617 | 145 | 3.6 | 600 | conifer | ECM | non-N-fixing | mineral | NA | 21.2 | 0.74 | 28.5 | 7.4 | 0.26 | NA | NA | 5.79 | NA | NA | NA | NA | 6.37 |
| Wang et al. 2017 | 45.717 | 126.617 | 145 | 3.6 | 600 | conifer | ECM | non-N-fixing | mineral | NA | 44.7 | 1.52 | 29.5 | 16.5 | 0.56 | NA | NA | 10.95 | NA | NA | NA | NA | 7.11 |
| Wang et al. 2017 | 45.717 | 126.617 | 145 | 3.6 | 600 | conifer | ECM | non-N-fixing | mineral | NA | 37.4 | 1.48 | 25.4 | 13.7 | 0.54 | NA | NA | 8.38 | NA | NA | NA | NA | 6.6 |
| Wang et al. 2017 | 45.717 | 126.617 | 145 | 3.6 | 600 | boradleaf | AM | non-N-fixing | mineral | NA | 44.2 | 1.5 | 29.5 | 16.8 | 0.57 | NA | NA | 6.18 | NA | NA | NA | NA | 7.01 |
| Wang et al. 2017 | 45.717 | 126.617 | 145 | 3.6 | 600 | boradleaf | AM | non-N-fixing | mineral | NA | 34.5 | 1.39 | 24.5 | 12.5 | 0.51 | NA | NA | 2.7 | NA | NA | NA | NA | 6.32 |
| Wang et al. 2017 | 45.717 | 126.617 | 145 | 3.6 | 600 | conifer | ECM | non-N-fixing | mineral | NA | 35.9 | 1.24 | 29.1 | 13.1 | 0.45 | NA | NA | 6.33 | NA | NA | NA | NA | 6.36 |
| Wang et al. 2017 | 45.717 | 126.617 | 145 | 3.6 | 600 | boradleaf | ECM | non-N-fixing | mineral | NA | 28 | 0.92 | 30.5 | 9.76 | 0.32 | NA | NA | 6.1 | NA | NA | NA | NA | 6.6 |
| Wang et al. 2017 | 45.717 | 126.617 | 145 | 3.6 | 600 | boradleaf | AM | non-N-fixing | mineral | NA | 41.4 | 1.54 | 27 | 15.4 | 0.57 | NA | NA | 8.27 | NA | NA | NA | NA | 6.93 |
| Wang et al. 2017 | 45.717 | 126.617 | 145 | 3.6 | 600 | boradleaf | ECM | non-N-fixing | mineral | NA | 31.9 | 1.19 | 27.1 | 11.1 | 0.41 | NA | NA | 5.86 | NA | NA | NA | NA | 6.8 |
| Wang et al. 2017 | 45.717 | 126.617 | 145 | 3.6 | 600 | conifer | ECM | non-N-fixing | mineral | NA | 38.7 | 1.42 | 27.3 | 15 | 0.55 | NA | NA | 4.86 | NA | NA | NA | NA | 6.39 |
| Wang et al. 2017 | 45.717 | 126.617 | 145 | 3.6 | 600 | boradleaf | AM | non-N-fixing | mineral | NA | 44.4 | 1.67 | 26.6 | 16.2 | 0.61 | NA | NA | 8.79 | NA | NA | NA | NA | 6.58 |
| Wang et al. 2017 | 45.717 | 126.617 | 145 | 3.6 | 600 | conifer | ECM | non-N-fixing | mineral | NA | 37.2 | 1.26 | 29.8 | 13.7 | 0.46 | NA | NA | 7.67 | NA | NA | NA | NA | 6.48 |
| Alriksson & Eriksson 1998 | 63.9 | 20.5 | 35 | 2.9 | 662 | conifer | ECM | non-N-fixing | forest floor | NA | NA | NA | 26.7 | 320 | 12 | NA | NA | NA | NA | NA | NA | NA | 5 |
| Alriksson & Eriksson 1998 | 63.9 | 20.5 | 35 | 2.9 | 662 | conifer | ECM | non-N-fixing | forest floor | NA | NA | NA | 29.2 | 350 | 12 | NA | NA | NA | NA | NA | NA | NA | 4.9 |
| Alriksson & Eriksson 1998 | 63.9 | 20.5 | 35 | 2.9 | 662 | conifer | ECM | non-N-fixing | forest floor | NA | NA | NA | 30.7 | 430 | 14 | NA | NA | NA | NA | NA | NA | NA | 4.4 |
| Alriksson & Eriksson 1998 | 63.9 | 20.5 | 35 | 2.9 | 662 | conifer | ECM | non-N-fixing | forest floor | NA | NA | NA | 25.3 | 380 | 15 | NA | NA | NA | NA | NA | NA | NA | 4.5 |
| Alriksson & Eriksson 1998 | 63.9 | 20.5 | 35 | 2.9 | 662 | boradleaf | ECM | non-N-fixing | forest floor | NA | NA | NA | 30.7 | 430 | 14 | NA | NA | NA | NA | NA | NA | NA | 5.2 |
| Alriksson & Eriksson 1998 | 63.9 | 20.5 | 35 | 2.9 | 662 | conifer | ECM | non-N-fixing | mineral | NA | NA | NA | 15.9 | 59 | 3.7 | NA | NA | NA | NA | NA | NA | NA | 4.6 |
| Alriksson & Eriksson 1998 | 63.9 | 20.5 | 35 | 2.9 | 662 | conifer | ECM | non-N-fixing | mineral | NA | NA | NA | 15.1 | 53 | 3.5 | NA | NA | NA | NA | NA | NA | NA | 5 |
| Alriksson & Eriksson 1998 | 63.9 | 20.5 | 35 | 2.9 | 662 | conifer | ECM | non-N-fixing | mineral | NA | NA | NA | 16.5 | 61 | 3.7 | NA | NA | NA | NA | NA | NA | NA | 4.6 |
| Alriksson & Eriksson 1998 | 63.9 | 20.5 | 35 | 2.9 | 662 | conifer | ECM | non-N-fixing | mineral | NA | NA | NA | 15.6 | 53 | 3.4 | NA | NA | NA | NA | NA | NA | NA | 4.7 |
| Alriksson & Eriksson 1998 | 63.9 | 20.5 | 35 | 2.9 | 662 | boradleaf | ECM | non-N-fixing | mineral | NA | NA | NA | 17.6 | 65 | 3.7 | NA | NA | NA | NA | NA | NA | NA | 4.8 |
| Alriksson & Eriksson 1998 | 63.9 | 20.5 | 35 | 2.9 | 662 | conifer | ECM | non-N-fixing | mineral | NA | NA | NA | 16.9 | 9.3 | 0.55 | NA | NA | NA | NA | NA | NA | NA | 5.4 |
| Alriksson & Eriksson 1998 | 63.9 | 20.5 | 35 | 2.9 | 662 | conifer | ECM | non-N-fixing | mineral | NA | NA | NA | 17.2 | 11 | 0.64 | NA | NA | NA | NA | NA | NA | NA | 5.4 |
| Alriksson & Eriksson 1998 | 63.9 | 20.5 | 35 | 2.9 | 662 | conifer | ECM | non-N-fixing | mineral | NA | NA | NA | 15.9 | 14 | 0.88 | NA | NA | NA | NA | NA | NA | NA | 5.3 |
| Alriksson & Eriksson 1998 | 63.9 | 20.5 | 35 | 2.9 | 662 | conifer | ECM | non-N-fixing | mineral | NA | NA | NA | 15.1 | 13 | 0.86 | NA | NA | NA | NA | NA | NA | NA | 5.4 |
| Alriksson & Eriksson 1998 | 63.9 | 20.5 | 35 | 2.9 | 662 | boradleaf | ECM | non-N-fixing | mineral | NA | NA | NA | 15.4 | 6.3 | 0.41 | NA | NA | NA | NA | NA | NA | NA | 5.5 |
| Neirynck et al. 2000 | 50.7 | 4.3 | 149 | 9.9 | 780 | boradleaf | ECM | non-N-fixing | mineral | NA | 16.57 | 1.29 | 12.7 | 41 | 3.19 | NA | NA | NA | NA | NA | NA | NA | 4.45 |
| Neirynck et al. 2000 | 50.7 | 4.3 | 149 | 9.9 | 780 | boradleaf | ECM | non-N-fixing | mineral | NA | 19.51 | 1.56 | 13.5 | 43 | 3.43 | NA | NA | NA | NA | NA | NA | NA | 3.63 |
| Neirynck et al. 2000 | 50.7 | 4.3 | 149 | 9.9 | 780 | boradleaf | ECM | non-N-fixing | mineral | NA | 22.35 | 1.71 | 13.5 | 56 | 4.28 | NA | NA | NA | NA | NA | NA | NA | 3.9 |
| Neirynck et al. 2000 | 50.7 | 4.283 | 149 | 9.9 | 780 | boradleaf | AM | non-N-fixing | mineral | NA | 12.89 | 1.34 | 11.5 | 44 | 4.59 | NA | NA | NA | NA | NA | NA | NA | 3.85 |
| Neirynck et al. 2000 | 50.7 | 4.283 | 149 | 9.9 | 780 | boradleaf | ECM | non-N-fixing | mineral | NA | 20.33 | 1.48 | 14.1 | 53 | 3.87 | NA | NA | NA | NA | NA | NA | NA | 3.65 |
| Neirynck et al. 2000 | 50.7 | 4.267 | 149 | 9.9 | 780 | boradleaf | AM | non-N-fixing | mineral | NA | 21.17 | 2.04 | 11.7 | 52 | 5 | NA | NA | NA | NA | NA | NA | NA | 3.95 |
| Neirynck et al. 2000 | 50.7 | 4.267 | 149 | 9.9 | 780 | boradleaf | ECM | non-N-fixing | mineral | NA | 19.32 | 1.28 | 16.6 | 47 | 3.11 | NA | NA | NA | NA | NA | NA | NA | 3.82 |
| Chodak et al. 2015 | 50.267 | 19.433 | 290 | 8 | 700 | boradleaf | AM | N-fixing | forest floor | NA | NA | NA | 18 | 384 | 21.7 | NA | NA | NA | NA | NA | NA | NA | 3.8 |
| Chodak et al. 2015 | 50.267 | 19.433 | 290 | 8 | 700 | boradleaf | ECM | non-N-fixing | forest floor | NA | NA | NA | 26 | 381 | 14.8 | NA | NA | NA | NA | NA | NA | NA | 4.6 |
| Chodak et al. 2015 | 50.267 | 19.433 | 290 | 8 | 700 | conifer | ECM | non-N-fixing | forest floor | NA | NA | NA | 36 | 338 | 9.4 | NA | NA | NA | NA | NA | NA | NA | 3.2 |
| Chodak et al. 2015 | 50.267 | 19.433 | 290 | 8 | 700 | conifer | ECM | non-N-fixing | forest floor | NA | NA | NA | 38 | 359 | 9.5 | NA | NA | NA | NA | NA | NA | NA | 3.3 |
| Chodak et al. 2015 | 50.267 | 19.433 | 290 | 8 | 700 | boradleaf | AM | N-fixing | mineral | NA | NA | NA | 15 | 16.1 | 1.07 | NA | NA | NA | NA | NA | NA | NA | 3.5 |
| Chodak et al. 2015 | 50.267 | 19.433 | 290 | 8 | 700 | boradleaf | ECM | non-N-fixing | mineral | NA | NA | NA | 24 | 7.9 | 0.33 | NA | NA | NA | NA | NA | NA | NA | 4.1 |
| Chodak et al. 2015 | 50.267 | 19.433 | 290 | 8 | 700 | conifer | ECM | non-N-fixing | mineral | NA | NA | NA | 33 | 5.4 | 0.17 | NA | NA | NA | NA | NA | NA | NA | 3.8 |
| Chodak et al. 2015 | 50.267 | 19.433 | 290 | 8 | 700 | conifer | ECM | non-N-fixing | mineral | NA | NA | NA | 34 | 6.1 | 0.18 | NA | NA | NA | NA | NA | NA | NA | 4.1 |
| Kaur et al. 2000 | 29.983 | 76.85 | 256 | 23.9 | 700 | boradleaf | AM | non-N-fixing | mineral | NA | NA | NA | 8.2 | 6.8 | 0.83 | NA | NA | NA | NA | NA | NA | NA | NA |
| Kaur et al. 2000 | 29.983 | 76.85 | 256 | 23.9 | 700 | boradleaf | ECM | non-N-fixing | mineral | NA | NA | NA | 8.3 | 4.8 | 0.58 | NA | NA | NA | NA | NA | NA | NA | NA |
| Kaur et al. 2000 | 29.983 | 76.85 | 256 | 23.9 | 700 | boradleaf | ECM | non-N-fixing | mineral | NA | NA | NA | 8.3 | 5 | 0.6 | NA | NA | NA | NA | NA | NA | NA | NA |
| Kaur et al. 2000 | 29.983 | 76.85 | 256 | 23.9 | 700 | boradleaf | AM | non-N-fixing | mineral | NA | NA | NA | 8.8 | 5.7 | 0.65 | NA | NA | NA | NA | NA | NA | NA | NA |
| Kaur et al. 2000 | 29.983 | 76.85 | 256 | 23.9 | 700 | boradleaf | ECM | non-N-fixing | mineral | NA | NA | NA | 10.9 | 3.8 | 0.35 | NA | NA | NA | NA | NA | NA | NA | NA |
| Kaur et al. 2000 | 29.983 | 76.85 | 256 | 23.9 | 700 | boradleaf | ECM | non-N-fixing | mineral | NA | NA | NA | 7.1 | 4.2 | 0.59 | NA | NA | NA | NA | NA | NA | NA | NA |
| Kaur et al. 2000 | 29.983 | 76.85 | 256 | 23.9 | 700 | boradleaf | AM | non-N-fixing | mineral | NA | NA | NA | 8.4 | 4.8 | 0.57 | NA | NA | NA | NA | NA | NA | NA | NA |
| Kaur et al. 2000 | 29.983 | 76.85 | 256 | 23.9 | 700 | boradleaf | ECM | non-N-fixing | mineral | NA | NA | NA | 11.2 | 2.9 | 0.26 | NA | NA | NA | NA | NA | NA | NA | NA |
| Kaur et al. 2000 | 29.983 | 76.85 | 256 | 23.9 | 700 | boradleaf | ECM | non-N-fixing | mineral | NA | NA | NA | 10 | 3.2 | 0.32 | NA | NA | NA | NA | NA | NA | NA | NA |
| Galicia & Garcia-Oliva 2004 | 19.5 | -105.083 | 106 | 24 | 746 | boradleaf | AM | non-N-fixing | mineral | NA | NA | NA | 8.6 | 31.74 | 3.67 | NA | NA | 740 | 830 | 86 | NA | NA | NA |
| Galicia & Garcia-Oliva 2004 | 19.5 | -105.083 | 106 | 24 | 746 | boradleaf | AM | non-N-fixing | mineral | NA | NA | NA | 11.3 | 35.42 | 3.14 | NA | NA | 530 | 330 | 87 | NA | NA | NA |
| Gartzia-Bengoetxea et al. 2016 | 43.167 | -2.85 | 410 | 14 | 1200 | boradleaf | ECM | non-N-fixing | forest floor | NA | NA | NA | 38.79 | 490.2 | 12.9 | NA | NA | 650 | NA | NA | NA | NA | NA |
| Gartzia-Bengoetxea et al. 2016 | 43.167 | -2.85 | 410 | 14 | 1200 | boradleaf | ECM | non-N-fixing | forest floor | NA | NA | NA | 35.31 | 494.4 | 14.4 | NA | NA | 690 | NA | NA | NA | NA | NA |
| Gartzia-Bengoetxea et al. 2016 | 43.167 | -2.85 | 410 | 14 | 1200 | conifer | ECM | non-N-fixing | forest floor | NA | NA | NA | 40.81 | 493.4 | 12.3 | NA | NA | 660 | NA | NA | NA | NA | NA |
| Gartzia-Bengoetxea et al. 2016 | 43.167 | -2.85 | 410 | 14 | 1200 | boradleaf | ECM | non-N-fixing | forest floor | NA | NA | NA | 38.4 | 512.6 | 13.6 | NA | NA | 560 | NA | NA | NA | NA | NA |
| Gartzia-Bengoetxea et al. 2016 | 43.167 | -2.85 | 410 | 14 | 1200 | boradleaf | ECM | non-N-fixing | mineral | NA | NA | NA | 8.6 | 17.2 | 2 | NA | NA | 1.2 | NA | NA | NA | NA | 5.4 |
| Gartzia-Bengoetxea et al. 2016 | 43.167 | -2.85 | 410 | 14 | 1200 | boradleaf | ECM | non-N-fixing | mineral | NA | NA | NA | 10.3 | 41.2 | 4 | NA | NA | 1.7 | NA | NA | NA | NA | 4.6 |
| Gartzia-Bengoetxea et al. 2016 | 43.167 | -2.85 | 410 | 14 | 1200 | conifer | ECM | non-N-fixing | mineral | NA | NA | NA | 10.5 | 21 | 2 | NA | NA | 0.6 | NA | NA | NA | NA | 4.5 |
| Gartzia-Bengoetxea et al. 2016 | 43.167 | -2.85 | 410 | 14 | 1200 | boradleaf | ECM | non-N-fixing | mineral | NA | NA | NA | 11.4 | 68.4 | 6 | NA | NA | 3.9 | NA | NA | NA | NA | 6.2 |
| Yu et al. 2008 | 42.717 | 122.367 | 246 | 6.2 | 450 | conifer | ECM | non-N-fixing | mineral | NA | NA | NA | 1.62 | 3.9 | 0.395 | 0.51 | 0.74 | 101 | 82.09 | 10.59 | NA | NA | 6.37 |
| Yu et al. 2008 | 42.717 | 122.367 | 246 | 6.2 | 450 | conifer | ECM | non-N-fixing | mineral | NA | NA | NA | 1.65 | 3.97 | 0.415 | 0.61 | 0.93 | 111 | 82.04 | 9.86 | NA | NA | 6.67 |
| Yu et al. 2008 | 42.717 | 122.367 | 246 | 6.2 | 450 | conifer | ECM | non-N-fixing | mineral | NA | NA | NA | 1.68 | 4.73 | 0.327 | 0.57 | 0.47 | 82 | 73.77 | 8.94 | NA | NA | 6.62 |
| Yang et al. 2018 | 46.267 | 131.35 | 283 | 3.4 | 620 | conifer | ECM | non-N-fixing | mineral | NA | NA | NA | 5.38125 | 16.267 | 2.983 | NA | NA | 170.523 | NA | NA | NA | NA | 5.37 |
| Yang et al. 2018 | 46.267 | 131.35 | 283 | 3.4 | 620 | conifer | ECM | non-N-fixing | mineral | NA | NA | NA | 11.2875 | 37.158 | 3.311 | NA | NA | 291.779 | NA | NA | NA | NA | 6.2 |
| Yang et al. 2018 | 46.267 | 131.35 | 283 | 3.4 | 620 | conifer | ECM | non-N-fixing | mineral | NA | NA | NA | 6.16875 | 18.664 | 3.033 | NA | NA | 290.37 | NA | NA | NA | NA | 5.57 |
| Yang et al. 2018 | 46.267 | 131.35 | 283 | 3.4 | 620 | conifer | ECM | non-N-fixing | mineral | NA | NA | NA | 9.23125 | 20.206 | 2.194 | NA | NA | 189.132 | NA | NA | NA | NA | 5.66 |
| Wang et al. 2016 | 25.05 | 102.767 | 1995 | 15 | 979 | conifer | ECM | non-N-fixing | mineral | NA | 10.67 | NA | NA | NA | NA | NA | NA | NA | NA | NA | NA | NA | NA |
| Wang et al. 2016 | 25.05 | 102.767 | 1995 | 15 | 979 | boradleaf | ECM | non-N-fixing | mineral | NA | 12.46 | NA | NA | NA | NA | NA | NA | NA | NA | NA | NA | NA | NA |
| Wotherspoon et al. 2014 | 43.533 | -80.2 | 332 | 7.2 | 833 | boradleaf | ECM | non-N-fixing | mineral | NA | NA | NA | NA | 19.1 | NA | NA | NA | NA | NA | NA | NA | NA | NA |
| Wotherspoon et al. 2014 | 43.533 | -80.2 | 332 | 7.2 | 833 | conifer | ECM | non-N-fixing | mineral | NA | NA | NA | NA | 19 | NA | NA | NA | NA | NA | NA | NA | NA | NA |
| Wotherspoon et al. 2014 | 43.533 | -80.2 | 332 | 7.2 | 833 | boradleaf | ECM | non-N-fixing | mineral | NA | NA | NA | NA | 18.7 | NA | NA | NA | NA | NA | NA | NA | NA | NA |
| Wotherspoon et al. 2014 | 43.533 | -80.2 | 332 | 7.2 | 833 | boradleaf | AM | non-N-fixing | mineral | NA | NA | NA | NA | 17.7 | NA | NA | NA | NA | NA | NA | NA | NA | NA |
| Wotherspoon et al. 2014 | 43.533 | -80.2 | 332 | 7.2 | 833 | conifer | AM | non-N-fixing | mineral | NA | NA | NA | NA | 20.4 | NA | NA | NA | NA | NA | NA | NA | NA | NA |
| Frouz et al. 2009 | 50.233 | 12.683 | 469 | 6.8 | 650 | boradleaf | AM | N-fixing | forest floor | NA | NA | NA | NA | 290 | NA | NA | NA | NA | NA | NA | NA | NA | NA |
| Frouz et al. 2009 | 50.233 | 12.683 | 469 | 6.8 | 650 | boradleaf | ECM | non-N-fixing | forest floor | NA | NA | NA | NA | 236 | NA | NA | NA | NA | NA | NA | NA | NA | NA |
| Frouz et al. 2009 | 50.233 | 12.683 | 469 | 6.8 | 650 | boradleaf | ECM | non-N-fixing | forest floor | NA | NA | NA | NA | 211 | NA | NA | NA | NA | NA | NA | NA | NA | NA |
| Frouz et al. 2009 | 50.233 | 12.683 | 469 | 6.8 | 650 | conifer | ECM | non-N-fixing | forest floor | NA | NA | NA | NA | 308 | NA | NA | NA | NA | NA | NA | NA | NA | NA |
| Frouz et al. 2009 | 50.233 | 12.683 | 469 | 6.8 | 650 | boradleaf | AM | N-fixing | mineral | NA | 27.69 | NA | NA | 71 | NA | NA | NA | NA | NA | NA | NA | NA | NA |
| Frouz et al. 2009 | 50.233 | 12.683 | 469 | 6.8 | 650 | boradleaf | ECM | non-N-fixing | mineral | NA | 34.58 | NA | NA | 95 | NA | NA | NA | NA | NA | NA | NA | NA | NA |
| Frouz et al. 2009 | 50.233 | 12.683 | 469 | 6.8 | 650 | boradleaf | ECM | non-N-fixing | mineral | NA | 25.66 | NA | NA | 67 | NA | NA | NA | NA | NA | NA | NA | NA | NA |
| Frouz et al. 2009 | 50.233 | 12.683 | 469 | 6.8 | 650 | conifer | ECM | non-N-fixing | mineral | NA | 15.08 | NA | NA | 42 | NA | NA | NA | NA | NA | NA | NA | NA | NA |
| Mueller et al. 2012 | 51.233 | 18.1 | 179 | 8.2 | 573 | conifer | ECM | non-N-fixing | forest floor | NA | 10.24 | NA | NA | 8.3 | NA | NA | NA | NA | NA | NA | NA | NA | NA |
| Mueller et al. 2012 | 51.233 | 18.1 | 179 | 8.2 | 573 | boradleaf | AM | non-N-fixing | forest floor | NA | 1.429 | NA | NA | 8.8 | NA | NA | NA | NA | NA | NA | NA | NA | NA |
| Mueller et al. 2012 | 51.233 | 18.1 | 179 | 8.2 | 573 | boradleaf | AM | non-N-fixing | forest floor | NA | 2.381 | NA | NA | 11.9 | NA | NA | NA | NA | NA | NA | NA | NA | NA |
| Mueller et al. 2012 | 51.233 | 18.1 | 179 | 8.2 | 573 | boradleaf | ECM | non-N-fixing | forest floor | NA | 8.81 | NA | NA | 10.9 | NA | NA | NA | NA | NA | NA | NA | NA | NA |
| Mueller et al. 2012 | 51.233 | 18.1 | 179 | 8.2 | 573 | boradleaf | ECM | non-N-fixing | forest floor | NA | 20.24 | NA | NA | 12.3 | NA | NA | NA | NA | NA | NA | NA | NA | NA |
| Mueller et al. 2012 | 51.233 | 18.1 | 179 | 8.2 | 573 | boradleaf | ECM | non-N-fixing | forest floor | NA | 8.571 | NA | NA | 10.9 | NA | NA | NA | NA | NA | NA | NA | NA | NA |
| Mueller et al. 2012 | 51.233 | 18.1 | 179 | 8.2 | 573 | conifer | ECM | non-N-fixing | forest floor | NA | 13.33 | NA | NA | 8.2 | NA | NA | NA | NA | NA | NA | NA | NA | NA |
| Mueller et al. 2012 | 51.233 | 18.1 | 179 | 8.2 | 573 | conifer | ECM | non-N-fixing | forest floor | NA | 17.62 | NA | NA | 16.7 | NA | NA | NA | NA | NA | NA | NA | NA | NA |
| Mueller et al. 2012 | 51.233 | 18.1 | 179 | 8.2 | 573 | conifer | ECM | non-N-fixing | forest floor | NA | 24.52 | NA | NA | 7.7 | NA | NA | NA | NA | NA | NA | NA | NA | NA |
| Mueller et al. 2012 | 51.233 | 18.1 | 179 | 8.2 | 573 | conifer | ECM | non-N-fixing | forest floor | NA | 20.48 | NA | NA | 10.4 | NA | NA | NA | NA | NA | NA | NA | NA | NA |
| Mueller et al. 2012 | 51.233 | 18.1 | 179 | 8.2 | 573 | conifer | ECM | non-N-fixing | forest floor | NA | 11.19 | NA | NA | 13.4 | NA | NA | NA | NA | NA | NA | NA | NA | NA |
| Mueller et al. 2012 | 51.233 | 18.1 | 179 | 8.2 | 573 | boradleaf | ECM | non-N-fixing | forest floor | NA | 7.857 | NA | NA | 11 | NA | NA | NA | NA | NA | NA | NA | NA | NA |
| Mueller et al. 2012 | 51.233 | 18.1 | 179 | 8.2 | 573 | boradleaf | ECM | non-N-fixing | forest floor | NA | 22.62 | NA | NA | 8.5 | NA | NA | NA | NA | NA | NA | NA | NA | NA |
| Mueller et al. 2012 | 51.233 | 18.1 | 179 | 8.2 | 573 | boradleaf | ECM | non-N-fixing | forest floor | NA | 2.381 | NA | NA | 7.8 | NA | NA | NA | NA | NA | NA | NA | NA | NA |
| Mueller et al. 2012 | 51.233 | 18.1 | 179 | 8.2 | 573 | conifer | ECM | non-N-fixing | mineral | NA | 19.9 | NA | NA | NA | NA | NA | NA | NA | NA | NA | NA | NA | NA |
| Mueller et al. 2012 | 51.233 | 18.1 | 179 | 8.2 | 573 | boradleaf | AM | non-N-fixing | mineral | NA | 14.2 | NA | NA | NA | NA | NA | NA | NA | NA | NA | NA | NA | NA |
| Mueller et al. 2012 | 51.233 | 18.1 | 179 | 8.2 | 573 | boradleaf | AM | non-N-fixing | mineral | NA | 27 | NA | NA | NA | NA | NA | NA | NA | NA | NA | NA | NA | NA |
| Mueller et al. 2012 | 51.233 | 18.1 | 179 | 8.2 | 573 | boradleaf | ECM | non-N-fixing | mineral | NA | 23.4 | NA | NA | NA | NA | NA | NA | NA | NA | NA | NA | NA | NA |
| Mueller et al. 2012 | 51.233 | 18.1 | 179 | 8.2 | 573 | boradleaf | ECM | non-N-fixing | mineral | NA | 25.5 | NA | NA | NA | NA | NA | NA | NA | NA | NA | NA | NA | NA |
| Mueller et al. 2012 | 51.233 | 18.1 | 179 | 8.2 | 573 | boradleaf | ECM | non-N-fixing | mineral | NA | 25.1 | NA | NA | NA | NA | NA | NA | NA | NA | NA | NA | NA | NA |
| Mueller et al. 2012 | 51.233 | 18.1 | 179 | 8.2 | 573 | conifer | ECM | non-N-fixing | mineral | NA | 19.7 | NA | NA | NA | NA | NA | NA | NA | NA | NA | NA | NA | NA |
| Mueller et al. 2012 | 51.233 | 18.1 | 179 | 8.2 | 573 | conifer | ECM | non-N-fixing | mineral | NA | 28.8 | NA | NA | NA | NA | NA | NA | NA | NA | NA | NA | NA | NA |
| Mueller et al. 2012 | 51.233 | 18.1 | 179 | 8.2 | 573 | conifer | ECM | non-N-fixing | mineral | NA | 18.5 | NA | NA | NA | NA | NA | NA | NA | NA | NA | NA | NA | NA |
| Mueller et al. 2012 | 51.233 | 18.1 | 179 | 8.2 | 573 | conifer | ECM | non-N-fixing | mineral | NA | 24.6 | NA | NA | NA | NA | NA | NA | NA | NA | NA | NA | NA | NA |
| Mueller et al. 2012 | 51.233 | 18.1 | 179 | 8.2 | 573 | conifer | ECM | non-N-fixing | mineral | NA | 25.1 | NA | NA | NA | NA | NA | NA | NA | NA | NA | NA | NA | NA |
| Mueller et al. 2012 | 51.233 | 18.1 | 179 | 8.2 | 573 | boradleaf | ECM | non-N-fixing | mineral | NA | 24.8 | NA | NA | NA | NA | NA | NA | NA | NA | NA | NA | NA | NA |
| Mueller et al. 2012 | 51.233 | 18.1 | 179 | 8.2 | 573 | boradleaf | ECM | non-N-fixing | mineral | NA | 20.5 | NA | NA | NA | NA | NA | NA | NA | NA | NA | NA | NA | NA |
| Mueller et al. 2012 | 51.233 | 18.1 | 179 | 8.2 | 573 | boradleaf | ECM | non-N-fixing | mineral | NA | 16.8 | NA | NA | NA | NA | NA | NA | NA | NA | NA | NA | NA | NA |
| Wei et al. 2010 | 38.817 | 110.383 | 1120 | 8.4 | 437 | conifer | ECM | non-N-fixing | mineral | NA | 2.826 | 0.46955 | 6 | 2.03 | 0.34 | NA | NA | NA | NA | NA | NA | NA | NA |
| Wei et al. 2010 | 38.817 | 110.383 | 1120 | 8.4 | 437 | boradleaf | ECM | non-N-fixing | mineral | NA | 3.547 | 0.69862 | 5.1 | 2.58 | 0.51 | NA | NA | NA | NA | NA | NA | NA | NA |
| Wei et al. 2010 | 38.817 | 110.383 | 1120 | 8.4 | 437 | conifer | ECM | non-N-fixing | mineral | NA | 1.878 | 0.33159 | 5.7 | 1.32 | 0.24 | NA | NA | NA | NA | NA | NA | NA | NA |
| Wei et al. 2010 | 38.817 | 110.383 | 1120 | 8.4 | 437 | boradleaf | ECM | non-N-fixing | mineral | NA | 2.386 | 0.43912 | 5.4 | 1.77 | 0.33 | NA | NA | NA | NA | NA | NA | NA | NA |
| Wei et al. 2010 | 38.817 | 110.383 | 1120 | 8.4 | 437 | conifer | ECM | non-N-fixing | mineral | NA | 2.172 | 0.46531 | 4.7 | 0.75 | 0.17 | NA | NA | NA | NA | NA | NA | NA | NA |
| Wei et al. 2010 | 38.817 | 110.383 | 1120 | 8.4 | 437 | boradleaf | ECM | non-N-fixing | mineral | NA | 2.511 | 0.58979 | 4.3 | 0.93 | 0.22 | NA | NA | NA | NA | NA | NA | NA | NA |
| Wei et al. 2010 | 38.817 | 110.383 | 1120 | 8.4 | 437 | conifer | ECM | non-N-fixing | mineral | NA | 1.59 | 0.41274 | 3.9 | 0.56 | 0.15 | NA | NA | NA | NA | NA | NA | NA | NA |
| Wei et al. 2010 | 38.817 | 110.383 | 1120 | 8.4 | 437 | boradleaf | ECM | non-N-fixing | mineral | NA | 1.364 | 0.39025 | 3.5 | 0.5 | 0.15 | NA | NA | NA | NA | NA | NA | NA | NA |
| Langenbruch et al. 2012 | 51.1 | 10.517 | 288 | 7.5 | 670 | boradlea f | ECM | non-N-fixing | forest floor | NA | 0.42 | 0.016 | 26.4 | NA | NA | NA | NA | NA | NA | NA | NA | NA | NA |
| Langenbruch et al. 2012 | 51.1 | 10.517 | 288 | 7.5 | 670 | boradlea f | AM | non-N-fixing | forest floor | NA | 0.16 | 0.007 | 24.1 | NA | NA | NA | NA | NA | NA | NA | NA | NA | NA |
| Langenbruch et al. 2012 | 51.1 | 10.517 | 288 | 7.5 | 670 | boradlea f | ECM | non-N-fixing | forest floor | NA | 0.2 | 0.009 | 21.7 | NA | NA | NA | NA | NA | NA | NA | NA | NA | NA |
| Langenbruch et al. 2012 | 51.1 | 10.517 | 288 | 7.5 | 670 | boradlea f | ECM | non-N-fixing | mineral | NA | 30.7 | 2.2 | 13.9 | NA | NA | NA | NA | NA | NA | NA | NA | NA | NA |
| Langenbruch et al. 2012 | 51.1 | 10.517 | 288 | 7.5 | 670 | boradlea f | AM | non-N-fixing | mineral | NA | 37.1 | 2.7 | 13.5 | NA | NA | NA | NA | NA | NA | NA | NA | NA | NA |
| Langenbruch et al. 2012 | 51.1 | 10.517 | 288 | 7.5 | 670 | boradlea f | ECM | non-N-fixing | mineral | NA | 27.1 | 2.1 | 13.1 | NA | NA | NA | NA | NA | NA | NA | NA | NA | NA |
| Wang & Zhong 2016 | 26.833 | 109.6 | 392 | 16.5 | 1200 | boradlea f | AM | non-N-fixing | mineral | NA | NA | NA | 12.3 | 24.7 | 2.01 | NA | NA | NA | NA | NA | NA | NA | NA |
| Wang & Zhong 2016 | 26.833 | 109.6 | 392 | 16.5 | 1200 | boradlea f | AM | non-N-fixing | mineral | NA | NA | NA | 11.8 | 22 | 1.87 | NA | NA | NA | NA | NA | NA | NA | NA |
| Wang & Zhong 2016 | 26.833 | 109.6 | 392 | 16.5 | 1200 | conifer | AM | non-N-fixing | mineral | NA | NA | NA | 13.2 | 23.3 | 1.77 | NA | NA | NA | NA | NA | NA | NA | NA |
| Wang & Zhong 2016 | 26.833 | 109.6 | 392 | 16.5 | 1200 | conifer | ECM | non-N-fixing | mineral | NA | NA | NA | 12.8 | 20.4 | 1.59 | NA | NA | NA | NA | NA | NA | NA | NA |
| Andivia et al. 2016 | 49.317 | 16.717 | 505 | 7.2 | 592 | boradlea f | ECM | non-N-fixing | forest floor | NA | 6.79 | NA | 22.29 | 329.8 | 16.6 | NA | NA | NA | NA | NA | NA | NA | NA |
| Andivia et al. 2016 | 49.317 | 16.717 | 505 | 7.2 | 592 | conifer | ECM | non-N-fixing | forest floor | NA | 26.25 | NA | 23.65 | 354.8 | 14.2 | NA | NA | NA | NA | NA | NA | NA | NA |
| Andivia et al. 2016 | 49.317 | 16.717 | 505 | 7.2 | 592 | boradlea f | ECM | non-N-fixing | mineral | NA | 41.73 | 1.43 | 21.94 | 27.9 | 1.3 | NA | NA | NA | NA | NA | NA | NA | NA |
| Andivia et al. 2016 | 49.317 | 16.717 | 505 | 7.2 | 592 | conifer | ECM | non-N-fixing | mineral | NA | 41.93 | 1.32 | 21.73 | 26.1 | 1.2 | NA | NA | NA | NA | NA | NA | NA | NA |
| Wang et al. 2015 | 45.4 | 127.47 | 486 | 2.8 | 629 | boradlea f | ECM | non-N-fixing | mineral | NA | 93.06 | 15.77 | NA | 19.885 | 3.37 | NA | NA | NA | NA | NA | NA | NA | 4.64 |
| Wang et al. 2015 | 45.4 | 127.47 | 486 | 2.8 | 629 | boradlea f | AM | non-N-fixing | mineral | NA | 87.76 | 14.73 | NA | 21.096 | 3.54 | NA | NA | NA | NA | NA | NA | NA | 4.77 |
| Wang et al. 2015 | 45.4 | 127.47 | 486 | 2.8 | 629 | boradlea f | AM | non-N-fixing | mineral | NA | 94.46 | 16.12 | NA | 21.274 | 3.63 | NA | NA | NA | NA | NA | NA | NA | 4.75 |
| Wang et al. 2015 | 45.4 | 127.47 | 486 | 2.8 | 629 | conifer | ECM | non-N-fixing | mineral | NA | 79.08 | 14.16 | NA | 17.651 | 3.16 | NA | NA | NA | NA | NA | NA | NA | 4.69 |
| Wang et al. 2015 | 45.4 | 127.47 | 486 | 2.8 | 629 | conifer | ECM | non-N-fixing | mineral | NA | 84.76 | 14.69 | NA | 19.62 | 3.4 | NA | NA | NA | NA | NA | NA | NA | 4.79 |
| Wang et al. 2015 | 45.4 | 127.47 | 486 | 2.8 | 629 | boradlea f | ECM | non-N-fixing | mineral | NA | 23.88 | NA | NA | 25.402 | NA | NA | NA | NA | 281.984 | NA | NA | NA | NA |
| Wang et al. 2015 | 45.4 | 127.47 | 486 | 2.8 | 629 | boradlea f | AM | non-N-fixing | mineral | NA | 26.48 | NA | NA | 27.586 | NA | NA | NA | NA | 484.073 | NA | NA | NA | NA |
| Wang et al. 2015 | 45.4 | 127.47 | 486 | 2.8 | 629 | boradlea f | AM | non-N-fixing | mineral | NA | 26.15 | NA | NA | 27.241 | NA | NA | NA | NA | 476.24 | NA | NA | NA | NA |
| Wang et al. 2015 | 45.4 | 127.47 | 486 | 2.8 | 629 | conifer | ECM | non-N-fixing | mineral | NA | 23.38 | NA | NA | 25.977 | NA | NA | NA | NA | 283.551 | NA | NA | NA | NA |
| Wang et al. 2015 | 45.4 | 127.47 | 486 | 2.8 | 629 | conifer | ECM | non-N-fixing | mineral | NA | 24.47 | NA | NA | 24.713 | NA | NA | NA | NA | 338.381 | NA | NA | NA | NA |
| Finzi et al. 1998 | 42 | -73.25 | 277 | 7.3 | 1218 | boradleaf | AM | non-N-fixing | forest floor | NA | 81 | 5.55 | 15 | NA | NA | NA | NA | NA | NA | NA | NA | NA | NA |
| Finzi et al. 1998 | 42 | -73.25 | 277 | 7.3 | 1218 | boradleaf | AM | non-N-fixing | forest floor | NA | 81 | 5.632 | 14.8 | NA | NA | NA | NA | NA | NA | NA | NA | NA | NA |
| Finzi et al. 1998 | 42 | -73.25 | 277 | 7.3 | 1218 | boradleaf | AM | non-N-fixing | forest floor | NA | 87 | 5.896 | 15.3 | NA | NA | NA | NA | NA | NA | NA | NA | NA | NA |
| Finzi et al. 1998 | 42 | -73.25 | 277 | 7.3 | 1218 | boradleaf | ECM | non-N-fixing | forest floor | NA | 82 | 4.629 | 18.5 | NA | NA | NA | NA | NA | NA | NA | NA | NA | NA |
| Finzi et al. 1998 | 42 | -73.25 | 277 | 7.3 | 1218 | boradleaf | ECM | non-N-fixing | forest floor | NA | 94 | 5.077 | 19.2 | NA | NA | NA | NA | NA | NA | NA | NA | NA | NA |
| Finzi et al. 1998 | 42 | -73.25 | 277 | 7.3 | 1218 | conifer | ECM | non-N-fixing | forest floor | NA | 108 | 5.776 | 19.5 | NA | NA | NA | NA | NA | NA | NA | NA | NA | NA |
| Finzi et al. 1998 | 42 | -73.25 | 277 | 7.3 | 1218 | boradleaf | AM | non-N-fixing | mineral | NA | 37.31 | NA | 14.1 | NA | NA | NA | NA | NA | NA | NA | NA | NA | NA |
| Finzi et al. 1998 | 42 | -73.25 | 277 | 7.3 | 1218 | boradleaf | AM | non-N-fixing | mineral | NA | 36.43 | NA | 13.9 | NA | NA | NA | NA | NA | NA | NA | NA | NA | NA |
| Finzi et al. 1998 | 42 | -73.25 | 277 | 7.3 | 1218 | boradleaf | AM | non-N-fixing | mineral | NA | 39 | NA | 17.1 | NA | NA | NA | NA | NA | NA | NA | NA | NA | NA |
| Finzi et al. 1998 | 42 | -73.25 | 277 | 7.3 | 1218 | boradleaf | ECM | non-N-fixing | mineral | NA | 35.51 | NA | 17.2 | NA | NA | NA | NA | NA | NA | NA | NA | NA | NA |
| Finzi et al. 1998 | 42 | -73.25 | 277 | 7.3 | 1218 | boradleaf | ECM | non-N-fixing | mineral | NA | 37.58 | NA | 19.4 | NA | NA | NA | NA | NA | NA | NA | NA | NA | NA |
| Finzi et al. 1998 | 42 | -73.25 | 277 | 7.3 | 1218 | conifer | ECM | non-N-fixing | mineral | NA | 39.48 | NA | 19.4 | NA | NA | NA | NA | NA | NA | NA | NA | NA | NA |
| Lu et al. 2017 | -25.933 | 153.083 | 25 | 21.1 | 1445 | conifer | ECM | non-N-fixing | mineral | NA | 19.9 | 0.55 | 36.1 | 19.7 | 0.54 | NA | NA | 25.72 | 116.29 | 15.07 | NA | NA | 4.51 |
| Lu et al. 2017 | -25.933 | 153.083 | 25 | 21.1 | 1445 | conifer | AM | non-N-fixing | mineral | NA | 12.97 | 0.6 | 21.68 | 10.9 | 0.5 | NA | NA | 28.94 | 62.5 | 8.49 | NA | NA | 6.03 |
| Lu et al. 2017 | -25.933 | 153.083 | 25 | 21.1 | 1445 | conifer | AM | non-N-fixing | mineral | NA | 11.48 | 0.6 | 18.92 | 9.9 | 0.52 | NA | NA | 31.35 | 147.63 | 24.59 | NA | NA | 6.45 |
| Ndlovu et al. 2013 | 0 | 38 | 684 | 23.3 | 658 | boradleaf | AM | non-N-fixing | mineral | NA | NA | NA | NA | 2.034 | NA | NA | NA | NA | NA | NA | NA | NA | 5.99 |
| Ndlovu et al. 2013 | 0 | 38 | 684 | 23.3 | 658 | boradleaf | AM | non-N-fixing | mineral | NA | NA | NA | NA | 2.6827 | NA | NA | NA | NA | NA | NA | NA | NA | 6.41 |
| Ndlovu et al. 2013 | 0 | 38 | 684 | 23.3 | 658 | boradleaf | AM | non-N-fixing | mineral | NA | NA | NA | NA | 2.8527 | NA | NA | NA | NA | NA | NA | NA | NA | 6.43 |
| Ndlovu et al. 2013 | -0.07 | 37.18 | 2470 | 14 | 1176 | boradleaf | AM | non-N-fixing | mineral | NA | NA | NA | NA | 2.4533 | NA | NA | NA | NA | NA | NA | NA | NA | 6.1 |
| Ndlovu et al. 2013 | -0.07 | 37.18 | 2470 | 14 | 1176 | boradleaf | AM | non-N-fixing | mineral | NA | NA | NA | NA | 2.9077 | NA | NA | NA | NA | NA | NA | NA | NA | 6.38 |
| Ndlovu et al. 2013 | -0.07 | 37.18 | 2470 | 14 | 1176 | boradleaf | AM | non-N-fixing | mineral | NA | NA | NA | NA | 2.561 | NA | NA | NA | NA | NA | NA | NA | NA | 6.31 |
| Ndlovu et al. 2013 | 0.05 | 37.65 | 1595 | 18.7 | 1709 | boradleaf | AM | non-N-fixing | mineral | NA | NA | NA | NA | 2.5258 | NA | NA | NA | NA | NA | NA | NA | NA | 5.65 |
| Ndlovu et al. 2013 | 0.05 | 37.65 | 1595 | 18.7 | 1709 | boradleaf | AM | non-N-fixing | mineral | NA | NA | NA | NA | 2.9697 | NA | NA | NA | NA | NA | NA | NA | NA | 6.18 |
| Ndlovu et al. 2013 | 0.05 | 37.65 | 1595 | 18.7 | 1709 | boradleaf | AM | N-fixing | mineral | NA | NA | NA | NA | 2.9204 | NA | NA | NA | NA | NA | NA | NA | NA | 6.16 |
| Ndlovu et al. 2013 | -0.26 | 37.28 | 3466 | 8.1 | 1734 | boradleaf | AM | non-N-fixing | mineral | NA | NA | NA | NA | 2.3149 | NA | NA | NA | NA | NA | NA | NA | NA | 6.69 |
| Ndlovu et al. 2013 | -0.26 | 37.28 | 3466 | 8.1 | 1734 | boradleaf | AM | non-N-fixing | mineral | NA | NA | NA | NA | 2.1419 | NA | NA | NA | NA | NA | NA | NA | NA | 6.73 |
| Ndlovu et al. 2013 | -0.26 | 37.28 | 3466 | 8.1 | 1734 | boradleaf | AM | non-N-fixing | mineral | NA | NA | NA | NA | 2.0417 | NA | NA | NA | NA | NA | NA | NA | NA | 6.67 |
| Aponte et al. 2014 | 36.517 | 37.083 | 478 | 15.5 | 1117 | boradleaf | ECM | non-N-fixing | mineral | NA | NA | NA | NA | 39 | 2.8 | 8.1 | 2.5 | NA | 378 | 46 | 7.9 | NA | NA |
| Aponte et al. 2014 | 36.517 | 37.083 | 478 | 15.5 | 1117 | conifer | ECM | non-N-fixing | mineral | NA | NA | NA | NA | 30 | 2.2 | 19 | 7.9 | NA | 63 | 6.9 | 4 | NA | NA |
| Zhou et al. 2017 | -25.933 | 153.083 | 26 | 21.1 | 1287 | conifer | ECM | non-N-fixing | mineral | NA | NA | NA | 31.8 | 13.81 | 0.44 | NA | NA | NA | NA | NA | NA | NA | 4.58 |
| Zhou et al. 2017 | -25.933 | 153.083 | 26 | 21.1 | 1287 | conifer | AM | non-N-fixing | mineral | NA | NA | NA | 23.1 | 10.13 | 0.43 | NA | NA | NA | NA | NA | NA | NA | 5.64 |
| Zhou et al. 2017 | -25.933 | 153.083 | 26 | 21.1 | 1287 | conifer | AM | non-N-fixing | mineral | NA | NA | NA | 21.2 | 8.92 | 0.42 | NA | NA | NA | NA | NA | NA | NA | 6.01 |
| Zhou et al. 2017 | -25.933 | 153.083 | 26 | 21.1 | 1287 | boradleaf | ECM | non-N-fixing | mineral | NA | NA | NA | 29.8 | 26.11 | 0.87 | NA | NA | NA | NA | NA | NA | NA | 4.49 |
| Chaiyasen et al. 2017 | 16.033 | 100.783 | 82 | 27.8 | 1213 | boradleaf | AM | non-N-fixing | mineral | NA | NA | NA | NA | 29.6 | NA | NA | NA | 100.33 | NA | NA | NA | NA | 5.56 |
| Chaiyasen et al. 2017 | 16.2 | 100.967 | 199 | 26.9 | 1226 | boradleaf | AM | non-N-fixing | mineral | NA | NA | NA | NA | 28.2 | NA | NA | NA | 57 | NA | NA | NA | NA | 6.36 |
| Chaiyasen et al. 2017 | 14.267 | 101.117 | 9 | 27.7 | 1342 | boradleaf | AM | non-N-fixing | mineral | NA | NA | NA | NA | 36.3 | NA | NA | NA | 368.67 | NA | NA | NA | NA | 6.68 |
| Chaiyasen et al. 2017 | 18.967 | 99.25 | 567 | 25.1 | 1120 | boradleaf | AM | non-N-fixing | mineral | NA | NA | NA | NA | 39.6 | NA | NA | NA | 128.67 | NA | NA | NA | NA | 5.7 |
| Chaiyasen et al. 2017 | 19.233 | 99.483 | 656 | 24.7 | 1158 | boradleaf | AM | non-N-fixing | mineral | NA | NA | NA | NA | 29.8 | NA | NA | NA | 49 | NA | NA | NA | NA | 5.23 |
| Chodak & Niklinska 2010 | 50.267 | 19.433 | 300 | 8 | 700 | conifer | ECM | non-N-fixing | forest floor | NA | NA | NA | 35.45 | 339.78 | 9.459 | NA | NA | NA | 4.68184 | NA | NA | 32.9721 | NA |
| Chodak & Niklinska 2010 | 50.267 | 19.433 | 300 | 8 | 700 | boradleaf | AM | N-fixing | forest floor | NA | NA | NA | 17.6 | 385.62 | 21.79 | NA | NA | NA | 6.01856 | NA | NA | 36.2454 | NA |
| Chodak & Niklinska 2010 | 50.267 | 19.433 | 300 | 8 | 700 | boradleaf | ECM | non-N-fixing | forest floor | NA | NA | NA | 25.64 | 382.92 | 14.91 | NA | NA | NA | 9.81824 | NA | NA | 60.5847 | NA |
| Chodak & Niklinska 2010 | 50.267 | 19.433 | 300 | 8 | 700 | conifer | ECM | non-N-fixing | forest floor | NA | NA | NA | 37.46 | 358.65 | 9.573 | NA | NA | NA | 6.05014 | NA | NA | 46.5425 | NA |

| | | | | | | | | | | | | | | | | | | | | | | | | |
|---|---|---|---|---|---|---|---|---|---|---|---|---|---|---|---|---|---|---|---|---|---|---|---|---|
| Chodak & Niklinska 2010 | 50.267 | 19.433 | 300 | 8 | 700 | conifer | ECM | non-N-fixing | mineral | NA | NA | NA | 32.55 | 5.3933 | 0.168 | NA | NA | NA | 0.100201 | NA | NA | 0.585561 | NA |
| Chodak & Niklinska 2010 | 50.267 | 19.433 | 300 | 8 | 700 | boradleaf | AM | N-fixing | mineral | NA | NA | NA | 14.93 | 16.18 | 1.064 | NA | NA | NA | 0.148072 | NA | NA | 0.818182 | NA |
| Chodak & Niklinska 2010 | 50.267 | 19.433 | 300 | 8 | 700 | boradleaf | ECM | non-N-fixing | mineral | NA | NA | NA | 23.67 | 7.8202 | 0.318 | NA | NA | NA | 0.213426 | NA | NA | 1.17914 | NA |
| Chodak & Niklinska 2010 | 50.267 | 19.433 | 300 | 8 | 700 | conifer | ECM | non-N-fixing | mineral | NA | NA | NA | 34.06 | 6.2023 | 0.176 | NA | NA | NA | 0.121896 | NA | NA | 0.57754 | NA |
| Liang et al. 2006 | 46.217 | -88.3 | 420 | 18 | 578 | boradleaf | ECM | non-N-fixing | mineral | NA | 41.43 | 3.08 | 13.46 | 31.03 | 2.31 | NA | NA | NA | NA | NA | NA | NA | NA |
| Liang et al. 2006 | 46.217 | -88.3 | 420 | 18 | 578 | boradleaf | AM | non-N-fixing | mineral | NA | 37.33 | 2.52 | 14.7 | 27.96 | 1.89 | NA | NA | NA | NA | NA | NA | NA | NA |
| Liang et al. 2006 | 46.217 | -88.3 | 420 | 18 | 1200 | conifer | ECM | non-N-fixing | mineral | NA | 53.76 | 3.23 | 16.71 | 29.14 | 1.75 | NA | NA | NA | NA | NA | NA | NA | NA |
| Wang et al. 2013 | 26.875 | 109.668 | 318 | 16.5 | 1200 | conifer | ECM | non-N-fixing | mineral | NA | 45.88 | 3.41 | NA | 20.3 | 1.51 | NA | NA | 0.24 | 550 | NA | NA | 75.9322 | 4.27 |
| Wang et al. 2013 | 26.875 | 109.668 | 318 | 16.5 | 1200 | boradleaf | AM | non-N-fixing | mineral | NA | 66.24 | 5.14 | NA | 27.6 | 2.14 | NA | NA | 0.32 | 415.385 | NA | NA | 67.7966 | 4.46 |
| Wang et al. 2013 | 26.875 | 109.668 | 318 | 16.5 | 1785 | boradleaf | AM | non-N-fixing | mineral | NA | 55.93 | 3.95 | NA | 23.9 | 1.69 | NA | NA | 0.21 | 484.615 | NA | NA | 67.3446 | 4.38 |
| Deng et al. 2010 | 28.25 | 116.917 | 46 | 17.8 | 1785 | conifer | ECM | non-N-fixing | forest floor | NA | NA | NA | NA | 35.83 | 2.11 | NA | NA | 390 | 740 | NA | NA | NA | 4.16 |
| Deng et al. 2010 | 28.25 | 116.917 | 46 | 17.8 | 1785 | boradleaf | AM | non-N-fixing | forest floor | NA | NA | NA | NA | 31.2 | 2.07 | NA | NA | 510 | 500 | NA | NA | NA | 4.75 |
| Deng et al. 2010 | 28.25 | 116.917 | 46 | 17.8 | 1785 | boradleaf | AM | N-fixing | forest floor | NA | NA | NA | NA | 45.46 | 3.05 | NA | NA | 380 | 830 | NA | NA | NA | 4.07 |
| Deng et al. 2010 | 28.25 | 116.917 | 46 | 17.8 | 1785 | conifer | ECM | non-N-fixing | mineral | NA | NA | NA | NA | 11.6 | 1.11 | NA | NA | 340 | 230 | NA | NA | NA | 4.19 |
| Deng et al. 2010 | 28.25 | 116.917 | 46 | 17.8 | 1785 | boradleaf | AM | non-N-fixing | mineral | NA | NA | NA | NA | 11.94 | 1.04 | NA | NA | 430 | 200 | NA | NA | NA | 4.65 |
| Deng et al. 2010 | 28.25 | 116.917 | 46 | 17.8 | 1785 | boradleaf | AM | N-fixing | mineral | NA | NA | NA | NA | 17.26 | 1.25 | NA | NA | 350 | 260 | NA | NA | NA | 4.3 |
| Kara et al. 2016 | 40.567 | 39.283 | 1120 | 9.4 | 461 | conifer | ECM | non-N-fixing | mineral | NA | NA | NA | NA | 14.9 | NA | NA | NA | NA | 165.865 | NA | NA | 20.2314 | 7.65 |
| Kara et al. 2016 | 40.567 | 39.283 | 1120 | 9.4 | 461 | boradleaf | ECM | N-fixing | mineral | NA | NA | NA | NA | 18.5 | NA | NA | NA | NA | 211.158 | NA | NA | 18.644 | 7.68 |
| Vesterdal et al. 1998 | 54.783 | 11.367 | 18 | 8.7 | 590 | conifer | ECM | non-N-fixing | forest floor | NA | 2.85 | 0.072 | 39.6 | NA | NA | NA | NA | NA | NA | NA | NA | NA | 4.88 |
| Vesterdal et al. 1998 | 54.783 | 11.367 | 18 | 8.7 | 590 | boradleaf | ECM | non-N-fixing | forest floor | NA | 2.57 | 0.093 | 27.6 | NA | NA | NA | NA | NA | NA | NA | NA | NA | 5.02 |
| Vesterdal et al. 1998 | 54.783 | 11.367 | 18 | 8.7 | 590 | conifer | ECM | non-N-fixing | forest floor | NA | 16.11 | 0.431 | 37.4 | NA | NA | NA | NA | NA | NA | NA | NA | NA | 3.58 |
| Vesterdal et al. 1998 | 54.783 | 11.367 | 18 | 8.7 | 590 | conifer | ECM | non-N-fixing | forest floor | NA | 4.02 | 0.155 | 25.9 | NA | NA | NA | NA | NA | NA | NA | NA | NA | 4 |
| Vesterdal et al. 1998 | 54.783 | 11.367 | 18 | 8.7 | 590 | boradleaf | ECM | non-N-fixing | forest floor | NA | 0.72 | 0.026 | 27.7 | NA | NA | NA | NA | NA | NA | NA | NA | NA | 4.13 |
| Vesterdal et al. 1998 | 54.783 | 11.367 | 18 | 8.7 | 590 | conifer | ECM | non-N-fixing | forest floor | NA | 6.33 | 0.249 | 25.4 | NA | NA | NA | NA | NA | NA | NA | NA | NA | 4.42 |
| Vesterdal et al. 1998 | 54.783 | 11.367 | 18 | 8.7 | 590 | conifer | ECM | non-N-fixing | forest floor | NA | 12.76 | 0.422 | 30.2 | NA | NA | NA | NA | NA | NA | NA | NA | NA | 3.87 |
| Vesterdal et al. 1998 | 54.233 | 11.483 | 18 | 8.9 | 582 | conifer | ECM | non-N-fixing | forest floor | NA | 1.88 | 0.048 | 39.2 | NA | NA | NA | NA | NA | NA | NA | NA | NA | 5 |
| Vesterdal et al. 1998 | 54.233 | 11.483 | 18 | 8.9 | 582 | boradleaf | ECM | non-N-fixing | forest floor | NA | 2.92 | 0.101 | 28.9 | NA | NA | NA | NA | NA | NA | NA | NA | NA | 5.05 |
| Vesterdal et al. 1998 | 54.233 | 11.483 | 18 | 8.9 | 582 | conifer | ECM | non-N-fixing | forest floor | NA | 16.67 | 0.434 | 38.4 | NA | NA | NA | NA | NA | NA | NA | NA | NA | 3.46 |
| Vesterdal et al. 1998 | 54.233 | 11.483 | 18 | 8.9 | 582 | conifer | ECM | non-N-fixing | forest floor | NA | 2.34 | 0.085 | 27.5 | NA | NA | NA | NA | NA | NA | NA | NA | NA | 4.39 |
| Vesterdal et al. 1998 | 54.233 | 11.483 | 18 | 8.9 | 582 | boradleaf | ECM | non-N-fixing | forest floor | NA | 1.51 | 0.055 | 27.5 | NA | NA | NA | NA | NA | NA | NA | NA | NA | 4.27 |
| Vesterdal et al. 1998 | 54.233 | 11.483 | 18 | 8.9 | 582 | conifer | ECM | non-N-fixing | forest floor | NA | 4.54 | 0.166 | 27.3 | NA | NA | NA | NA | NA | NA | NA | NA | NA | 5.02 |
| Vesterdal et al. 1998 | 54.233 | 11.483 | 18 | 8.9 | 582 | conifer | ECM | non-N-fixing | forest floor | NA | 7.21 | 0.263 | 27.4 | NA | NA | NA | NA | NA | NA | NA | NA | NA | 4.1 |
| Vesterdal et al. 1998 | 55.967 | 12.35 | 10 | 8.1 | 595 | conifer | ECM | non-N-fixing | forest floor | NA | 3.26 | 0.102 | 32 | NA | NA | NA | NA | NA | NA | NA | NA | NA | 4.62 |
| Vesterdal et al. 1998 | 55.967 | 12.35 | 10 | 8.1 | 595 | boradleaf | ECM | non-N-fixing | forest floor | NA | 3.43 | 0.135 | 25.4 | NA | NA | NA | NA | NA | NA | NA | NA | NA | 4.62 |
| Vesterdal et al. 1998 | 55.967 | 12.35 | 10 | 8.1 | 595 | conifer | ECM | non-N-fixing | forest floor | NA | 20.88 | 0.595 | 35.1 | NA | NA | NA | NA | NA | NA | NA | NA | NA | 3.51 |
| Vesterdal et al. 1998 | 55.967 | 12.35 | 10 | 8.1 | 595 | conifer | ECM | non-N-fixing | forest floor | NA | 7.46 | 0.288 | 25.9 | NA | NA | NA | NA | NA | NA | NA | NA | NA | 3.89 |
| Vesterdal et al. 1998 | 55.967 | 12.35 | 10 | 8.1 | 595 | boradleaf | ECM | non-N-fixing | forest floor | NA | 2.25 | 0.072 | 31.3 | NA | NA | NA | NA | NA | NA | NA | NA | NA | 4.2 |
| Vesterdal et al. 1998 | 55.967 | 12.35 | 10 | 8.1 | 595 | conifer | ECM | non-N-fixing | forest floor | NA | 11.09 | 0.394 | 28.1 | NA | NA | NA | NA | NA | NA | NA | NA | NA | 4.27 |
| Vesterdal et al. 1998 | 55.967 | 12.35 | 10 | 8.1 | 595 | conifer | ECM | non-N-fixing | forest floor | NA | 7.44 | 0.238 | 31.3 | NA | NA | NA | NA | NA | NA | NA | NA | NA | 4.08 |
| Vesterdal et al. 1998 | 56.467 | 10.533 | 36 | 7.8 | 659 | conifer | ECM | non-N-fixing | forest floor | NA | 12.68 | 0.585 | 21.7 | NA | NA | NA | NA | NA | NA | NA | NA | NA | 3.46 |
| Vesterdal et al. 1998 | 56.467 | 10.533 | 36 | 7.8 | 659 | boradleaf | ECM | non-N-fixing | forest floor | NA | 15.2 | 0.672 | 22.6 | NA | NA | NA | NA | NA | NA | NA | NA | NA | 3.73 |
| Vesterdal et al. 1998 | 56.467 | 10.533 | 36 | 7.8 | 659 | conifer | ECM | non-N-fixing | forest floor | NA | 18.25 | 0.517 | 35.3 | NA | NA | NA | NA | NA | NA | NA | NA | NA | 3.36 |
| Vesterdal et al. 1998 | 56.467 | 10.533 | 36 | 7.8 | 659 | conifer | ECM | non-N-fixing | forest floor | NA | 11.43 | 0.54 | 21.2 | NA | NA | NA | NA | NA | NA | NA | NA | NA | 3.54 |
| Vesterdal et al. 1998 | 56.467 | 10.533 | 36 | 7.8 | 659 | boradleaf | ECM | non-N-fixing | forest floor | NA | 8.19 | 0.401 | 20.4 | NA | NA | NA | NA | NA | NA | NA | NA | NA | 4.04 |
| Vesterdal et al. 1998 | 56.467 | 10.533 | 36 | 7.8 | 659 | conifer | ECM | non-N-fixing | forest floor | NA | 20.79 | 0.887 | 23.4 | NA | NA | NA | NA | NA | NA | NA | NA | NA | 3.34 |
| Vesterdal et al. 1998 | 56.467 | 10.533 | 36 | 7.8 | 659 | conifer | ECM | non-N-fixing | forest floor | NA | 16.56 | 0.716 | 23.1 | NA | NA | NA | NA | NA | NA | NA | NA | NA | 3.71 |
| Vesterdal et al. 1998 | 56.713 | 10.017 | 33 | 7.7 | 648 | conifer | ECM | non-N-fixing | forest floor | NA | 2.38 | 0.072 | 33.1 | NA | NA | NA | NA | NA | NA | NA | NA | NA | 4.92 |
| Vesterdal et al. 1998 | 56.713 | 10.017 | 33 | 7.7 | 648 | boradleaf | ECM | non-N-fixing | forest floor | NA | 3.5 | 0.131 | 26.7 | NA | NA | NA | NA | NA | NA | NA | NA | NA | 5.28 |
| Vesterdal et al. 1998 | 56.713 | 10.017 | 33 | 7.7 | 648 | conifer | ECM | non-N-fixing | forest floor | NA | 14.9 | 0.446 | 33.4 | NA | NA | NA | NA | NA | NA | NA | NA | NA | 3.57 |
| Vesterdal et al. 1998 | 56.713 | 10.017 | 33 | 7.7 | 648 | conifer | ECM | non-N-fixing | forest floor | NA | 5.85 | 0.214 | 27.3 | NA | NA | NA | NA | NA | NA | NA | NA | NA | 4.41 |
| Vesterdal et al. 1998 | 56.713 | 10.017 | 33 | 7.7 | 648 | boradleaf | ECM | non-N-fixing | forest floor | NA | 1.08 | 0.037 | 29.2 | NA | NA | NA | NA | NA | NA | NA | NA | NA | 3.36 |
| Vesterdal et al. 1998 | 56.713 | 10.017 | 33 | 7.7 | 648 | conifer | ECM | non-N-fixing | forest floor | NA | 5.97 | 0.23 | 26 | NA | NA | NA | NA | NA | NA | NA | NA | NA | 4.85 |
| Vesterdal et al. 1998 | 56.713 | 10.017 | 33 | 7.7 | 648 | conifer | ECM | non-N-fixing | forest floor | NA | 6.68 | 0.194 | 34.4 | NA | NA | NA | NA | NA | NA | NA | NA | NA | 4.5 |
| Vesterdal et al. 1998 | 55.133 | 8.833 | 9 | 8.3 | 839 | conifer | ECM | non-N-fixing | forest floor | NA | 2.69 | 0.1 | 26.9 | NA | NA | NA | NA | NA | NA | NA | NA | NA | 4.75 |
| Vesterdal et al. 1998 | 55.133 | 8.833 | 9 | 8.3 | 839 | boradleaf | ECM | non-N-fixing | forest floor | NA | 3.86 | 0.115 | 33.6 | NA | NA | NA | NA | NA | NA | NA | NA | NA | 4.89 |
| Vesterdal et al. 1998 | 55.133 | 8.833 | 9 | 8.3 | 839 | conifer | ECM | non-N-fixing | forest floor | NA | 18.19 | 0.504 | 36.1 | NA | NA | NA | NA | NA | NA | NA | NA | NA | 3.51 |
| Vesterdal et al. 1998 | 55.133 | 8.833 | 9 | 8.3 | 839 | conifer | ECM | non-N-fixing | forest floor | NA | 9.48 | 0.342 | 27.7 | NA | NA | NA | NA | NA | NA | NA | NA | NA | 3.93 |
| Vesterdal et al. 1998 | 55.133 | 8.833 | 9 | 8.3 | 839 | boradleaf | ECM | non-N-fixing | forest floor | NA | 0.73 | 0.02 | 36.5 | NA | NA | NA | NA | NA | NA | NA | NA | NA | 3.98 |
| Vesterdal et al. 1998 | 55.133 | 8.833 | 9 | 8.3 | 839 | conifer | ECM | non-N-fixing | forest floor | NA | 9.14 | 0.311 | 29.4 | NA | NA | NA | NA | NA | NA | NA | NA | NA | 4.38 |
| Vesterdal et al. 1998 | 55.133 | 8.833 | 9 | 8.3 | 839 | conifer | ECM | non-N-fixing | forest floor | NA | 7.43 | 0.257 | 28.9 | NA | NA | NA | NA | NA | NA | NA | NA | NA | 4.15 |
| Vesterdal et al. 1998 | 56.283 | 8.433 | 9 | 8.7 | 862 | conifer | ECM | non-N-fixing | forest floor | NA | 14.95 | 0.54 | 27.7 | NA | NA | NA | NA | NA | NA | NA | NA | NA | 3.61 |
| Vesterdal et al. 1998 | 56.283 | 8.433 | 9 | 8.7 | 862 | boradleaf | ECM | non-N-fixing | forest floor | NA | 19.26 | 0.866 | 22.2 | NA | NA | NA | NA | NA | NA | NA | NA | NA | 3.82 |
| Vesterdal et al. 1998 | 56.283 | 8.433 | 9 | 8.7 | 862 | conifer | ECM | non-N-fixing | forest floor | NA | 16.16 | 0.533 | 30.3 | NA | NA | NA | NA | NA | NA | NA | NA | NA | 3.36 |
| Vesterdal et al. 1998 | 56.283 | 8.433 | 9 | 8.7 | 862 | conifer | ECM | non-N-fixing | forest floor | NA | 15.92 | 0.685 | 23.2 | NA | NA | NA | NA | NA | NA | NA | NA | NA | 3.7 |
| Vesterdal et al. 1998 | 56.283 | 8.433 | 9 | 8.7 | 862 | boradleaf | ECM | non-N-fixing | forest floor | NA | 6.1 | 0.301 | 20.3 | NA | NA | NA | NA | NA | NA | NA | NA | NA | 3.79 |
| Vesterdal et al. 1998 | 56.283 | 8.433 | 9 | 8.7 | 862 | conifer | ECM | non-N-fixing | forest floor | NA | 19.78 | 0.797 | 24.8 | NA | NA | NA | NA | NA | NA | NA | NA | NA | 3.53 |
| Vesterdal et al. 1998 | 56.283 | 8.433 | 9 | 8.7 | 862 | conifer | ECM | non-N-fixing | forest floor | NA | 20.69 | 0.818 | 25.3 | NA | NA | NA | NA | NA | NA | NA | NA | NA | 3.52 |
| Bell et al. 2017 | 37.4285 | -83.1709 | 368 | 12.1 | 1385 | conifer | ECM | non-N-fixing | mineral | NA | NA | NA | 0.21 | 0.22 | 1.04 | NA | NA | 6.79 | NA | NA | NA | NA | 6.2 |
| Bell et al. 2017 | 37.4285 | -83.1709 | 368 | 12.1 | 1385 | conifer | ECM | non-N-fixing | mineral | NA | NA | NA | 0.17 | 0.34 | 1.96 | NA | NA | 7.79 | NA | NA | NA | NA | 5.72 |
| Bolat et al. 2015 | 41.617 | 32.75 | 602 | 9.1 | 490 | boradleaf | ECM | N-fixing | mineral | NA | 7.11 | 1.89 | 3.92 | 6.4 | 1.7 | NA | NA | NA | 314.943 | 43.848 | NA | 7.263 | 8.26 |
| Sariyildiz et al. 2015 | 41.383 | 33.767 | 870 | 9.5 | 490 | conifer | ECM | non-N-fixing | mineral | NA | 48.3 | 2.37 | 21 | 41.3 | 2 | NA | NA | NA | NA | NA | NA | NA | 5.4 |
| Sariyildiz et al. 2015 | 41.383 | 33.767 | 870 | 9.5 | 490 | conifer | ECM | non-N-fixing | mineral | NA | 78.8 | 4.2 | 19 | 32.3 | 1.5 | NA | NA | NA | NA | NA | NA | NA | 5.4 |
| Sariyildiz et al. 2015 | 41.383 | 33.767 | 870 | 9 | 490 | boradleaf | ECM | non-N-fixing | mineral | NA | 44.5 | 4.98 | 15 | 36.7 | 2.5 | NA | NA | NA | NA | NA | NA | NA | 5.7 |
| Sariyildiz et al. 2015 | 41.383 | 33.767 | 870 | 9 | 490 | boradleaf | ECM | non-N-fixing | mineral | NA | 72.9 | 5.77 | 13 | 28.4 | 2.2 | NA | NA | NA | NA | NA | NA | NA | 5.7 |
| Sariyildiz et al. 2015 | 41.383 | 33.767 | 870 | 8.5 | 490 | conifer | ECM | non-N-fixing | mineral | NA | 41.1 | 5.19 | 8 | 31.8 | 4.4 | NA | NA | NA | NA | NA | NA | NA | 5.5 |
| Sariyildiz et al. 2015 | 41.383 | 33.767 | 870 | 8.5 | 550 | conifer | ECM | non-N-fixing | mineral | NA | 67 | 9.57 | 7 | 25.1 | 3.9 | NA | NA | NA | NA | NA | NA | NA | 5.5 |
| Sariyildiz et al. 2015 | 41.383 | 33.767 | 870 | 11.5 | 550 | conifer | ECM | non-N-fixing | mineral | NA | 43 | 3.32 | 13 | 35.4 | 2.7 | NA | NA | NA | NA | NA | NA | NA | 5.9 |
| Sariyildiz et al. 2015 | 41.383 | 33.767 | 870 | 11.5 | 550 | conifer | ECM | non-N-fixing | mineral | NA | 70.9 | 5.93 | 12 | 28 | 2.3 | NA | NA | NA | NA | NA | NA | NA | 5.9 |
| Sariyildiz et al. 2015 | 41.383 | 33.767 | 870 | 11.5 | 550 | conifer | ECM | non-N-fixing | mineral | NA | 38.1 | 3.8 | 9 | 31.4 | 3.1 | NA | NA | NA | NA | NA | NA | NA | NA |
| Sariyildiz et al. 2015 | 41.383 | 33.767 | 870 | 11.5 | 568 | conifer | ECM | non-N-fixing | mineral | NA | 61.3 | 6.82 | 10 | 24.2 | 2.7 | NA | NA | NA | NA | NA | NA | NA | NA |
| Oostra et al. 2006 | 56.033 | 12.683 | 40 | 8.2 | 568 | boradleaf | AM | non-N-fixing | forest floor | NA | NA | NA | 19.1 | 158 | 8.2 | NA | NA | NA | NA | NA | NA | NA | 5.77 |
| Oostra et al. 2006 | 56.033 | 12.683 | 40 | 8.2 | 568 | boradleaf | ECM | non-N-fixing | forest floor | NA | NA | NA | 21.8 | 171 | 7.8 | NA | NA | NA | NA | NA | NA | NA | 4.97 |
| Oostra et al. 2006 | 56.033 | 12.683 | 40 | 8.2 | 568 | boradleaf | AM | non-N-fixing | forest floor | NA | NA | NA | 20 | 118 | 5.9 | NA | NA | NA | NA | NA | NA | NA | 5.87 |
| Oostra et al. 2006 | 56.033 | 12.683 | 40 | 8.2 | 568 | boradleaf | ECM | non-N-fixing | forest floor | NA | NA | NA | 21.7 | 241 | 11 | NA | NA | NA | NA | NA | NA | NA | 5.13 |
| Oostra et al. 2006 | 56.033 | 12.683 | 40 | 8.2 | 568 | boradleaf | ECM | non-N-fixing | forest floor | NA | NA | NA | 22.6 | 234 | 10.3 | NA | NA | NA | NA | NA | NA | NA | 4.96 |
| Oostra et al. 2006 | 56.033 | 12.683 | 40 | 8.2 | 568 | conifer | ECM | non-N-fixing | forest floor | NA | NA | NA | 22.4 | 287 | 12.8 | NA | NA | NA | NA | NA | NA | NA | 4.23 |
| Oostra et al. 2006 | 56.033 | 12.683 | 40 | 8.2 | 568 | boradleaf | AM | non-N-fixing | mineral | NA | NA | NA | 15.2 | 49 | 3.2 | NA | NA | NA | NA | NA | NA | NA | 4.88 |
| Oostra et al. 2006 | 56.033 | 12.683 | 40 | 8.2 | 568 | boradleaf | ECM | non-N-fixing | mineral | NA | NA | NA | 15.5 | 43 | 2.8 | NA | NA | NA | NA | NA | NA | NA | 4.23 |
| Oostra et al. 2006 | 56.033 | 12.683 | 40 | 8.2 | 568 | boradleaf | AM | non-N-fixing | mineral | NA | NA | NA | 14.6 | 51 | 3.5 | NA | NA | NA | NA | NA | NA | NA | 5.87 |
| Oostra et al. 2006 | 56.033 | 12.683 | 40 | 8.2 | 568 | boradleaf | ECM | non-N-fixing | mineral | NA | NA | NA | 15.7 | 45 | 2.8 | NA | NA | NA | NA | NA | NA | NA | 4.48 |
| Oostra et al. 2006 | 56.033 | 12.683 | 40 | 8.2 | 568 | boradleaf | ECM | non-N-fixing | mineral | NA | NA | NA | 17.1 | 64 | 3.8 | NA | NA | NA | NA | NA | NA | NA | 4.53 |
| Oostra et al. 2006 | 56.033 | 12.683 | 40 | 8.2 | 568 | conifer | ECM | non-N-fixing | mineral | NA | NA | NA | 20.6 | 54 | 2.6 | NA | NA | NA | NA | NA | NA | NA | 3.76 |
| Cheng et al. 2013 | 33.433 | 108.433 | 1746 | 8.3 | 830 | conifer | ECM | non-N-fixing | mineral | NA | 40.05 | 2.07 | 19.32 | 37.08 | 1.92 | 14.01 | 3.02 | 400 | 772.355 | 97.674 | NA | 13.1969 | 6.06 |
| Cheng et al. 2013 | 33.433 | 108.433 | 1746 | 8.3 | 830 | boradleaf | ECM | non-N-fixing | mineral | NA | 23.53 | 1.5 | 15.71 | 26.44 | 1.69 | 20.87 | 2.65 | 340 | 845.531 | 123.01 | NA | 9.897672 | 3.95 |
| Cheng et al. 2013 | 22.45 | 108.467 | 1746 | 21.6 | 1576 | conifer | ECM | non-N-fixing | mineral | NA | 20.08 | 1.44 | 13.98 | 18.09 | 1.3 | 18.86 | 3.13 | 320 | 505.695 | 61.884 | NA | 6.59844 | 5.93 |
| Cheng et al. 2013 | 22.45 | 108.467 | 1746 | 21.6 | 1576 | conifer | ECM | non-N-fixing | mineral | NA | 51.82 | 4.08 | 12.74 | 48.89 | 3.85 | 20.1 | 2.57 | 1140 | 914.721 | 80.648 | NA | 15.45427 | 5.86 |
| Yu et al. 2015 | 34.483 | 107.983 | 699 | 10.8 | 601.6 | boradleaf | AM | non-N-fixing | mineral | NA | NA | NA | 10.55 | 11.93 | 1.13 | 4.77 | 1.03 | 380 | NA | NA | NA | NA | 7.72 |
| Kooch et al. 2018 | 36.35 | 53 | 559 | 15.5 | 808 | boradleaf | AM | non-N-fixing | mineral | NA | 31.3 | 5.38 | 5.79 | 16.3 | 2.8 | 574.917 | 58.9232 | NA | 35.3431 | 18.634 | NA | NA | 6.86 |
| Kooch et al. 2018 | 36.35 | 53 | 559 | 15.5 | 808 | boradleaf | AM | non-N-fixing | mineral | NA | 31.37 | 4.35 | 7.3 | 16.6 | 2.3 | 584.257 | 56.7924 | NA | 29.2209 | 17.398 | NA | NA | 6.67 |
| Kooch et al. 2018 | 36.35 | 53 | 559 | 15.5 | 808 | conifer | AM | non-N-fixing | mineral | NA | 57.3 | 4.58 | 12.69 | 20 | 1.6 | 620.406 | 53.0178 | NA | 17.6471 | 11.338 | NA | NA | 6.16 |
| Kooch et al. 2018 | 36.35 | 53 | 559 | 15.5 | 808 | conifer | ECM | non-N-fixing | mineral | NA | 44.89 | 2.09 | 20.28 | 25.8 | 1.2 | 623.684 | 52.5252 | NA | 16.0684 | 10.253 | NA | NA | 5.84 |
| Lejon et al. 2005 | 47.3 | 4.067 | 618 | 9 | 1300 | boradleaf | ECM | non-N-fixing | mineral | NA | NA | NA | 18.2 | 67 | 3.7 | NA | NA | NA | 904 | NA | NA | NA | 3.6 |
| Lejon et al. 2005 | 47.3 | 4.067 | 618 | 9 | 1300 | conifer | ECM | non-N-fixing | mineral | NA | NA | NA | 19.5 | 72 | 3.6 | NA | NA | NA | 773 | NA | NA | NA | 3.6 |
| Lejon et al. 2005 | 47.3 | 4.067 | 618 | 9 | 1300 | conifer | ECM | non-N-fixing | mineral | NA | NA | NA | 19.1 | 77 | 4 | NA | NA | NA | 699 | NA | NA | NA | 3.6 |
| Li et al. 2015 | 24.717 | 102.567 | 1974 | 14.7 | 918 | boradleaf | ECM | non-N-fixing | mineral | NA | NA | NA | 40.58 | 22.72 | 0.56 | NA | NA | 9182.24 | NA | NA | NA | NA | 6.9 |
| Li et al. 2015 | 24.717 | 102.567 | 1974 | 14.7 | 918 | conifer | AM | non-N-fixing | mineral | NA | NA | NA | 22.3 | 16.28 | 0.73 | NA | NA | 7945.04 | NA | NA | NA | NA | 6.2 |
| Li et al. 2015 | 24.717 | 102.567 | 1974 | 14.7 | 918 | conifer | ECM | non-N-fixing | mineral | NA | NA | NA | 23.39 | 18.01 | 0.77 | NA | NA | 3914.22 | NA | NA | NA | NA | 5.7 |
| Li et al. 2015 | 24.717 | 102.567 | 1974 | 14.7 | 918 | boradleaf | ECM | non-N-fixing | mineral | NA | NA | NA | 33.14 | 24.86 | 0.75 | NA | NA | 1746.71 | NA | NA | NA | NA | 6.2 |
| Li et al. 2015 | 24.717 | 102.567 | 1974 | 14.7 | 918 | conifer | AM | non-N-fixing | mineral | NA | NA | NA | 24.31 | 33.79 | 1.39 | NA | NA | 1829 | NA | NA | NA | NA | 6.2 |
| Li et al. 2015 | 24.717 | 102.567 | 1974 | 14.7 | 918 | conifer | ECM | non-N-fixing | mineral | NA | NA | NA | 21.06 | 20.85 | 0.99 | NA | NA | 21234.7 | NA | NA | NA | NA | 6.2 |
| Li et al. 2015 | 24.717 | 102.567 | 1974 | 14.7 | 918 | boradleaf | ECM | non-N-fixing | mineral | NA | NA | NA | 33.45 | 47.5 | 1.42 | NA | NA | 7892.97 | NA | NA | NA | NA | 6.5 |
| Li et al. 2015 | 24.717 | 102.567 | 1974 | 14.7 | 918 | conifer | AM | non-N-fixing | mineral | NA | NA | NA | 36.39 | 56.04 | 1.54 | NA | NA | 9139.15 | NA | NA | NA | NA | 6.9 |
| Li et al. 2015 | 24.717 | 102.567 | 1974 | 14.7 | 918 | conifer | ECM | non-N-fixing | mineral | NA | NA | NA | 30.04 | 27.04 | 0.9 | NA | NA | 2158.34 | NA | NA | NA | NA | 5.7 |
| Galicia & Garcia-Oliva 2011 | 19.5 | -105.083 | 62 | 24 | 746 | boradleaf | AM | non-N-fixing | forest floor | NA | NA | NA | 22 | 317 | 16.4 | NA | NA | 2650 | NA | NA | NA | NA | NA |
| Galicia & Garcia-Oliva 2011 | 19.5 | -105.083 | 62 | 24 | 746 | boradleaf | AM | non-N-fixing | forest floor | NA | NA | NA | 23 | 301 | 13.4 | NA | NA | 1770 | NA | NA | NA | NA | NA |
| Galicia & Garcia-Oliva 2011 | 19.5 | -105.083 | 62 | 24 | 746 | boradleaf | AM | non-N-fixing | mineral | NA | NA | NA | 14.3 | 40.1 | 2.8 | NA | NA | 590 | 774 | 115 | NA | NA | NA |
| Galicia & Garcia-Oliva 2011 | 19.5 | -105.083 | 62 | 24 | 746 | boradleaf | AM | non-N-fixing | mineral | NA | NA | NA | 10.4 | 30.3 | 2.9 | NA | NA | 470 | 662 | 66 | NA | NA | NA |
| Lucas-Borja et al. 2012 | 40.017 | -1.983 | 1199 | 11.9 | 595 | conifer | ECM | non-N-fixing | mineral | NA | NA | NA | 28.61 | 102.3 | 4.4 | NA | NA | NA | NA | NA | NA | 48.59 | 6.82 |
| Lucas-Borja et al. 2012 | 40.017 | -1.983 | 1199 | 11.9 | 595 | conifer | ECM | non-N-fixing | mineral | NA | NA | NA | 31.23 | 153 | 5.2 | NA | NA | NA | NA | NA | NA | 82.93 | 5.71 |
| Luo et al. 2016 | 22.167 | 106.7 | 231 | 21 | 1400 | boradleaf | AM | non-N-fixing | mineral | NA | NA | NA | 13.1 | 18.8 | 1.43 | 8.08 | 2.32 | NA | NA | NA | NA | NA | 4.51 |
| Luo et al. 2016 | 22.167 | 106.7 | 231 | 21 | 1400 | conifer | ECM | non-N-fixing | mineral | NA | NA | NA | 16.94 | 17.07 | 1.01 | 5.1 | 1.65 | NA | NA | NA | NA | NA | 4.67 |
| Luo et al. 2016 | 22.167 | 106.7 | 231 | 21 | 1400 | boradleaf | AM | non-N-fixing | forest floor | 3.895 | NA | NA | 14.13 | 483.6 | 34.2 | NA | NA | NA | NA | NA | NA | NA | NA |
| Luo et al. 2016 | 22.167 | 106.7 | 231 | 21 | 1400 | conifer | ECM | non-N-fixing | forest floor | 3.367 | NA | NA | 36.67 | 509.7 | 13.9 | NA | NA | NA | NA | NA | NA | NA | NA |
| Mareschal et al. 2010 | 47.3 | 4.083 | 657 | 9 | 638 | conifer | ECM | non-N-fixing | mineral | NA | NA | NA | NA | 66.54 | NA | NA | NA | NA | NA | NA | NA | NA | 4.15 |
| Mareschal et al. 2010 | 47.3 | 4.083 | 657 | 9 | 638 | conifer | ECM | non-N-fixing | mineral | NA | NA | NA | NA | 67.22 | NA | NA | NA | NA | NA | NA | NA | NA | 4.04 |
| Mareschal et al. 2010 | 47.3 | 4.083 | 657 | 9 | 638 | conifer | ECM | non-N-fixing | mineral | NA | NA | NA | NA | 69.27 | NA | NA | NA | NA | NA | NA | NA | NA | 4.03 |
| Mareschal et al. 2010 | 47.3 | 4.083 | 657 | 9 | 638 | boradleaf | ECM | non-N-fixing | mineral | NA | NA | NA | NA | 79.51 | NA | NA | NA | NA | NA | NA | NA | NA | 3.77 |
| Melvin & Goodale 2013 | 42.45 | -76.417 | 448 | 7.8 | 430 | conifer | ECM | non-N-fixing | forest floor | 4.55 | 12.63 | 0.32 | 28.9 | NA | NA | NA | NA | NA | NA | NA | NA | NA | NA |
| Melvin & Goodale 2013 | 42.45 | -76.417 | 448 | 7.8 | 430 | boradleaf | ECM | non-N-fixing | forest floor | 4.4 | 4.63 | 0.09 | 30.2 | NA | NA | NA | NA | NA | NA | NA | NA | NA | NA |
| Melvin & Goodale 2013 | 42.45 | -76.417 | 448 | 7.8 | 430 | boradleaf | AM | non-N-fixing | forest floor | 4.51 | 5.19 | 0.13 | 26.3 | NA | NA | NA | NA | NA | NA | NA | NA | NA | NA |
| Melvin & Goodale 2013 | 42.45 | -76.417 | 448 | 7.8 | 430 | conifer | ECM | non-N-fixing | mineral | NA | NA | NA | 19.46 | 97.7 | 4.96 | NA | NA | NA | NA | NA | NA | NA | 3.77 |
| Melvin & Goodale 2013 | 42.45 | -76.417 | 448 | 7.8 | 430 | boradleaf | ECM | non-N-fixing | mineral | NA | NA | NA | 14.76 | 45.95 | 3.04 | NA | NA | NA | NA | NA | NA | NA | 4.15 |
| Melvin & Goodale 2013 | 42.45 | -76.417 | 448 | 7.8 | 483 | boradleaf | AM | non-N-fixing | mineral | NA | NA | NA | 13.37 | 44.38 | 3.33 | NA | NA | NA | NA | NA | NA | NA | 4.24 |
| Thomas & Prescott 2000 | 50.8 | -119.433 | 981 | 5.9 | 483 | boradleaf | ECM | non-N-fixing | forest floor | 2.148 | NA | NA | 24.37 | 449.5 | 18.5 | NA | NA | 1400 | NA | NA | NA | NA | NA |
| Thomas & Prescott 2000 | 50.8 | -119.433 | 981 | 5.9 | 483 | conifer | ECM | non-N-fixing | forest floor | 2.376 | NA | NA | 21.88 | 455.1 | 20.9 | NA | NA | 1400 | NA | NA | NA | NA | NA |
| Thomas & Prescott 2000 | 50.8 | -119.433 | 981 | 5.9 | 483 | conifer | ECM | non-N-fixing | forest floor | 3.773 | NA | NA | 28.51 | 464.6 | 16.5 | NA | NA | 1200 | NA | NA | NA | NA | NA |
| Laik et al. 2009 | 25.983 | 85.8 | 70 | 24.8 | 1300 | boradleaf | AM | N-fixing | mineral | 5.15 | 10.95 | NA | NA | 5.33 | NA | NA | NA | NA | 134.42 | NA | NA | NA | 8.58 |
| Laik et al. 2009 | 25.983 | 85.8 | 70 | 24.8 | 1300 | boradleaf | AM | N-fixing | mineral | 8.32 | 16.48 | NA | NA | 8.08 | NA | NA | NA | NA | 110.39 | NA | NA | NA | 8.38 |
| Laik et al. 2009 | 25.983 | 85.8 | 70 | 24.8 | 1300 | boradleaf | AM | N-fixing | mineral | 6.25 | 13.04 | NA | NA | 6.21 | NA | NA | NA | NA | 113.27 | NA | NA | NA | 8.47 |
| Laik et al. 2009 | 25.983 | 85.8 | 70 | 24.8 | 1300 | boradleaf | ECM | non-N-fixing | mineral | 8.46 | 17.09 | NA | NA | 8.26 | NA | NA | NA | NA | 180.86 | NA | NA | NA | 8.37 |
| Laik et al. 2009 | 25.983 | 85.8 | 70 | 24.8 | 1300 | boradleaf | AM | non-N-fixing | mineral | 5.85 | 12.26 | NA | NA | 5.8 | NA | NA | NA | NA | 127.83 | NA | NA | NA | 8.56 |
| Laik et al. 2009 | 25.983 | 85.8 | 70 | 24.8 | 1300 | boradleaf | AM | non-N-fixing | mineral | 8.2 | 17.34 | NA | NA | 8.5 | NA | NA | NA | NA | 194.4 | NA | NA | NA | 8.37 |
| Giardina et al. 2001 | 39.64 | -104.33 | 1705 | 9.2 | 382 | boradleaf | ECM | non-N-fixing | mineral | NA | NA | NA | 18 | 63 | 3.5 | NA | NA | 650 | NA | NA | NA | NA | 6.03 |
| Giardina et al. 2001 | 39.64 | -104.33 | 1705 | 9.2 | 382 | boradleaf | ECM | non-N-fixing | mineral | NA | NA | NA | 20.4 | 94 | 4.6 | NA | NA | 858 | NA | NA | NA | NA | 5.97 |
| Giardina et al. 2001 | 39.64 | -104.33 | 1705 | 9.2 | 382 | boradleaf | ECM | non-N-fixing | mineral | NA | NA | NA | 12 | 53 | 4.4 | NA | NA | 760 | NA | NA | NA | NA | 6.2 |
| Giardina et al. 2001 | 39.64 | -104.33 | 1705 | 9.2 | 382 | boradleaf | ECM | non-N-fixing | mineral | NA | NA | NA | 18.3 | 77 | 4.2 | NA | NA | 769 | NA | NA | NA | NA | 5.97 |
| Giardina et al. 2001 | 39.64 | -104.33 | 1705 | 9.2 | 382 | boradleaf | ECM | non-N-fixing | mineral | NA | NA | NA | 18.8 | 47 | 2.5 | NA | NA | 670 | NA | NA | NA | NA | 6.53 |
| Giardina et al. 2001 | 39.64 | -104.33 | 1705 | 9.2 | 382 | boradleaf | ECM | non-N-fixing | mineral | NA | NA | NA | 16.8 | 32 | 1.9 | NA | NA | 577 | NA | NA | NA | NA | 5.87 |
| Giardina et al. 2001 | 39.64 | -104.33 | 1705 | 9.2 | 382 | conifer | ECM | non-N-fixing | mineral | NA | NA | NA | 20.4 | 47 | 2.3 | NA | NA | 500 | NA | NA | NA | NA | 4.97 |
| Giardina et al. 2001 | 39.64 | -104.33 | 1705 | 9.2 | 382 | conifer | ECM | non-N-fixing | mineral | NA | NA | NA | 20.8 | 27 | 1.3 | NA | NA | 350 | NA | NA | NA | NA | 5.2 |
| Giardina et al. 2001 | 39.64 | -104.33 | 1705 | 9.2 | 382 | conifer | ECM | non-N-fixing | mineral | NA | NA | NA | 12.2 | 11 | 0.9 | NA | NA | 135 | NA | NA | NA | NA | 4.87 |
| Giardina et al. 2001 | 39.64 | -104.33 | 1705 | 9.2 | 382 | conifer | ECM | non-N-fixing | mineral | NA | NA | NA | 22.5 | 18 | 0.8 | NA | NA | 343 | NA | NA | NA | NA | 5.37 |
| Giardina et al. 2001 | 39.64 | -104.33 | 1705 | 9.2 | 382 | conifer | ECM | non-N-fixing | mineral | NA | NA | NA | 12.7 | 14 | 1.1 | NA | NA | 443 | NA | NA | NA | NA | 5.53 |
| Giardina et al. 2001 | 39.64 | -104.33 | 1705 | 9.2 | 382 | conifer | ECM | non-N-fixing | mineral | NA | NA | NA | 18.9 | 17 | 0.9 | NA | NA | 226 | NA | NA | NA | NA | 5.57 |
| Sevgi et al. 2011 | 41.15 | 28.91 | 31 | 12.8 | 1074 | conifer | ECM | non-N-fixing | forest floor | NA | NA | NA | NA | 349 | NA | NA | NA | NA | NA | NA | NA | NA | NA |
| Sevgi et al. 2011 | 41.15 | 28.91 | 31 | 12.8 | 1074 | boradleaf | ECM | non-N-fixing | forest floor | NA | NA | NA | NA | 310 | NA | NA | NA | NA | NA | NA | NA | NA | NA |
| Sevgi et al. 2011 | 41.15 | 28.91 | 31 | 12.8 | 1074 | conifer | ECM | non-N-fixing | forest floor | NA | NA | NA | NA | 293 | NA | NA | NA | NA | NA | NA | NA | NA | NA |
| Sevgi et al. 2011 | 41.15 | 28.91 | 31 | 12.8 | 1074 | conifer | ECM | non-N-fixing | forest floor | NA | NA | NA | NA | 255 | NA | NA | NA | NA | NA | NA | NA | NA | NA |
| Sevgi et al. 2011 | 41.15 | 28.91 | 31 | 12.8 | 1074 | conifer | ECM | non-N-fixing | forest floor | NA | NA | NA | NA | 208 | NA | NA | NA | NA | NA | NA | NA | NA | NA |
| Sevgi et al. 2011 | 41.15 | 28.91 | 31 | 12.8 | 1074 | conifer | ECM | non-N-fixing | forest floor | NA | NA | NA | NA | 341 | NA | NA | NA | NA | NA | NA | NA | NA | NA |
| Sevgi et al. 2011 | 41.15 | 28.91 | 31 | 12.8 | 1074 | conifer | ECM | non-N-fixing | mineral | NA | 23.77 | NA | NA | 44.3 | NA | NA | NA | NA | NA | NA | NA | NA | 4.63 |
| Sevgi et al. 2011 | 41.15 | 28.91 | 31 | 12.8 | 1074 | boradleaf | ECM | non-N-fixing | mineral | NA | 20.83 | NA | NA | 43.3 | NA | NA | NA | NA | NA | NA | NA | NA | 5.29 |
| Sevgi et al. 2011 | 41.15 | 28.91 | 31 | 12.8 | 1074 | conifer | ECM | non-N-fixing | mineral | NA | 19.89 | NA | NA | 34.5 | NA | NA | NA | NA | NA | NA | NA | NA | 5.72 |
| Sevgi et al. 2011 | 41.15 | 28.91 | 31 | 12.8 | 1074 | conifer | ECM | non-N-fixing | mineral | NA | 29.31 | NA | NA | 54.4 | NA | NA | NA | NA | NA | NA | NA | NA | 6.04 |
| Sevgi et al. 2011 | 41.15 | 28.91 | 31 | 12.8 | 1074 | conifer | ECM | non-N-fixing | mineral | NA | 21.96 | NA | NA | 38.3 | NA | NA | NA | NA | NA | NA | NA | NA | 4.84 |
| Sevgi et al. 2011 | 41.15 | 28.91 | 31 | 12.8 | 1074 | conifer | ECM | non-N-fixing | mineral | NA | 28.31 | NA | NA | 50.7 | NA | NA | NA | NA | NA | NA | NA | NA | 4.63 |
| Laganiere et al. 2011 | 49.133 | -78.767 | 305 | 0.7 | 890 | conifer | ECM | non-N-fixing | mineral | NA | 46.3 | 2.5 | 18.8 | 37.5 | 2 | NA | NA | NA | NA | NA | NA | NA | 4.3 |
| Laganiere et al. 2011 | 49.133 | -78.767 | 305 | 0.7 | 890 | boradleaf | ECM | non-N-fixing | mineral | NA | 34.7 | 2.4 | 14.6 | 26.6 | 1.8 | NA | NA | NA | NA | NA | NA | NA | 4.7 |
| Taylor et al. 2016 | 33.883 | -83.35 | 170 | 16.5 | 1270 | boradleaf | AM | non-N-fixing | mineral | NA | NA | NA | 15.08 | 28.652 | 1.9 | 1.42 | 0.04244 | NA | NA | NA | NA | NA | 6.1 |
| Taylor et al. 2016 | 33.883 | -83.35 | 170 | 16.5 | 1270 | boradleaf | AM | non-N-fixing | mineral | NA | NA | NA | 16.09 | 30.571 | 1.9 | 0.85 | 0.02208 | NA | NA | NA | NA | NA | 6.52 |
| Taylor et al. 2016 | 33.883 | -83.35 | 170 | 16.5 | 1270 | boradleaf | AM | non-N-fixing | mineral | NA | NA | NA | 18.64 | 26.096 | 1.4 | 0.52 | 0.02848 | NA | NA | NA | NA | NA | 5.84 |
| Taylor et al. 2016 | 33.883 | -83.35 | 170 | 16.5 | 1270 | conifer | AM | non-N-fixing | mineral | NA | NA | NA | 20.93 | 39.767 | 1.9 | 1.61 | 0.01727 | NA | NA | NA | NA | NA | 5.73 |
| Taylor et al. 2016 | 33.883 | -83.35 | 170 | 16.5 | 1270 | boradleaf | ECM | non-N-fixing | mineral | NA | NA | NA | 15.8 | 31.6 | 2 | 1.22 | 0.01709 | NA | NA | NA | NA | NA | 6.02 |
| Taylor et al. 2016 | 33.883 | -83.35 | 170 | 16.5 | 1270 | boradleaf | ECM | non-N-fixing | mineral | NA | NA | NA | 19.35 | 40.635 | 2.1 | 0.59 | 0.01403 | NA | NA | NA | NA | NA | 5.33 |
| Taylor et al. 2016 | 33.883 | -83.35 | 170 | 16.5 | 1270 | boradleaf | ECM | non-N-fixing | mineral | NA | NA | NA | 17.92 | 26.88 | 1.5 | 1.06 | 0.01423 | NA | NA | NA | NA | NA | 5.83 |
| Taylor et al. 2016 | 33.883 | -83.35 | 170 | 16.5 | 1270 | conifer | ECM | non-N-fixing | mineral | NA | NA | NA | 23.06 | 34.59 | 1.5 | 1.42 | 0.00918 | NA | NA | NA | NA | NA | 5.36 |
| Wan et al. 2015 | 26.8 | 117.967 | 277 | 19.3 | 1669 | boradleaf | AM | non-N-fixing | forest floor | 9.5 | NA | NA | 24.4 | 469.3 | 19.3 | NA | NA | NA | NA | NA | NA | NA | NA |
| Wan et al. 2015 | 26.8 | 117.967 | 277 | 19.3 | 1669 | conifer | AM | non-N-fixing | forest floor | 4.3 | NA | NA | 57.7 | 513.6 | 8.9 | NA | NA | NA | NA | NA | NA | NA | NA |
| Wan et al. 2015 | 26.8 | 117.967 | 277 | 19.3 | 1669 | boradleaf | AM | non-N-fixing | mineral | NA | NA | NA | 17.9 | 39.5 | 2.2 | 13 | 0.49 | NA | NA | NA | NA | NA | NA |
| Wan et al. 2015 | 26.8 | 117.967 | 277 | 19.3 | 1669 | conifer | AM | non-N-fixing | mineral | NA | NA | NA | 16.2 | 30.2 | 1.9 | 12.2 | 3.86 | NA | NA | NA | NA | NA | NA |
| Ulery et al. 1995 | 34.2 | -118.15 | 428 | 14.4 | 678 | boradleaf | ECM | non-N-fixing | mineral | NA | 22.41 | 1.09 | NA | 34.8 | 1.7 | NA | NA | NA | NA | NA | NA | NA | 5.8 |
| Ulery et al. 1995 | 34.2 | -118.15 | 428 | 14.4 | 678 | conifer | ECM | non-N-fixing | mineral | NA | 1.39 | 0.12 | NA | 12.9 | 1.1 | NA | NA | NA | NA | NA | NA | NA | 4.9 |
| Ulery et al. 1995 | 34.2 | -118.15 | 428 | 14.4 | 678 | boradleaf | ECM | non-N-fixing | mineral | NA | 6.03 | 0.23 | NA | 43.7 | 1.7 | NA | NA | NA | NA | NA | NA | NA | 5.7 |
| Ulery et al. 1995 | 34.2 | -118.15 | 428 | 14.4 | 678 | boradleaf | AM | non-N-fixing | mineral | NA | 10.06 | 0.74 | NA | 54.7 | 4 | NA | NA | NA | NA | NA | NA | NA | 5.4 |
| Michelsen et al. 1993 | 8.867 | 38.533 | 2380 | 18.2 | 1019 | conifer | AM | non-N-fixing | mineral | 5.01 | 171.4 | 12.74 | NA | 36.06 | 2.68 | NA | NA | 2.6 | NA | NA | NA | NA | 4.91 |
| Michelsen et al. 1993 | 8.867 | 38.533 | 2380 | 18.2 | 1019 | boradleaf | ECM | non-N-fixing | mineral | 5.83 | 164.4 | 13.44 | NA | 33.53 | 2.74 | NA | NA | 4.2 | NA | NA | NA | NA | 5 |
| Michelsen et al. 1993 | 8.867 | 38.533 | 2380 | 18.2 | 1019 | conifer | AM | non-N-fixing | mineral | 6.53 | 153 | 12.97 | NA | 39.76 | 3.37 | NA | NA | 17.6 | NA | NA | NA | NA | 5.47 |
| Lemma et al. 2006 | 7.55 | 36.583 | 2200 | 19.4 | 1517 | conifer | AM | non-N-fixing | mineral | NA | 26.2 | NA | NA | 101.1 | NA | NA | NA | NA | NA | NA | NA | NA | 6.5 |
| Lemma et al. 2006 | 7.55 | 36.583 | 2200 | 19.4 | 1517 | conifer | ECM | non-N-fixing | mineral | NA | 14.5 | NA | NA | 56.3 | NA | NA | NA | NA | NA | NA | NA | NA | 5.6 |
| Lemma et al. 2006 | 7.55 | 36.583 | 2200 | 19.4 | 1517 | boradleaf | ECM | non-N-fixing | mineral | NA | 23.6 | NA | NA | 91.5 | NA | NA | NA | NA | NA | NA | NA | NA | 5.9 |
| Kasel et al. 2011 | -36.833 | 143.016 | 275 | 14 | 434 | boradleaf | AM | non-N-fixing | mineral | NA | 27.78 | NA | NA | 49.15 | NA | NA | NA | NA | NA | NA | NA | NA | NA |
| Kasel et al. 2011 | -36.833 | 143.016 | 275 | 14 | 434 | boradleaf | AM | non-N-fixing | mineral | NA | 40.91 | NA | NA | 78.92 | NA | NA | NA | NA | NA | NA | NA | NA | NA |
| Kasel et al. 2011 | -36.833 | 143.016 | 275 | 14 | 434 | conifer | ECM | non-N-fixing | mineral | NA | 32.16 | NA | NA | 49.85 | NA | NA | NA | NA | NA | NA | NA | NA | NA |
| Kasel et al. 2011 | -36.833 | 143.016 | 275 | 14 | 434 | boradleaf | ECM | non-N-fixing | mineral | NA | 26.91 | NA | NA | 52.62 | NA | NA | NA | NA | NA | NA | NA | NA | NA |
| Kasel et al. 2011 | -37.117 | 144.117 | 268 | 13.5 | 463 | boradleaf | AM | non-N-fixing | mineral | NA | 30.63 | NA | NA | 43.62 | NA | NA | NA | NA | NA | NA | NA | NA | NA |
| Kasel et al. 2011 | -37.117 | 144.117 | 268 | 13.5 | 463 | boradleaf | AM | non-N-fixing | mineral | NA | 44.84 | NA | NA | 64.38 | NA | NA | NA | NA | NA | NA | NA | NA | NA |
| Kasel et al. 2011 | -37.117 | 144.117 | 268 | 13.5 | 463 | conifer | ECM | non-N-fixing | mineral | NA | 29.97 | NA | NA | 40.15 | NA | NA | NA | NA | NA | NA | NA | NA | NA |
| Kasel et al. 2011 | -37.117 | 144.117 | 268 | 13.5 | 463 | boradleaf | ECM | non-N-fixing | mineral | NA | 35.66 | NA | NA | 51.23 | NA | NA | NA | NA | NA | NA | NA | NA | NA |
| Kasel et al. 2011 | -36.933 | 144.283 | 551 | 12.9 | 487 | boradleaf | AM | non-N-fixing | mineral | NA | 45.5 | NA | NA | 114.09 | NA | NA | NA | NA | NA | NA | NA | NA | NA |
| Kasel et al. 2011 | -36.933 | 144.283 | 551 | 12.9 | 487 | boradleaf | AM | non-N-fixing | mineral | NA | 51.84 | NA | NA | 130.78 | NA | NA | NA | NA | NA | NA | NA | NA | NA |
| Kasel et al. 2011 | -36.933 | 144.283 | 551 | 12.9 | 487 | conifer | ECM | non-N-fixing | mineral | NA | 46.81 | NA | NA | 86.26 | NA | NA | NA | NA | NA | NA | NA | NA | NA |
| Kasel et al. 2011 | -36.933 | 144.283 | 551 | 12.9 | 487 | boradleaf | ECM | non-N-fixing | mineral | NA | 35.66 | NA | NA | 77.22 | NA | NA | NA | NA | NA | NA | NA | NA | NA |
| Kasel et al. 2011 | -37.033 | 144.75 | 383 | 13.4 | 482 | boradleaf | AM | non-N-fixing | mineral | NA | 30.63 | NA | NA | 54.26 | NA | NA | NA | NA | NA | NA | NA | NA | NA |
| Kasel et al. 2011 | -37.033 | 144.75 | 383 | 13.4 | 482 | boradleaf | AM | non-N-fixing | mineral | NA | 43.09 | NA | NA | 84.17 | NA | NA | NA | NA | NA | NA | NA | NA | NA |
| Kasel et al. 2011 | -37.033 | 144.75 | 383 | 13.4 | 482 | conifer | ECM | non-N-fixing | mineral | NA | 35.22 | NA | NA | 59.13 | NA | NA | NA | NA | NA | NA | NA | NA | NA |
| Kasel et al. 2011 | -37.033 | 144.75 | 383 | 13.4 | 482 | boradleaf | ECM | non-N-fixing | mineral | NA | 36.09 | NA | NA | 72.35 | NA | NA | NA | NA | NA | NA | NA | NA | NA |
| Lu et al. 2012 | -25.933 | 153.083 | 1287 | 21.3 | 1287 | conifer | ECM | non-N-fixing | mineral | NA | 20.18 | 0.56 | 35.9 | 19.4 | 0.54 | 22.4 | 0.03 | NA | NA | NA | NA | NA | 4.5 |
| Lu et al. 2012 | -25.933 | 153.083 | 1287 | 21.3 | 1287 | conifer | AM | non-N-fixing | mineral | NA | 14.5 | 0.66 | 21.9 | 12.5 | 0.57 | 22.9 | 2.38 | NA | NA | NA | NA | NA | 6 |
| Lu et al. 2012 | -25.933 | 153.083 | 1287 | 21.3 | 1287 | conifer | AM | non-N-fixing | mineral | NA | 12.53 | 0.6 | 20.8 | 10.8 | 0.52 | 21.4 | 8.02 | NA | NA | NA | NA | NA | 6.2 |
| Quideau et al. 1998 | 34.283 | -117.633 | 678 | 14.4 | 678 | conifer | ECM | non-N-fixing | forest floor | 5.18 | 5.18 | 0.14 | NA | NA | NA | NA | NA | NA | NA | NA | NA | NA | NA |
| Quideau et al. 1998 | 34.283 | -117.633 | 678 | 14.4 | 678 | boradleaf | ECM | non-N-fixing | forest floor | 4.9 | 4.9 | 0.11 | NA | NA | NA | NA | NA | NA | NA | NA | NA | NA | NA |
| Quideau et al. 1998 | 34.283 | -117.633 | 678 | 14.4 | 678 | boradleaf | AM | non-N-fixing | forest floor | 4.08 | 4.08 | 0.08 | NA | NA | NA | NA | NA | NA | NA | NA | NA | NA | NA |
| Quideau et al. 1998 | 34.283 | -117.633 | 678 | 14.4 | 678 | boradleaf | ECM | non-N-fixing | forest floor | 3.22 | 3.22 | 0.1 | NA | NA | NA | NA | NA | NA | NA | NA | NA | NA | NA |
| Quideau et al. 1998 | 34.283 | -117.633 | 678 | 14.4 | 678 | conifer | ECM | non-N-fixing | mineral | NA | 20.3 | 0.44 | NA | NA | NA | NA | NA | NA | NA | NA | NA | NA | NA |
| Quideau et al. 1998 | 34.283 | -117.633 | 678 | 14.4 | 678 | boradleaf | ECM | non-N-fixing | mineral | NA | 37.6 | 1.12 | NA | NA | NA | NA | NA | NA | NA | NA | NA | NA | NA |
| Quideau et al. 1998 | 34.283 | -117.633 | 678 | 14.4 | 678 | boradleaf | AM | non-N-fixing | mineral | NA | 13.98 | 1.19 | NA | NA | NA | NA | NA | NA | NA | NA | NA | NA | NA |
| Quideau et al. 1998 | 34.283 | -117.633 | 678 | 14.4 | 678 | boradleaf | ECM | non-N-fixing | mineral | NA | 9.02 | 0.59 | NA | NA | NA | NA | NA | NA | NA | NA | NA | NA | NA |
| Templer et al. 2003 | 41.983 | -74.383 | 842 | 4.3 | 1530 | boradleaf | ECM | non-N-fixing | mineral | NA | NA | NA | 18.36 | 343.1 | 18.8 | 69.4 | 11 | NA | 236.082 | NA | NA | NA | NA |
| Templer et al. 2003 | 41.983 | -74.383 | 842 | 4.3 | 1530 | conifer | ECM | non-N-fixing | mineral | NA | NA | NA | 20.04 | 465.5 | 23.3 | 61.5 | 6.3 | NA | 186.566 | NA | NA | NA | NA |
| Templer et al. 2003 | 41.983 | -74.383 | 842 | 4.3 | 1530 | boradleaf | ECM | non-N-fixing | mineral | NA | NA | NA | 20.74 | 358.6 | 17.4 | 56.4 | 1.3 | NA | 141.813 | NA | NA | NA | NA |
| Templer et al. 2003 | 41.983 | -74.383 | 842 | 4.3 | 1530 | boradleaf | AM | non-N-fixing | mineral | NA | NA | NA | 17.29 | 427.1 | 24.9 | 51.2 | 53.6 | NA | 128.013 | NA | NA | NA | NA |
| Templer et al. 2003 | 41.983 | -74.383 | 842 | 4.3 | 1530 | boradleaf | ECM | non-N-fixing | mineral | NA | NA | NA | 19.54 | 434.7 | 22.3 | 81 | 8 | NA | 233.248 | NA | NA | NA | NA |
| Grayston & Prescott 2005 | 48.817 | -124.783 | 853 | 8.8 | 3943 | conifer | AM | non-N-fixing | mineral | NA | NA | NA | 37 | 483 | 13.6 | NA | NA | NA | NA | NA | NA | NA | NA |
| Grayston & Prescott 2005 | 48.817 | -124.783 | 853 | 8.8 | 3943 | conifer | ECM | non-N-fixing | mineral | NA | NA | NA | 47 | 484 | 11.4 | NA | NA | NA | NA | NA | NA | NA | NA |
| Grayston & Prescott 2005 | 48.817 | -124.783 | 853 | 8.8 | 3943 | conifer | ECM | non-N-fixing | mineral | NA | NA | NA | 34 | 474 | 14.4 | NA | NA | NA | NA | NA | NA | NA | NA |
| Grayston & Prescott 2005 | 48.817 | -124.783 | 853 | 8.8 | 3943 | conifer | ECM | non-N-fixing | mineral | NA | NA | NA | 35.6 | 454 | 13.2 | NA | NA | NA | NA | NA | NA | NA | NA |
| Nsabimana et al. 2009 | -2.6 | 29.733 | 1743 | 19.1 | 1246.4 | boradleaf | AM | non-N-fixing | mineral | NA | 81.9 | 5.4 | 15.17 | NA | NA | NA | NA | NA | NA | NA | NA | NA | 5.3 |
| Nsabimana et al. 2009 | -2.6 | 29.733 | 1743 | 19.1 | 1246.4 | boradleaf | AM | non-N-fixing | mineral | NA | 92.7 | 5.9 | 15.71 | NA | NA | NA | NA | NA | NA | NA | NA | NA | 5.2 |
| Nsabimana et al. 2009 | -2.6 | 29.733 | 1743 | 19.1 | 1246.4 | boradleaf | AM | non-N-fixing | mineral | NA | 83.9 | 5.5 | 15.25 | NA | NA | NA | NA | NA | NA | NA | NA | NA | 5.4 |
| Nsabimana et al. 2009 | -2.6 | 29.733 | 1743 | 19.1 | 1246.4 | boradleaf | AM | non-N-fixing | mineral | NA | 85.4 | 5.6 | 15.25 | NA | NA | NA | NA | NA | NA | NA | NA | NA | 5.2 |
| Nsabimana et al. 2009 | -2.6 | 29.733 | 1743 | 19.1 | 1246.4 | boradleaf | ECM | non-N-fixing | mineral | NA | 98.4 | 4.9 | 20.08 | NA | NA | NA | NA | NA | NA | NA | NA | NA | 4.2 |
| Nsabimana et al. 2009 | -2.6 | 29.733 | 1743 | 19.1 | 1246.4 | boradleaf | ECM | non-N-fixing | mineral | NA | 64.5 | 3.1 | 20.81 | NA | NA | NA | NA | NA | NA | NA | NA | NA | 3.9 |
| Nsabimana et al. 2009 | -2.6 | 29.733 | 1743 | 19.1 | 1246.4 | boradleaf | ECM | non-N-fixing | mineral | NA | 83.5 | 5 | 16.7 | NA | NA | NA | NA | NA | NA | NA | NA | NA | 5.2 |
| Nsabimana et al. 2009 | -2.6 | 29.733 | 1743 | 19.1 | 1246.4 | boradleaf | ECM | non-N-fixing | mineral | NA | 67.7 | 3.1 | 21.84 | NA | NA | NA | NA | NA | NA | NA | NA | NA | 3.5 |
| Nsabimana et al. 2009 | -2.6 | 29.733 | 1743 | 19.1 | 1246.4 | boradleaf | ECM | non-N-fixing | mineral | NA | 88.7 | 5.7 | 15.56 | NA | NA | NA | NA | NA | NA | NA | NA | NA | 5.2 |
| Nsabimana et al. 2009 | -2.6 | 29.733 | 1743 | 19.1 | 1246.4 | boradleaf | ECM | non-N-fixing | mineral | NA | 57.3 | 3 | 19.1 | NA | NA | NA | NA | NA | NA | NA | NA | NA | 4.2 |
| Nsabimana et al. 2009 | -2.6 | 29.733 | 1743 | 19.1 | 1246.4 | boradleaf | ECM | non-N-fixing | mineral | NA | 109.3 | 5.7 | 19.18 | NA | NA | NA | NA | NA | NA | NA | NA | NA | 4.7 |
| Nsabimana et al. 2009 | -2.6 | 29.733 | 1743 | 19.1 | 1246.4 | boradleaf | ECM | non-N-fixing | mineral | NA | 85.7 | 4.4 | 19.48 | NA | NA | NA | NA | NA | NA | NA | NA | NA | 4.6 |
| Hoogmoed et al. 2014 | -36.86 | 145.58 | 243 | 14.1 | 650 | boradleaf | AM | non-N-fixing | forest floor | 2.28 | NA | NA | 32.1 | 439 | 14.4 | NA | NA | NA | NA | NA | NA | NA | NA |
| Hoogmoed et al. 2014 | -36.86 | 145.58 | 243 | 14.1 | 650 | boradleaf | AM | N-fixing | forest floor | 1.86 | NA | NA | 35.4 | 446 | 13.3 | NA | NA | NA | NA | NA | NA | NA | NA |
| Hoogmoed et al. 2014 | -36.86 | 145.58 | 243 | 14.1 | 650 | boradleaf | ECM | non-N-fixing | forest floor | 1.88 | NA | NA | 45.6 | 443 | 10.5 | NA | NA | NA | NA | NA | NA | NA | NA |
| Hoogmoed et al. 2014 | -36.86 | 145.58 | 243 | 14.1 | 650 | boradleaf | ECM | non-N-fixing | forest floor | 2.75 | NA | NA | 47.9 | 445 | 9.7 | NA | NA | NA | NA | NA | NA | NA | NA |
| Hoogmoed et al. 2014 | -36.86 | 145.58 | 243 | 14.1 | 650 | boradleaf | AM | non-N-fixing | mineral | NA | 7.58 | 0.647 | 11.8 | NA | NA | NA | NA | NA | NA | NA | NA | NA | NA |
| Hoogmoed et al. 2014 | -36.86 | 145.58 | 243 | 14.1 | 650 | boradleaf | AM | N-fixing | mineral | NA | 4.39 | 0.336 | 13.1 | NA | NA | NA | NA | NA | NA | NA | NA | NA | NA |
| Hoogmoed et al. 2014 | -36.86 | 145.58 | 243 | 14.1 | 650 | boradleaf | ECM | non-N-fixing | mineral | NA | 3.83 | 0.301 | 12.6 | NA | NA | NA | NA | NA | NA | NA | NA | NA | NA |
| Hoogmoed et al. 2014 | -36.86 | 145.58 | 243 | 14.1 | 650 | boradleaf | ECM | non-N-fixing | mineral | NA | 4.84 | 0.358 | 13.6 | NA | NA | NA | NA | NA | NA | NA | NA | NA | NA |
| Mertens et al. 2007 | 51.917 | 4.5 | 25 | 9.6 | 785 | boradleaf | ECM | non-N-fixing | mineral | 2.8 | 2.35 | NA | NA | NA | NA | NA | NA | NA | NA | NA | NA | NA | 5.6 |
| Mertens et al. 2007 | 51.917 | 4.5 | 25 | 9.6 | 785 | boradleaf | AM | non-N-fixing | mineral | 3.1 | 1.71 | NA | NA | NA | NA | NA | NA | NA | NA | NA | NA | NA | 7.1 |
| Mertens et al. 2007 | 51.917 | 4.5 | 25 | 9.6 | 785 | boradleaf | AM | non-N-fixing | mineral | 3.8 | 1.9 | NA | NA | NA | NA | NA | NA | NA | NA | NA | NA | NA | 6.8 |
| Mertens et al. 2007 | 51.917 | 4.5 | 25 | 9.6 | 785 | boradleaf | ECM | non-N-fixing | mineral | 2.2 | 2.32 | NA | NA | NA | NA | NA | NA | NA | NA | NA | NA | NA | 6.5 |
| Yin et al. 2014 | 25.25 | 116.917 | 1056 | 17.8 | 1785 | conifer | ECM | non-N-fixing | forest floor | NA | NA | NA | 18.1 | 34.12 | 1.88 | NA | NA | NA | 490 | NA | NA | NA | 4.05 |
| Yin et al. 2014 | 25.25 | 116.917 | 1056 | 17.8 | 1785 | boradleaf | AM | non-N-fixing | forest floor | NA | NA | NA | 19.7 | 32.25 | 1.64 | NA | NA | NA | 420 | NA | NA | NA | 4.55 |
| Yin et al. 2014 | 25.25 | 116.917 | 1056 | 17.8 | 1785 | boradleaf | AM | N-fixing | forest floor | NA | NA | NA | 15.4 | 41.68 | 2.71 | NA | NA | NA | 440 | NA | NA | NA | 3.86 |
| Yin et al. 2014 | 25.25 | 116.917 | 1056 | 17.8 | 1785 | conifer | ECM | non-N-fixing | mineral | NA | 8.03 | 0.55 | 14.7 | 13.05 | 0.89 | NA | NA | NA | 200 | NA | NA | NA | 4.38 |
| Yin et al. 2014 | 25.25 | 116.917 | 1056 | 17.8 | 1785 | boradleaf | AM | non-N-fixing | mineral | NA | 8.11 | 0.57 | 14.2 | 12.77 | 0.9 | NA | NA | NA | 200 | NA | NA | NA | 4.38 |
| Yin et al. 2014 | 25.25 | 116.917 | 1056 | 17.8 | 1785 | boradleaf | AM | N-fixing | mineral | NA | 9.96 | 0.61 | 16.3 | 17.79 | 1.09 | NA | NA | NA | 190 | NA | NA | NA | 4.17 |
| Smith et al. 2002 | -2 | -54 | 2 | 26 | 1900 | conifer | ECM | non-N-fixing | forest floor | 5.135 | 5.42 | 0.062 | NA | NA | NA | NA | NA | NA | NA | NA | NA | NA | NA |
| Smith et al. 2002 | -2 | -54 | 2 | 26 | 1900 | boradleaf | AM | non-N-fixing | forest floor | 4.27 | 4.92 | 0.149 | NA | NA | NA | NA | NA | NA | NA | NA | NA | NA | NA |
| Smith et al. 2002 | -2 | -54 | 2 | 26 | 1900 | boradleaf | ECM | non-N-fixing | forest floor | 3.804 | 3.595 | 0.051 | NA | NA | NA | NA | NA | NA | NA | NA | NA | NA | NA |
| Smith et al. 2002 | -2 | -54 | 2 | 26 | 1900 | conifer | ECM | non-N-fixing | mineral | NA | 92 | 4.957 | 18.66 | 50 | 2.68 | NA | NA | NA | NA | NA | NA | NA | NA |
| Smith et al. 2002 | -2 | -54 | 2 | 26 | 1900 | boradleaf | AM | non-N-fixing | mineral | NA | 91 | 5.804 | 15.85 | 49.6 | 3.13 | NA | NA | NA | NA | NA | NA | NA | NA |
| Smith et al. 2002 | -2 | -54 | 2 | 26 | 1900 | boradleaf | ECM | non-N-fixing | mineral | NA | 116 | 7.586 | 15.34 | 62.9 | 4.1 | NA | NA | NA | NA | NA | NA | NA | NA |
| Lemenih et al. 2004 | 7.333 | 38.75 | 2106 | 20 | 1200 | conifer | AM | non-N-fixing | mineral | NA | 53.72 | 5.02 | 10.71 | 63.2 | 5.9 | NA | NA | 7.9 | NA | NA | NA | NA | 6.2 |
| Lemenih et al. 2004 | 7.333 | 38.75 | 2106 | 20 | 1200 | boradleaf | ECM | non-N-fixing | mineral | NA | 37.54 | 3.65 | 10.29 | 39.1 | 3.8 | NA | NA | 5.7 | NA | NA | NA | NA | 5.4 |
| Zheng et al. 2005 | 27.083 | 112.3 | 114 | 17.9 | 1237 | conifer | ECM | non-N-fixing | mineral | NA | NA | NA | 14.83 | 8.01 | 0.54 | NA | NA | 0.53 | 119.913 | NA | NA | NA | 4.22 |
| Zheng et al. 2005 | 27.083 | 112.3 | 114 | 17.9 | 1237 | conifer | AM | non-N-fixing | mineral | NA | NA | NA | 14.62 | 10.67 | 0.73 | NA | NA | 0.95 | 129.22 | NA | NA | NA | 4.37 |
| Zheng et al. 2005 | 27.083 | 112.3 | 114 | 17.9 | 1237 | boradleaf | AM | non-N-fixing | mineral | NA | NA | NA | 14.3 | 10.58 | 0.74 | NA | NA | 1.22 | 141.224 | NA | NA | NA | 4.54 |
| Raich et al. 2007 | 10.083 | -84.083 | 1993 | 25.8 | 4000 | boradleaf | AM | non-N-fixing | forest floor | 11.7 | 5.9 | 0.169 | 34.91 | NA | NA | NA | NA | NA | NA | NA | NA | NA | NA |
| Raich et al. 2007 | 10.083 | -84.083 | 1993 | 25.8 | 4000 | boradleaf | AM | N-fixing | forest floor | 10.3 | 5.3 | 0.208 | 25.48 | NA | NA | NA | NA | NA | NA | NA | NA | NA | NA |
| Raich et al. 2007 | 10.083 | -84.083 | 1993 | 25.8 | 4000 | conifer | ECM | non-N-fixing | forest floor | 10 | 5 | 0.137 | 36.5 | NA | NA | NA | NA | NA | NA | NA | NA | NA | NA |
| Raich et al. 2007 | 10.083 | -84.083 | 1993 | 25.8 | 4000 | boradleaf | AM | non-N-fixing | forest floor | 7.4 | 3.8 | 0.122 | 31.15 | NA | NA | NA | NA | NA | NA | NA | NA | NA | NA |
| Raich et al. 2007 | 10.083 | -84.083 | 1993 | 25.8 | 4000 | boradleaf | AM | non-N-fixing | forest floor | 11.4 | 5.6 | 0.147 | 38.1 | NA | NA | NA | NA | NA | NA | NA | NA | NA | NA |
| Raich et al. 2007 | 10.083 | -84.083 | 1993 | 25.8 | 4000 | boradleaf | AM | non-N-fixing | forest floor | 10.1 | 4.5 | 0.162 | 27.78 | NA | NA | NA | NA | NA | NA | NA | NA | NA | NA |
| Zheng et al. 2008 | 27.083 | 112.3 | 143 | 17.9 | 1237 | conifer | ECM | non-N-fixing | forest floor | NA | NA | NA | NA | 576.61 | NA | NA | NA | NA | NA | NA | NA | NA | NA |
| Zheng et al. 2008 | 27.083 | 112.3 | 143 | 17.9 | 1237 | conifer | AM | non-N-fixing | forest floor | NA | NA | NA | NA | 438.3 | NA | NA | NA | NA | NA | NA | NA | NA | NA |
| Zheng et al. 2008 | 27.083 | 112.3 | 143 | 17.9 | 1237 | boradleaf | AM | non-N-fixing | forest floor | NA | NA | NA | NA | 492.8 | NA | NA | NA | NA | NA | NA | NA | NA | NA |
| Johnsen et al. 2013 | 30.82 | -84.76 | 34 | 18.9 | 1257 | boradleaf | AM | non-N-fixing | mineral | NA | 62.2 | 2.82 | 24 | 79.4 | 3.7 | NA | NA | NA | NA | NA | NA | NA | NA |
| Johnsen et al. 2013 | 30.82 | -84.76 | 34 | 18.9 | 1257 | conifer | ECM | non-N-fixing | mineral | NA | 59.3 | 2.02 | 141 | 72.5 | 2.5 | NA | NA | NA | NA | NA | NA | NA | NA |
| Tang & Li 2014 | 25.667 | 101.857 | 1181 | 21.6 | 634.3 | boradleaf | AM | N-fixing | mineral | 3.22 | 18.29 | 1.74 | 10.51 | 7.67 | 0.73 | NA | NA | 14.74 | 184.69 | 18.28 | 9.14 | 4.08 | 6.2 |
| Tang & Li 2014 | 25.667 | 101.857 | 1181 | 21.6 | 634.3 | boradleaf | AM | non-N-fixing | mineral | 2.9 | 16.66 | 1.64 | 10.19 | 7.03 | 0.69 | NA | NA | 14.6 | 180.46 | 18.51 | 9.32 | 4.08 | 6.22 |
| Tang & Li 2014 | 25.667 | 101.857 | 1181 | 21.6 | 634.3 | boradleaf | ECM | N-fixing | mineral | 0.95 | 10.99 | 0.94 | 11.67 | 4.55 | 0.39 | NA | NA | 13.25 | 170.83 | 17.44 | 7.98 | 3.84 | 6.3 |
| Tang & Li 2014 | 25.667 | 101.857 | 1181 | 21.6 | 634.3 | boradleaf | AM | non-N-fixing | mineral | 1.91 | 13.88 | 1.02 | 13.6 | 5.71 | 0.42 | NA | NA | 13.54 | 168.59 | 17.08 | 7.94 | 3.84 | 6.24 |
| Tang & Li 2014 | 25.667 | 101.857 | 1181 | 21.6 | 634.3 | boradleaf | ECM | non-N-fixing | mineral | 1.41 | 12.32 | 0.92 | 13.34 | 5.07 | 0.38 | NA | NA | 13.88 | 166.89 | 16.68 | 7.68 | 3.84 | 6.3 |
| Jiang & Xu 2006 | 30.233 | 119.7 | 114 | 15.9 | 1424 | conifer | ECM | non-N-fixing | mineral | 2.05 | NA | NA | NA | 15.97 | NA | NA | NA | NA | 260 | NA | NA | NA | NA |
| Jiang & Xu 2006 | 30.233 | 119.7 | 114 | 15.9 | 1424 | conifer | AM | non-N-fixing | mineral | NA | NA | NA | NA | 16 | NA | NA | NA | NA | 257 | NA | NA | NA | NA |
| Yan et al. 2014 | 28.117 | 113.033 | 59 | 17.2 | 1422 | boradleaf | AM | non-N-fixing | mineral | 3.26 | NA | NA | 9.1 | 0.92 | 0.101 | NA | NA | NA | NA | NA | NA | NA | 4.1 |
| Yan et al. 2014 | 28.117 | 113.033 | 59 | 17.2 | 1422 | conifer | AM | non-N-fixing | mineral | 3.66 | NA | NA | 18.2 | 2.31 | 0.127 | NA | NA | NA | NA | NA | NA | NA | 4 |
| Yan et al. 2014 | 28.117 | 113.033 | 59 | 17.2 | 1422 | conifer | ECM | non-N-fixing | mineral | 3.41 | NA | NA | 9.9 | 0.95 | 0.096 | NA | NA | NA | NA | NA | NA | NA | 4.1 |
| Yan et al. 2014 | 28.117 | 113.033 | 59 | 17.2 | 1422 | conifer | ECM | non-N-fixing | mineral | 3.79 | NA | NA | 8.7 | 1.43 | 0.164 | NA | NA | NA | NA | NA | NA | NA | 4 |
| Devi et al. 2013 | 30.833 | 77.133 | 1179 | 19 | 1100 | boradleaf | ECM | non-N-fixing | forest floor | 3.098 | 5.86 | NA | NA | NA | NA | NA | NA | NA | NA | NA | NA | NA | NA |
| Devi et al. 2013 | 30.833 | 77.133 | 1179 | 19 | 1100 | conifer | ECM | non-N-fixing | forest floor | 6.79 | 7.19 | NA | NA | NA | NA | NA | NA | NA | NA | NA | NA | NA | NA |
| Devi et al. 2013 | 30.833 | 77.133 | 1179 | 19 | 1100 | boradleaf | AM | N-fixing | forest floor | 4.52 | 12.69 | NA | NA | NA | NA | NA | NA | NA | NA | NA | NA | NA | NA |
| Devi et al. 2013 | 30.833 | 77.133 | 1179 | 19 | 1100 | boradleaf | AM | N-fixing | forest floor | 3.33 | 5.66 | NA | NA | NA | NA | NA | NA | NA | NA | NA | NA | NA | NA |
| Devi et al. 2013 | 30.833 | 77.133 | 1179 | 19 | 1100 | boradleaf | AM | N-fixing | forest floor | 2.99 | 8.34 | NA | NA | NA | NA | NA | NA | NA | NA | NA | NA | NA | NA |
| Devi et al. 2013 | 30.833 | 77.133 | 1179 | 19 | 1100 | boradleaf | ECM | N-fixing | forest floor | 2.88 | 6.6 | NA | NA | NA | NA | NA | NA | NA | NA | NA | NA | NA | NA |
| Devi et al. 2013 | 30.833 | 77.133 | 1179 | 19 | 1100 | boradleaf | ECM | non-N-fixing | forest floor | 3.6 | 7.7 | NA | NA | NA | NA | NA | NA | NA | NA | NA | NA | NA | NA |
| Devi et al. 2013 | 30.833 | 77.133 | 1179 | 19 | 1100 | boradleaf | AM | non-N-fixing | forest floor | 4.03 | 3.01 | NA | NA | NA | NA | NA | NA | NA | NA | NA | NA | NA | NA |
| Devi et al. 2013 | 30.833 | 77.133 | 1179 | 19 | 1100 | boradleaf | ECM | non-N-fixing | mineral | NA | 60.7 | NA | NA | NA | NA | NA | NA | NA | NA | NA | NA | NA | NA |
| Devi et al. 2013 | 30.833 | 77.133 | 1179 | 19 | 1100 | conifer | ECM | non-N-fixing | mineral | NA | 53.41 | NA | NA | NA | NA | NA | NA | NA | NA | NA | NA | NA | NA |
| Devi et al. 2013 | 30.833 | 77.133 | 1179 | 19 | 1100 | boradleaf | AM | N-fixing | mineral | NA | 64.06 | NA | NA | NA | NA | NA | NA | NA | NA | NA | NA | NA | NA |
| Devi et al. 2013 | 30.833 | 77.133 | 1179 | 19 | 1100 | boradleaf | AM | N-fixing | mineral | NA | 50.12 | NA | NA | NA | NA | NA | NA | NA | NA | NA | NA | NA | NA |
| Devi et al. 2013 | 30.833 | 77.133 | 1179 | 19 | 1100 | boradleaf | AM | N-fixing | mineral | NA | 49.16 | NA | NA | NA | NA | NA | NA | NA | NA | NA | NA | NA | NA |
| Devi et al. 2013 | 30.833 | 77.133 | 1179 | 19 | 1100 | boradleaf | ECM | N-fixing | mineral | NA | 49.8 | NA | NA | NA | NA | NA | NA | NA | NA | NA | NA | NA | NA |
| Devi et al. 2013 | 30.833 | 77.133 | 1179 | 19 | 1100 | boradleaf | ECM | non-N-fixing | mineral | NA | 40.03 | NA | NA | NA | NA | NA | NA | NA | NA | NA | NA | NA | NA |
| Devi et al. 2013 | 30.833 | 77.133 | 1179 | 19 | 1100 | boradleaf | AM | non-N-fixing | mineral | NA | 49.8 | NA | NA | NA | NA | NA | NA | NA | NA | NA | NA | NA | NA |
| Devi et al. 2013 | 30.833 | 77.133 | 1179 | 19 | 1100 | boradleaf | ECM | non-N-fixing | mineral | NA | 165 | NA | NA | NA | NA | NA | NA | NA | NA | NA | NA | NA | NA |
| Devi et al. 2013 | 30.833 | 77.133 | 1179 | 19 | 1100 | conifer | ECM | non-N-fixing | mineral | NA | 165 | NA | NA | NA | NA | NA | NA | NA | NA | NA | NA | NA | NA |
| Devi et al. 2013 | 30.833 | 77.133 | 1179 | 19 | 1100 | boradleaf | AM | N-fixing | mineral | NA | 18 | NA | NA | NA | NA | NA | NA | NA | NA | NA | NA | NA | NA |
| Devi et al. 2013 | 30.833 | 77.133 | 1179 | 19 | 1100 | boradleaf | AM | N-fixing | mineral | NA | 195 | NA | NA | NA | NA | NA | NA | NA | NA | NA | NA | NA | NA |
| Devi et al. 2013 | 30.833 | 77.133 | 1179 | 19 | 1100 | boradleaf | AM | N-fixing | mineral | NA | 163 | NA | NA | NA | NA | NA | NA | NA | NA | NA | NA | NA | NA |
| Devi et al. 2013 | 30.833 | 77.133 | 1179 | 19 | 1100 | boradleaf | ECM | N-fixing | mineral | NA | 213 | NA | NA | NA | NA | NA | NA | NA | NA | NA | NA | NA | NA |
| Devi et al. 2013 | 30.833 | 77.133 | 1179 | 19 | 1100 | boradleaf | ECM | non-N-fixing | mineral | NA | 164 | NA | NA | NA | NA | NA | NA | NA | NA | NA | NA | NA | NA |
| Devi et al. 2013 | 30.833 | 77.133 | 1179 | 19 | 1100 | boradleaf | AM | non-N-fixing | mineral | NA | 207 | NA | NA | NA | NA | NA | NA | NA | NA | NA | NA | NA | NA |
| Jahed et al. 2014 | 36.317 | 51.85 | 1827 | 16.4 | 997 | boradleaf | AM | non-N-fixing | mineral | NA | NA | NA | 10.18 | 26.26 | 2.58 | NA | NA | 20.83 | NA | NA | NA | NA | 6.64 |
| Jahed et al. 2014 | 36.317 | 51.85 | 1827 | 16.4 | 997 | conifer | ECM | non-N-fixing | mineral | NA | NA | NA | 12.12 | 34.67 | 2.86 | NA | NA | 14.08 | NA | NA | NA | NA | 6.28 |
| Riestra et al. 2012 | -36.6 | -64.17 | 169 | 18.9 | 1257 | conifer | ECM | non-N-fixing | mineral | NA | 13.41 | NA | NA | 24.3 | NA | NA | NA | NA | NA | NA | NA | NA | NA |
| Riestra et al. 2012 | -36.6 | -64.17 | 169 | 18.9 | 1257 | boradleaf | ECM | non-N-fixing | mineral | NA | 10.31 | NA | NA | 17.9 | NA | NA | NA | NA | NA | NA | NA | NA | NA |
| Riestra et al. 2012 | -36.6 | -64.17 | 169 | 18.9 | 1257 | conifer | ECM | non-N-fixing | mineral | NA | 8.26 | NA | NA | 13.5 | NA | NA | NA | NA | NA | NA | NA | NA | NA |
| Riestra et al. 2012 | -36.6 | -64.17 | 169 | 18.9 | 1257 | boradleaf | AM | non-N-fixing | mineral | NA | 15.5 | NA | NA | 32.7 | NA | NA | NA | NA | NA | NA | NA | NA | NA |
| Ovington 1954 | 51.075 | 0.45 | 25 | 9.9 | 705 | conifer | ECM | non-N-fixing | forest floor | NA | NA | NA | 37.8 | 540.2 | 14.3 | NA | NA | NA | NA | NA | NA | NA | 4.1 |
| Ovington 1954 | 51.075 | 0.45 | 25 | 9.9 | 705 | conifer | ECM | non-N-fixing | forest floor | NA | NA | NA | 30.5 | 514.7 | 16.9 | NA | NA | NA | NA | NA | NA | NA | 4.6 |
| Ovington 1954 | 51.075 | 0.45 | 25 | 9.9 | 705 | conifer | ECM | non-N-fixing | forest floor | NA | NA | NA | 24.3 | 418.5 | 17.2 | NA | NA | NA | NA | NA | NA | NA | 4.7 |
| Ovington 1954 | 51.075 | 0.45 | 25 | 9.9 | 705 | conifer | ECM | non-N-fixing | forest floor | NA | NA | NA | 25.9 | 455.5 | 17.6 | NA | NA | NA | NA | NA | NA | NA | 4.6 |
| Ovington 1954 | 51.075 | 0.45 | 25 | 9.9 | 705 | conifer | AM | non-N-fixing | forest floor | NA | NA | NA | 32 | 486.8 | 15.2 | NA | NA | NA | NA | NA | NA | NA | 4.8 |
| Ovington 1954 | 51.075 | 0.45 | 25 | 9.9 | 705 | conifer | ECM | non-N-fixing | forest floor | NA | NA | NA | 27.3 | 398.6 | 14.6 | NA | NA | NA | NA | NA | NA | NA | 4.9 |
| Ovington 1954 | 51.075 | 0.45 | 25 | 9.9 | 705 | conifer | ECM | non-N-fixing | forest floor | NA | NA | NA | 26.9 | 493.1 | 18.3 | NA | NA | NA | NA | NA | NA | NA | 4.7 |
| Ovington 1954 | 51.075 | 0.45 | 25 | 9.9 | 705 | boradleaf | ECM | non-N-fixing | forest floor | NA | NA | NA | 22.8 | 462.7 | 20.3 | NA | NA | NA | NA | NA | NA | NA | 4.8 |
| Ovington 1954 | 51.075 | 0.45 | 25 | 9.9 | 705 | boradleaf | ECM | non-N-fixing | forest floor | NA | NA | NA | 26.3 | 476.4 | 18.1 | NA | NA | NA | NA | NA | NA | NA | 4.8 |
| Ovington 1954 | 51.075 | 0.45 | 25 | 9.9 | 705 | boradleaf | ECM | non-N-fixing | forest floor | NA | NA | NA | 25.8 | 482.3 | 18.7 | NA | NA | NA | NA | NA | NA | NA | 5.1 |
| Ovington 1956 | 51.075 | 0.45 | 25 | 9.9 | 705 | conifer | ECM | non-N-fixing | mineral | NA | NA | NA | 18.04 | 36.8 | 2.04 | NA | NA | NA | NA | NA | NA | NA | NA |
| Ovington 1956 | 51.075 | 0.45 | 25 | 9.9 | 705 | conifer | ECM | non-N-fixing | mineral | NA | NA | NA | 16.01 | 40.5 | 2.53 | NA | NA | NA | NA | NA | NA | NA | NA |
| Ovington 1956 | 51.075 | 0.45 | 25 | 9.9 | 705 | conifer | ECM | non-N-fixing | mineral | NA | NA | NA | 23.48 | 42.5 | 1.81 | NA | NA | NA | NA | NA | NA | NA | NA |
| Ovington 1956 | 51.075 | 0.45 | 25 | 9.9 | 705 | conifer | ECM | non-N-fixing | mineral | NA | NA | NA | 22.98 | 57 | 2.48 | NA | NA | NA | NA | NA | NA | NA | NA |
| Ovington 1956 | 51.075 | 0.45 | 25 | 9.9 | 705 | conifer | AM | non-N-fixing | mineral | NA | NA | NA | 17.37 | 42.2 | 2.43 | NA | NA | NA | NA | NA | NA | NA | NA |
| Ovington 1956 | 51.075 | 0.45 | 25 | 9.9 | 705 | conifer | ECM | non-N-fixing | mineral | NA | NA | NA | 19.31 | 42.1 | 2.18 | NA | NA | NA | NA | NA | NA | NA | NA |
| Ovington 1956 | 51.075 | 0.45 | 25 | 9.9 | 705 | conifer | ECM | non-N-fixing | mineral | NA | NA | NA | 17.93 | 33 | 1.84 | NA | NA | NA | NA | NA | NA | NA | NA |
| Ovington 1956 | 51.075 | 0.45 | 25 | 9.9 | 705 | boradleaf | ECM | non-N-fixing | mineral | NA | NA | NA | 15.45 | 47.9 | 3.1 | NA | NA | NA | NA | NA | NA | NA | NA |
| Ovington 1956 | 51.075 | 0.45 | 25 | 9.9 | 705 | boradleaf | ECM | non-N-fixing | mineral | NA | NA | NA | 16.75 | 45.4 | 2.71 | NA | NA | NA | NA | NA | NA | NA | NA |
| Ovington 1956 | 51.075 | 0.45 | 25 | 9.9 | 705 | boradleaf | ECM | non-N-fixing | mineral | NA | NA | NA | 19.22 | 39.4 | 2.05 | NA | NA | NA | NA | NA | NA | NA | NA |
| Defrieri et al. 2011 | 43 | -71.517 | 19 | 8.2 | 942 | conifer | ECM | non-N-fixing | mineral | NA | NA | NA | 6.97 | 23 | 3.3 | NA | NA | 24 | NA | NA | NA | NA | NA |
| Defrieri et al. 2011 | 43 | -71.517 | 19 | 8.2 | 942 | boradleaf | AM | non-N-fixing | mineral | NA | NA | NA | 11.67 | 35 | 3 | NA | NA | 26 | NA | NA | NA | NA | NA |
| Asadiyan et al. 2013 | 36.217 | 36.167 | 85 | 11.9 | 858 | boradleaf | ECM | non-N-fixing | mineral | NA | NA | NA | 9.95 | 54.5 | 5.7 | 22 | 17.43 | NA | NA | NA | NA | NA | 6.54 |
| Asadiyan et al. 2013 | 36.217 | 36.167 | 85 | 11.9 | 858 | conifer | ECM | non-N-fixing | mineral | NA | NA | NA | 10.02 | 30.1 | 3.2 | 24.19 | 17.72 | NA | NA | NA | NA | NA | 6.95 |
| Asadiyan et al. 2013 | 36.217 | 36.167 | 85 | 11.9 | 858 | boradleaf | AM | non-N-fixing | mineral | NA | NA | NA | 7.86 | 61.4 | 8.5 | 45.4 | 21.09 | NA | NA | NA | NA | NA | 7.4 |
| Kooch et al. 2012 | 36.467 | 52.233 | 58 | 16.8 | 733 | boradleaf | AM | non-N-fixing | mineral | NA | 103.2 | 9.96 | 11.35 | 19.7 | 1.9 | NA | NA | NA | NA | NA | NA | NA | NA |
| Kooch et al. 2012 | 36.467 | 52.233 | 58 | 16.8 | 733 | boradleaf | ECM | non-N-fixing | mineral | NA | 121.4 | 7.71 | 15.07 | 24.08 | 1.53 | NA | NA | NA | NA | NA | NA | NA | NA |
| Kooch et al. 2012 | 36.467 | 52.233 | 58 | 16.8 | 733 | conifer | ECM | non-N-fixing | mineral | NA | 168.2 | 7.95 | 19.53 | 33.63 | 1.59 | NA | NA | NA | NA | NA | NA | NA | NA |
| Christiansen et al. 2010 | 55.95 | 9.617 | 75 | 7.3 | 825 | boradleaf | AM | non-N-fixing | forest floor | NA | 2 | 0.03 | 64 | NA | NA | NA | NA | NA | NA | NA | NA | NA | 5.4 |
| Christiansen et al. 2010 | 55.95 | 9.617 | 75 | 7.3 | 825 | boradleaf | ECM | non-N-fixing | forest floor | NA | 5.3 | 0.2 | 33 | NA | NA | NA | NA | NA | NA | NA | NA | NA | 4.5 |
| Christiansen et al. 2010 | 55.95 | 9.617 | 75 | 7.3 | 825 | boradleaf | ECM | non-N-fixing | forest floor | NA | 1.3 | 0.03 | 44 | NA | NA | NA | NA | NA | NA | NA | NA | NA | 4.9 |
| Christiansen et al. 2010 | 55.95 | 9.617 | 75 | 7.3 | 825 | boradleaf | AM | non-N-fixing | forest floor | NA | 1.7 | 0.03 | 55 | NA | NA | NA | NA | NA | NA | NA | NA | NA | 4.8 |
| Christiansen et al. 2010 | 55.95 | 9.617 | 75 | 7.3 | 825 | boradleaf | ECM | non-N-fixing | forest floor | NA | 4 | 0.2 | 27 | NA | NA | NA | NA | NA | NA | NA | NA | NA | 4.3 |
| Christiansen et al. 2010 | 55.95 | 9.617 | 75 | 7.3 | 825 | conifer | ECM | non-N-fixing | forest floor | NA | 8.9 | 0.3 | 29 | NA | NA | NA | NA | NA | NA | NA | NA | NA | 4.4 |
| Christiansen et al. 2010 | 55.417 | 12.05 | 47 | 7.8 | 631 | boradleaf | ECM | non-N-fixing | forest floor | NA | 5.9 | 0.2 | 27 | NA | NA | NA | NA | NA | NA | NA | NA | NA | 4.3 |
| Christiansen et al. 2010 | 55.417 | 12.05 | 47 | 7.8 | 631 | boradleaf | ECM | non-N-fixing | forest floor | NA | 1.5 | 0.04 | 34 | NA | NA | NA | NA | NA | NA | NA | NA | NA | 5 |
| Christiansen et al. 2010 | 55.417 | 12.05 | 47 | 7.8 | 631 | boradleaf | AM | non-N-fixing | forest floor | NA | 1 | 0.03 | 38 | NA | NA | NA | NA | NA | NA | NA | NA | NA | 4.9 |
| Christiansen et al. 2010 | 55.417 | 12.05 | 47 | 7.8 | 631 | boradleaf | ECM | non-N-fixing | forest floor | NA | 3.5 | 0.1 | 24 | NA | NA | NA | NA | NA | NA | NA | NA | NA | 4.6 |
| Christiansen et al. 2010 | 55.417 | 12.05 | 47 | 7.8 | 631 | conifer | ECM | non-N-fixing | forest floor | NA | 19.8 | 0.8 | 25 | NA | NA | NA | NA | NA | NA | NA | NA | NA | 3.5 |
| Christiansen et al. 2010 | 55.95 | 9.617 | 77 | 7.3 | 825 | boradleaf | AM | non-N-fixing | mineral | NA | 68 | 5.6 | 12 | NA | NA | NA | NA | NA | NA | NA | NA | NA | 4.8 |
| Christiansen et al. 2010 | 55.95 | 9.617 | 77 | 7.3 | 825 | boradleaf | ECM | non-N-fixing | mineral | NA | 52 | 4.2 | 12 | NA | NA | NA | NA | NA | NA | NA | NA | NA | 4.3 |
| Christiansen et al. 2010 | 55.95 | 9.617 | 77 | 7.3 | 825 | boradleaf | ECM | non-N-fixing | mineral | NA | 67 | 5.6 | 12 | NA | NA | NA | NA | NA | NA | NA | NA | NA | 4.7 |
| Christiansen et al. 2010 | 55.95 | 9.617 | 77 | 7.3 | 825 | boradleaf | AM | non-N-fixing | mineral | NA | 59 | 5 | 12 | NA | NA | NA | NA | NA | NA | NA | NA | NA | 4.7 |
| Christiansen et al. 2010 | 55.95 | 9.617 | 77 | 7.3 | 825 | boradleaf | ECM | non-N-fixing | mineral | NA | 72 | 5.7 | 13 | NA | NA | NA | NA | NA | NA | NA | NA | NA | 4.5 |
| Christiansen et al. 2010 | 55.95 | 9.617 | 77 | 7.3 | 825 | conifer | ECM | non-N-fixing | mineral | NA | 67 | 5.1 | 13 | NA | NA | NA | NA | NA | NA | NA | NA | NA | 4.5 |
| Christiansen et al. 2010 | 55.417 | 12.05 | 47 | 7.8 | 631 | boradleaf | ECM | non-N-fixing | mineral | NA | 61 | 3.4 | 18 | NA | NA | NA | NA | NA | NA | NA | NA | NA | 3.8 |
| Christiansen et al. 2010 | 55.417 | 12.05 | 47 | 7.8 | 631 | boradleaf | ECM | non-N-fixing | mineral | NA | 70 | 3.8 | 18 | NA | NA | NA | NA | NA | NA | NA | NA | NA | 3.9 |
| Christiansen et al. 2010 | 55.417 | 12.05 | 47 | 7.8 | 631 | boradleaf | AM | non-N-fixing | mineral | NA | 57 | 3.9 | 14 | NA | NA | NA | NA | NA | NA | NA | NA | NA | 4.4 |
| Christiansen et al. 2010 | 55.417 | 12.05 | 47 | 7.8 | 631 | boradleaf | ECM | non-N-fixing | mineral | NA | 75 | 4.1 | 18 | NA | NA | NA | NA | NA | NA | NA | NA | NA | 3.8 |
| Christiansen et al. 2010 | 55.417 | 12.05 | 47 | 7.8 | 631 | conifer | ECM | non-N-fixing | mineral | NA | 63 | 3.4 | 18 | NA | NA | NA | NA | NA | NA | NA | NA | NA | 3.7 |
| Wang et al. 2010 | 22.567 | 112.833 | 40 | 21.7 | 1700 | boradleaf | ECM | N-fixing | mineral | NA | 34.72 | 0.61 | 33.2 | 78.9 | 1.39 | NA | NA | NA | NA | NA | NA | NA | 3.81 |
| Wang et al. 2010 | 22.567 | 112.833 | 40 | 21.7 | 1700 | boradleaf | ECM | N-fixing | mineral | NA | 38.17 | 0.59 | 37.4 | 73.4 | 1.13 | NA | NA | NA | NA | NA | NA | NA | 3.85 |
| Wang et al. 2010 | 22.567 | 112.833 | 40 | 21.7 | 1700 | boradleaf | ECM | non-N-fixing | mineral | NA | 22.84 | 0.41 | 32.1 | 50.2 | 0.91 | NA | NA | NA | NA | NA | NA | NA | 3.88 |
| Wang et al. 2010 | 22.567 | 112.833 | 40 | 21.7 | 1700 | boradleaf | AM | non-N-fixing | mineral | NA | 24.95 | 0.42 | 33.8 | 49.4 | 0.84 | NA | NA | NA | NA | NA | NA | NA | 3.87 |
| Peng et al. 2013 | 25.383 | 101.583 | 40 | 21.5 | 700 | boradleaf | AM | N-fixing | mineral | NA | 38.34 | 2.38 | 16.14 | 14.2 | 0.88 | NA | NA | 23.2 | NA | NA | NA | NA | 6.85 |
| Peng et al. 2013 | 25.383 | 101.583 | 40 | 21.5 | 700 | boradlea f | ECM | non-N-fixing | mineral | NA | 27.4 | 1.42 | 19.23 | 10 | 0.52 | NA | NA | 21.8 | NA | NA | NA | NA | 5.41 |
| Wei et al. 2009 | 18.383 | 108.733 | 452 | 19.7 | 3000 | conifer | AM | non-N-fixing | mineral | NA | 44.54 | 2.18 | 20.47 | 17.4 | 0.85 | NA | NA | 0.77 | 109.8 | NA | NA | NA | 4.94 |
| Wei et al. 2009 | 18.383 | 108.733 | 452 | 19.7 | 3000 | conifer | ECM | non-N-fixing | mineral | NA | 41.18 | 2.09 | 19.73 | 14.4 | 0.73 | NA | NA | 0.83 | 86.8 | NA | NA | NA | 4.43 |
| Yang et al. 2004 | 26.183 | 117.433 | 204 | 19.1 | 1749 | conifer | AM | non-N-fixing | mineral | 5.468 | 40.56 | 2.69 | 15.1 | 16.9 | 1.12 | NA | NA | 4.7 | NA | NA | NA | NA | 4.8 |
| Yang et al. 2004 | 26.183 | 117.433 | 204 | 19.1 | 1749 | conifer | AM | non-N-fixing | mineral | 7.291 | 40 | 3.1 | 12.9 | 17.7 | 1.37 | NA | NA | 5.6 | NA | NA | NA | NA | 5.1 |
| Yang et al. 2004 | 26.183 | 117.433 | 204 | 19.1 | 1749 | boradlea f | AM | N-fixing | mineral | 5.6895 | 40.25 | 2.97 | 13.6 | 17.5 | 1.29 | NA | NA | 6.8 | NA | NA | NA | NA | 5.1 |
| Yang et al. 2004 | 26.183 | 117.433 | 204 | 19.1 | 1749 | boradlea f | ECM | non-N-fixing | mineral | 9.538 | 37.62 | 2.46 | 15.3 | 17.1 | 1.12 | NA | NA | 5.9 | NA | NA | NA | NA | 5.3 |
| Garcia-Montiel & Binkley 1998 | 19.5 | -155.25 | 1224 | 21 | 4000 | boradlea f | AM | N-fixing | mineral | NA | NA | NA | 12.08 | 125.07 | 10.27 | NA | NA | NA | NA | NA | NA | NA | NA |
| Garcia-Montiel & Binkley 1998 | 19.5 | -155.25 | 1224 | 21 | 4000 | boradlea f | ECM | non-N-fixing | mineral | NA | NA | NA | 15.22 | 112.68 | 7.39 | NA | NA | NA | NA | NA | NA | NA | NA |
| Garcia-Montiel & Binkley 1998 | 19.5 | -155.25 | 1224 | 21 | 4000 | boradlea f | AM | N-fixing | mineral | NA | NA | NA | 11.82 | 135.21 | 11.31 | NA | NA | NA | NA | NA | NA | NA | NA |
| Garcia-Montiel & Binkley 1998 | 19.5 | -155.25 | 1224 | 21 | 4000 | boradlea f | ECM | non-N-fixing | mineral | NA | NA | NA | 13.33 | 134.09 | 10.12 | NA | NA | NA | NA | NA | NA | NA | NA |
| Garcia-Montiel & Binkley 1998 | 19.5 | -155.25 | 1224 | 21 | 4000 | boradlea f | AM | N-fixing | mineral | NA | NA | NA | 11.95 | 125.07 | 10.41 | NA | NA | NA | NA | NA | NA | NA | NA |
| Garcia-Montiel & Binkley 1998 | 19.5 | -155.25 | 1224 | 21 | 4000 | boradlea f | ECM | non-N-fixing | mineral | NA | NA | NA | 13.71 | 120.56 | 8.93 | NA | NA | NA | NA | NA | NA | NA | NA |
| Jiang et al. 2010 | 26.733 | 115.067 | 82 | 18.6 | 1726 | conifer | ECM | non-N-fixing | mineral | NA | 13.77 | 0.81 | 16 | 8.5 | 0.5 | 8.9 | 1.4 | NA | 143.68 | 20.63 | NA | NA | 4.3 |
| Jiang et al. 2010 | 26.733 | 115.067 | 82 | 18.6 | 1726 | conifer | ECM | non-N-fixing | mineral | NA | 19.93 | 1.28 | 16 | 12.3 | 0.79 | 12.4 | 0.6 | NA | 74.21 | 9.76 | NA | NA | 4.4 |
| Kulakova 2012 | 49.417 | 46.767 | 24 | 7.6 | 298 | boradlea f | ECM | non-N-fixing | mineral | 3.55 | 45.4 | 3.48 | 13.1 | 103.2 | 7.9 | NA | NA | NA | NA | NA | NA | NA | NA |
| Kulakova 2012 | 49.417 | 46.767 | 24 | 7.6 | 298 | conifer | ECM | non-N-fixing | mineral | NA | 92 | 4.82 | 19.1 | 172 | 9 | NA | NA | NA | NA | NA | NA | NA | NA |
| Kulakova 2012 | 49.417 | 46.767 | 24 | 7.6 | 298 | boradlea f | AM | non-N-fixing | mineral | NA | 91.4 | 5.93 | 15.4 | 160.5 | 10.4 | NA | NA | NA | NA | NA | NA | NA | NA |
| Kulakova 2012 | 49.417 | 46.767 | 24 | 7.6 | 298 | boradlea f | AM | non-N-fixing | mineral | 4.1 | 20.1 | 1.35 | 14.8 | 34.1 | 2.3 | NA | NA | NA | NA | NA | NA | NA | NA |
| Laudicina et al. 2012 | 37.55 | 13.917 | 471 | 17.8 | 460 | boradlea f | ECM | non-N-fixing | mineral | NA | NA | NA | 12.1 | 23 | 1.9 | NA | NA | 77 | 779 | 60 | NA | NA | 7.8 |
| Laudicina et al. 2012 | 37.55 | 13.917 | 471 | 17.8 | 460 | boradlea f | ECM | non-N-fixing | mineral | NA | NA | NA | 8.8 | 14.1 | 1.6 | NA | NA | 68 | 671 | 53 | NA | NA | 7.6 |
| Laudicina et al. 2012 | 37.55 | 13.917 | 471 | 17.8 | 460 | conifer | ECM | non-N-fixing | mineral | NA | NA | NA | 10.3 | 12.3 | 1.2 | NA | NA | 54 | 586 | 44 | NA | NA | 7.9 |
| Laudicina et al. 2012 | 37.55 | 13.917 | 471 | 17.8 | 460 | conifer | AM | non-N-fixing | mineral | NA | NA | NA | 8.8 | 13.2 | 1.5 | NA | NA | 53 | 459 | 41 | NA | NA | 7.9 |
| Demessie et al. 2012 | 7.283 | 38.8 | 2284 | 15.5 | 973 | boradleaf | ECM | non-N-fixing | forest floor | 11.94 | NA | NA | 68 | 493 | 70 | NA | NA | NA | NA | NA | NA | NA | NA |
| Demessie et al. 2012 | 7.283 | 38.8 | 2284 | 15.5 | 973 | boradleaf | ECM | non-N-fixing | forest floor | 11.94 | NA | NA | 49.5 | 467 | 9 | NA | NA | NA | NA | NA | NA | NA | NA |
| Demessie et al. 2012 | 7.283 | 38.8 | 2284 | 15.5 | 973 | boradleaf | ECM | non-N-fixing | forest floor | 9.68 | NA | NA | 49.3 | 473 | 10 | NA | NA | NA | NA | NA | NA | NA | NA |
| Demessie et al. 2012 | 7.283 | 38.8 | 2284 | 15.5 | 973 | conifer | ECM | non-N-fixing | forest floor | 6.43 | NA | NA | 28.1 | 462 | 7 | NA | NA | NA | NA | NA | NA | NA | NA |
| Demessie et al. 2012 | 7.283 | 38.8 | 2284 | 15.5 | 973 | conifer | AM | non-N-fixing | forest floor | 6.07 | NA | NA | 64.8 | 390 | 14 | NA | NA | NA | NA | NA | NA | NA | NA |
| Demessie et al. 2012 | 7.283 | 38.8 | 2284 | 15.5 | 973 | conifer | AM | non-N-fixing | forest floor | 4.94 | NA | NA | 37.3 | 458 | 18 | NA | NA | NA | NA | NA | NA | NA | NA |
| Lemma 2012 | 7.1 | 38.617 | 1745 | 19.5 | 1240 | conifer | ECM | non-N-fixing | mineral | NA | NA | NA | 12.5 | 57.5 | 4.6 | NA | NA | NA | NA | NA | NA | NA | 6.2 |
| Lemma 2012 | 7.1 | 38.617 | 1745 | 19.5 | 1240 | conifer | AM | non-N-fixing | mineral | NA | NA | NA | 10.4 | 58.3 | 5.6 | NA | NA | NA | NA | NA | NA | NA | 6.4 |
| Li et al. 2014 | 30.3 | 119.4 | 285 | 14.8 | 1630 | conifer | AM | non-N-fixing | mineral | NA | NA | NA | 13.71 | 26.59 | 1.94 | 61.22 | 17.65 | 3.56 | NA | NA | NA | NA | 4.45 |
| Li et al. 2014 | 30.3 | 119.4 | 285 | 14.8 | 1630 | conifer | ECM | non-N-fixing | mineral | NA | NA | NA | 12.64 | 27.43 | 2.17 | 67.38 | 22.84 | 2.48 | NA | NA | NA | NA | 4.7 |
| Lugo et al. 1990 | 18.275 | -65.85 | 750 | 23 | 3800 | conifer | ECM | non-N-fixing | mineral | NA | 27.06 | 1.9 | 14.3 | 32.9 | 2.3 | NA | NA | 11 | NA | NA | NA | NA | 4.7 |
| Lugo et al. 1990 | 18.275 | -65.85 | 750 | 23 | 3800 | conifer | ECM | non-N-fixing | mineral | NA | 32.94 | 1.9 | 17 | 44.1 | 2.6 | NA | NA | 21 | NA | NA | NA | NA | 4.6 |
| Lugo et al. 1990 | 18.275 | -65.85 | 750 | 23 | 3800 | boradleaf | ECM | non-N-fixing | mineral | NA | 41.18 | 3.1 | 13.6 | 57.1 | 4.2 | NA | NA | 15 | NA | NA | NA | NA | 4.8 |
| Lugo et al. 1990 | 18.275 | -65.85 | 750 | 23 | 3800 | boradleaf | AM | non-N-fixing | mineral | NA | 36.47 | 3.2 | 12.5 | 52.4 | 4.2 | NA | NA | 36 | NA | NA | NA | NA | 4.5 |
| Lugo et al. 1990 | 18.275 | -65.85 | 750 | 23 | 3800 | boradleaf | AM | non-N-fixing | mineral | NA | 28.24 | 2.4 | 11.9 | 51.2 | 4.3 | NA | NA | 25 | NA | NA | NA | NA | 4.7 |
| Lugo et al. 1990 | 18.275 | -65.85 | 750 | 23 | 3800 | boradleaf | ECM | non-N-fixing | mineral | NA | 32.94 | 2.1 | 15.2 | 57.6 | 3.8 | NA | NA | 26 | NA | NA | NA | NA | 4.7 |
| Lugo et al. 1990 | 18.275 | -65.85 | 750 | 23 | 3800 | boradleaf | AM | non-N-fixing | mineral | NA | 30 | 2.9 | 10.4 | 51.8 | 5 | NA | NA | 28 | NA | NA | NA | NA | 4.4 |
| Lugo et al. 1990 | 18.275 | -65.85 | 750 | 23 | 3800 | boradleaf | AM | non-N-fixing | mineral | NA | 35.88 | 2.7 | 13.4 | 49.4 | 3.7 | NA | NA | 25 | NA | NA | NA | NA | 4.7 |
| Lugo et al. 1990 | 18.275 | -65.85 | 750 | 23 | 3800 | boradleaf | AM | non-N-fixing | mineral | NA | 23.53 | 2.2 | 11.5 | 58.8 | 5.1 | NA | NA | 31 | NA | NA | NA | NA | 4.6 |
| Lugo et al. 1990 | 18.275 | -65.85 | 750 | 23 | 3800 | boradleaf | AM | non-N-fixing | mineral | NA | 27.65 | 2.3 | 12.3 | 57.6 | 4.7 | NA | NA | 36 | NA | NA | NA | NA | 4.8 |
| Abiyu et al. 2011 | 11.2 | 39.667 | 2686 | 21 | 2350 | conifer | AM | non-N-fixing | mineral | NA | 22.95 | 2.61 | 32.8 | 25.5 | 2.9 | NA | NA | 7.8 | NA | NA | NA | NA | 5.8 |
| Abiyu et al. 2011 | 11.2 | 39.667 | 2686 | 21 | 2350 | boradleaf | ECM | non-N-fixing | mineral | NA | 21.33 | 2.43 | 32.8 | 23.7 | 2.7 | NA | NA | 8.2 | NA | NA | NA | NA | 5.6 |
| Phillips & Fahey 2006 | 42.45 | -76.417 | 450 | 7.8 | 874 | boradleaf | ECM | N-fixing | mineral | NA | NA | NA | NA | 211.8 | NA | NA | NA | NA | NA | NA | NA | NA | 3.94 |
| Phillips & Fahey 2006 | 42.45 | -76.417 | 450 | 7.8 | 874 | boradleaf | AM | non-N-fixing | mineral | NA | NA | NA | NA | 172.9 | NA | NA | NA | NA | NA | NA | NA | NA | 4.11 |
| Phillips & Fahey 2006 | 42.45 | -76.417 | 450 | 7.8 | 874 | boradleaf | AM | non-N-fixing | mineral | NA | NA | NA | NA | 58.8 | NA | NA | NA | NA | NA | NA | NA | NA | 4.67 |
| Phillips & Fahey 2006 | 42.45 | -76.417 | 450 | 7.8 | 874 | boradleaf | AM | non-N-fixing | mineral | NA | NA | NA | NA | 45.3 | NA | NA | NA | NA | NA | NA | NA | NA | 4.95 |
| Phillips & Fahey 2006 | 42.45 | -76.417 | 450 | 7.8 | 874 | boradleaf | AM | non-N-fixing | mineral | NA | NA | NA | NA | 80.6 | NA | NA | NA | NA | NA | NA | NA | NA | 4.16 |
| Phillips & Fahey 2006 | 42.45 | -76.417 | 450 | 7.8 | 874 | boradleaf | AM | non-N-fixing | mineral | NA | NA | NA | NA | 45.9 | NA | NA | NA | NA | NA | NA | NA | NA | 5.56 |
| Phillips & Fahey 2006 | 42.45 | -76.417 | 450 | 7.8 | 874 | conifer | ECM | non-N-fixing | mineral | NA | NA | NA | NA | 292.4 | NA | NA | NA | NA | NA | NA | NA | NA | 3.81 |
| Phillips & Fahey 2006 | 42.45 | -76.417 | 450 | 7.8 | 874 | boradleaf | ECM | non-N-fixing | mineral | NA | NA | NA | NA | 278.8 | NA | NA | NA | NA | NA | NA | NA | NA | 4.21 |
| Phillips & Fahey 2006 | 42.45 | -76.417 | 450 | 7.8 | 874 | conifer | ECM | non-N-fixing | mineral | NA | NA | NA | NA | 237.1 | NA | NA | NA | NA | NA | NA | NA | NA | 4.06 |
| Phillips & Fahey 2006 | 42.45 | -76.417 | 450 | 7.8 | 874 | boradleaf | ECM | non-N-fixing | mineral | NA | NA | NA | NA | 59.4 | NA | NA | NA | NA | NA | NA | NA | NA | 5.93 |
| Phillips & Fahey 2006 | 42.45 | -76.417 | 450 | 7.8 | 874 | boradleaf | ECM | non-N-fixing | mineral | NA | NA | NA | NA | 58.2 | NA | NA | NA | NA | NA | NA | NA | NA | 4.83 |
| Phillips & Fahey 2006 | 42.45 | -76.417 | 450 | 7.8 | 874 | conifer | ECM | non-N-fixing | mineral | NA | NA | NA | NA | 134.7 | NA | NA | NA | NA | NA | NA | NA | NA | 3.94 |
| Saha et al. 2007 | 25.683 | 91.917 | 961 | 19.5 | 2208 | conifer | ECM | non-N-fixing | mineral | 6.215 | 11.04 | NA | NA | 3.54 | NA | NA | NA | NA | NA | NA | NA | NA | NA |
| Saha et al. 2007 | 25.683 | 91.917 | 961 | 19.5 | 2208 | boradleaf | AM | non-N-fixing | mineral | 4.7375 | 10.53 | NA | NA | 3.22 | NA | NA | NA | NA | NA | NA | NA | NA | NA |
| Saha et al. 2007 | 25.683 | 91.917 | 961 | 19.5 | 2208 | boradleaf | AM | N-fixing | mineral | 3.4175 | 8.52 | NA | NA | 2.31 | NA | NA | NA | NA | NA | NA | NA | NA | NA |
| Saha et al. 2007 | 25.683 | 91.917 | 961 | 19.5 | 2208 | boradleaf | AM | non-N-fixing | mineral | 5.1225 | 10.58 | NA | NA | 3.36 | NA | NA | NA | NA | NA | NA | NA | NA | NA |
| Singh et al. 2000 | 26.75 | 80.883 | 126 | 25.7 | 920 | boradleaf | ECM | non-N-fixing | mineral | NA | 40.95 | 2.47 | 16.6 | 7.8 | 0.47 | NA | NA | NA | NA | NA | NA | NA | 8.95 |
| Singh et al. 2000 | 26.75 | 80.883 | 126 | 25.7 | 920 | boradleaf | AM | non-N-fixing | mineral | NA | 31.99 | 2.12 | 15.1 | 6.2 | 0.41 | NA | NA | NA | NA | NA | NA | NA | 8.71 |
| Russell et al. 2007 | 10.433 | -83.983 | 43 | 25.8 | 3960 | boradleaf | AM | non-N-fixing | mineral | NA | 51.03 | 3.79 | 13.5 | 49.3 | 3.66 | NA | NA | 3.3 | NA | NA | NA | NA | 4.31 |
| Russell et al. 2007 | 10.433 | -83.983 | 43 | 25.8 | 3960 | boradleaf | AM | N-fixing | mineral | NA | 40.05 | 3.16 | 12.7 | 44.5 | 3.51 | NA | NA | 4.2 | NA | NA | NA | NA | 4.15 |
| Russell et al. 2007 | 10.433 | -83.983 | 43 | 25.8 | 3960 | conifer | ECM | non-N-fixing | mineral | NA | 52.07 | 3.93 | 13.2 | 44.5 | 3.36 | NA | NA | 4.2 | NA | NA | NA | NA | 4.36 |
| Russell et al. 2007 | 10.433 | -83.983 | 43 | 25.8 | 3960 | boradleaf | AM | non-N-fixing | mineral | NA | 49.1 | 3.78 | 13 | 46.1 | 3.55 | NA | NA | 4.7 | NA | NA | NA | NA | 4.3 |
| Russell et al. 2007 | 10.433 | -83.983 | 43 | 25.8 | 3960 | boradleaf | AM | non-N-fixing | mineral | NA | 42.15 | 3.23 | 13.1 | 55.1 | 4.22 | NA | NA | 4.2 | NA | NA | NA | NA | 4.47 |
| Russell et al. 2007 | 10.433 | -83.983 | 43 | 25.8 | 3960 | boradleaf | AM | non-N-fixing | mineral | NA | 46.3 | 3.67 | 12.6 | 50.6 | 4.01 | NA | NA | 3.2 | NA | NA | NA | NA | 4.48 |
| Mellor et al. 2013 | 41.867 | -100.333 | 870 | 8.4 | 561 | conifer | AM | non-N-fixing | mineral | NA | 6.43 | 0.65 | 9.9 | NA | NA | NA | NA | NA | NA | NA | NA | NA | 7.12 |
| Mellor et al. 2013 | 41.867 | -100.333 | 870 | 8.4 | 561 | conifer | AM | non-N-fixing | mineral | NA | 3.7 | 0.61 | 6.1 | NA | NA | NA | NA | NA | NA | NA | NA | NA | 6.37 |
| Mellor et al. 2013 | 41.867 | -100.333 | 870 | 8.4 | 561 | conifer | ECM | non-N-fixing | mineral | NA | 4.06 | 0.34 | 11.9 | NA | NA | NA | NA | NA | NA | NA | NA | NA | 4.98 |
| Wang et al. 2010 | 22.167 | 106.833 | 373 | 21 | 1400 | conifer | ECM | non-N-fixing | mineral | 2.318 | 46.9 | 2.58 | 18.2 | NA | NA | 4.9 | 1.85 | NA | NA | NA | NA | NA | NA |
| Wang et al. 2010 | 22.167 | 106.833 | 373 | 21 | 1400 | boradleaf | ECM | non-N-fixing | mineral | 2.719 | 49.61 | 3.28 | 15.2 | NA | NA | 5.16 | 2.35 | NA | NA | NA | NA | NA | NA |
| Wang et al. 2010 | 22.167 | 106.833 | 373 | 21 | 1400 | boradleaf | AM | non-N-fixing | mineral | 2.423 | 54.51 | 3.29 | 16.7 | NA | NA | 5.39 | 2.35 | NA | NA | NA | NA | NA | NA |
| Wang et al. 2010 | 22.167 | 106.833 | 373 | 21 | 1400 | boradleaf | AM | non-N-fixing | mineral | 3.123 | 51.46 | 3.32 | 15.5 | NA | NA | 3.75 | 3.24 | NA | NA | NA | NA | NA | NA |
| Wang et al. 2010 | 22.167 | 106.833 | 373 | 21 | 1400 | conifer | ECM | non-N-fixing | mineral | NA | 29.2 | NA | NA | 26.8 | NA | NA | NA | NA | NA | NA | NA | NA | NA |
| Wang et al. 2010 | 22.167 | 106.833 | 373 | 21 | 1400 | boradleaf | ECM | non-N-fixing | mineral | NA | 32.6 | NA | NA | 29.7 | NA | NA | NA | NA | NA | NA | NA | NA | NA |
| Wang et al. 2010 | 22.167 | 106.833 | 373 | 21 | 1400 | boradleaf | AM | non-N-fixing | mineral | NA | 34.9 | NA | NA | 31.3 | NA | NA | NA | NA | NA | NA | NA | NA | NA |
| Wang et al. 2010 | 22.167 | 106.833 | 373 | 21 | 1400 | boradleaf | AM | non-N-fixing | mineral | NA | 34.4 | NA | NA | 31.8 | NA | NA | NA | NA | NA | NA | NA | NA | NA |
| Zhang et al. 2012 | 22.567 | 112.833 | 373 | 22.5 | 1534 | boradleaf | ECM | N-fixing | mineral | NA | NA | NA | 13.8 | 22.1 | 1.6 | 16 | 17.8 | 1.8 | 254 | 41.4 | NA | NA | 3.83 |
| Zhang et al. 2012 | 22.567 | 112.833 | 373 | 22.5 | 1534 | boradleaf | ECM | non-N-fixing | mineral | NA | NA | NA | 10.3 | 15.5 | 1.5 | 13.4 | 13.6 | 1.6 | 288 | 43.9 | NA | NA | 3.91 |
| Ushio et al. 2008 | 6.08333 | 116.55 | 1560 | 18 | 3080 | conifer | AM | non-N-fixing | forest floor | NA | NA | NA | 21.8 | 350 | 15.8 | NA | NA | NA | NA | NA | NA | NA | 3.84 |
| Ushio et al. 2008 | 6.08333 | 116.55 | 1560 | 18 | 3080 | conifer | AM | non-N-fixing | forest floor | NA | NA | NA | 25.4 | 456 | 18.1 | NA | NA | NA | NA | NA | NA | NA | 3.83 |
| Ushio et al. 2008 | 6.08333 | 116.55 | 1560 | 18 | 3080 | boradleaf | ECM | non-N-fixing | forest floor | NA | NA | NA | 23.4 | 358 | 15.2 | NA | NA | NA | NA | NA | NA | NA | 4.08 |
| Ushio et al. 2008 | 6.08333 | 116.55 | 1560 | 18 | 3080 | boradleaf | AM | non-N-fixing | forest floor | NA | NA | NA | 21.8 | 379 | 17.2 | NA | NA | NA | NA | NA | NA | NA | 4.02 |
| Ushio et al. 2008 | 6.08333 | 116.55 | 1560 | 18 | 3080 | boradleaf | ECM | non-N-fixing | forest floor | NA | NA | NA | 24.1 | 370 | 15.3 | NA | NA | NA | NA | NA | NA | NA | 4.13 |
| Wu et al. 2014 | 26.31667 | 117.6 | 455 | 20.1 | 1670 | boradleaf | ECM | non-N-fixing | mineral | 5.91 | NA | NA | 15.1 | 29.84 | 1.97 | NA | NA | NA | NA | NA | NA | NA | 4.4 |
| Wu et al. 2014 | 26.31667 | 117.6 | 455 | 20.1 | 1670 | conifer | AM | non-N-fixing | mineral | 3.4 | NA | NA | 15.2 | 22.91 | 1.51 | NA | NA | NA | NA | NA | NA | NA | 4.87 |
| Song et al. 2013 | 27.5 | 114.5 | 617 | 16.5 | 1591 | conifer | AM | non-N-fixing | mineral | NA | NA | NA | 14.68 | 21.433 | 1.46 | NA | NA | NA | NA | NA | NA | NA | 3.69 |
| Song et al. 2013 | 27.5 | 114.5 | 617 | 16.5 | 1591 | boradleaf | AM | non-N-fixing | mineral | NA | NA | NA | 12.28 | 21.981 | 1.79 | NA | NA | NA | NA | NA | NA | NA | 3.92 |
| Pereira et al. 2011 | 41.57 | -6.501944 | 750 | 12 | 555 | boradleaf | AM | non-N-fixing | mineral | NA | NA | NA | 12.56 | 12.15 | 0.96 | NA | NA | NA | NA | NA | NA | NA | 5.56 |
| Pereira et al. 2011 | 41.57 | -6.501944 | 750 | 12 | 555 | boradleaf | ECM | N-fixing | mineral | NA | NA | NA | 11.18 | 11.23 | 1 | NA | NA | NA | NA | NA | NA | NA | 5.62 |
| Chen et al. 2015 | 26.19167 | 117.4333 | 201 | 19.1 | 1749 | boradleaf | ECM | non-N-fixing | mineral | 4.51 | 21.45 | NA | NA | 19.597 | NA | NA | NA | NA | NA | NA | NA | NA | NA |
| Chen et al. 2015 | 26.19167 | 117.4333 | 201 | 19.1 | 1749 | boradleaf | AM | N-fixing | mineral | 2.57 | 23.06 | NA | NA | 20.536 | NA | NA | NA | NA | NA | NA | NA | NA | NA |
| Chen et al. 2015 | 26.19167 | 117.4333 | 201 | 19.1 | 1749 | conifer | AM | non-N-fixing | mineral | 3.19 | 19.74 | NA | NA | 17.816 | NA | NA | NA | NA | NA | NA | NA | NA | NA |
| Chen et al. 2015 | 26.19167 | 117.4333 | 201 | 19.1 | 1749 | conifer | AM | non-N-fixing | mineral | 2.15 | 20.61 | NA | NA | 18.383 | NA | NA | NA | NA | NA | NA | NA | NA | NA |
| Forrester et al. 2013 | -37.5833 | 149.1667 | 158 | 17.8 | 1009 | boradleaf | AM | non-N-fixing | mineral | 3.29 | 36.11 | 1.15 | 31.4 | NA | NA | NA | NA | NA | NA | NA | NA | NA | NA |
| Forrester et al. 2013 | -37.5833 | 149.1667 | 158 | 17.8 | 1009 | boradleaf | AM | non-N-fixing | mineral | NA | 81.07 | 1.987 | 40.8 | NA | NA | NA | NA | NA | NA | NA | NA | NA | NA |
| Forrester et al. 2013 | -37.5833 | 149.1667 | 158 | 17.8 | 1009 | boradleaf | ECM | non-N-fixing | mineral | 2.43 | 30.05 | 0.712 | 42.2 | NA | NA | NA | NA | NA | NA | NA | NA | NA | NA |
| Forrester et al. 2013 | -37.5833 | 149.1667 | 158 | 17.8 | 1009 | boradleaf | ECM | non-N-fixing | mineral | NA | 67.04 | 1.493 | 44.9 | NA | NA | NA | NA | NA | NA | NA | NA | NA | NA |
| Boyle et al. 1990 | 44.66667 | -123.3333 | 300 | 10.5 | 1300 | conifer | ECM | non-N-fixing | mineral | NA | 22.28 | 1.14 | 19.5 | 30.8 | 1.5 | NA | NA | NA | NA | NA | NA | NA | NA |
| Boyle et al. 1990 | 44.66667 | -123.3333 | 300 | 10.5 | 1300 | boradleaf | AM | non-N-fixing | mineral | NA | 28.14 | 1.59 | 17.5 | 44.5 | 2.5 | NA | NA | NA | NA | NA | NA | NA | NA |
| Boyle et al. 1990 | 44.66667 | -123.3333 | 300 | 10.5 | 1300 | conifer | ECM | non-N-fixing | mineral | NA | 31 | 1.52 | 20.2 | 39.3 | 1.9 | NA | NA | NA | NA | NA | NA | NA | NA |
| Boyle et al. 1990 | 44.66667 | -123.3333 | 300 | 10.5 | 1300 | boradleaf | AM | non-N-fixing | mineral | NA | 30.34 | 1.56 | 19.5 | 45.6 | 2.3 | NA | NA | NA | NA | NA | NA | NA | NA |
| Boyle et al. 1990 | 44.66667 | -123.3333 | 300 | 10.5 | 1300 | conifer | ECM | non-N-fixing | mineral | NA | 29.64 | 1.49 | 19.7 | 42.6 | 2.1 | NA | NA | NA | NA | NA | NA | NA | NA |
| Boyle et al. 1990 | 44.66667 | -123.3333 | 300 | 10.5 | 1300 | boradleaf | AM | non-N-fixing | mineral | NA | 35.97 | 1.71 | 21.2 | 53 | 2.5 | NA | NA | NA | NA | NA | NA | NA | NA |
| Boyle et al. 1990 | 44.66667 | -123.3333 | 300 | 10.5 | 1300 | conifer | ECM | non-N-fixing | mineral | NA | 37.22 | 1.79 | 20.7 | 49.1 | 2.3 | NA | NA | NA | NA | NA | NA | NA | NA |
| Boyle et al. 1990 | 44.66667 | -123.3333 | 300 | 10.5 | 1300 | boradleaf | AM | non-N-fixing | mineral | NA | 56.22 | 1.97 | 28.5 | 82.3 | 2.8 | NA | NA | NA | NA | NA | NA | NA | NA |
| Boyle et al. 1990 | 44.66667 | -123.3333 | 300 | 10.5 | 1300 | conifer | ECM | non-N-fixing | mineral | NA | 25.04 | 1.28 | 19.5 | 28.9 | 1.4 | NA | NA | NA | NA | NA | NA | NA | NA |
| Boyle et al. 1990 | 44.66667 | -123.3333 | 300 | 10.5 | 1300 | boradleaf | AM | non-N-fixing | mineral | NA | 29.37 | 1.76 | 17 | 35.8 | 2.1 | NA | NA | NA | NA | NA | NA | NA | NA |
| Jiang et al. 2012 | 26.7333 | 115.0667 | 121 | 18.6 | 1726 | conifer | ECM | non-N-fixing | mineral | NA | NA | NA | 16 | 8.5 | 0.55 | NA | NA | NA | NA | NA | NA | NA | NA |
| Jiang et al. 2012 | 26.7333 | 115.0667 | 121 | 18.6 | 1726 | conifer | ECM | non-N-fixing | mineral | NA | NA | NA | 15 | 12.3 | 0.79 | NA | NA | NA | NA | NA | NA | NA | NA |
| Jiang et al. 2012 | 26.7333 | 115.0667 | 121 | 18.6 | 1726 | boradleaf | AM | non-N-fixing | mineral | NA | NA | NA | 14 | 13.5 | 0.94 | NA | NA | NA | NA | NA | NA | NA | NA |
| Jiang et al. 2012 | 26.7333 | 115.0667 | 121 | 18.6 | 1726 | boradleaf | AM | non-N-fixing | mineral | NA | NA | NA | 14 | 21.9 | 1.55 | NA | NA | NA | NA | NA | NA | NA | NA |
| Perez-Bejarano et al. 2010 | 38.38333 | -0.98333 | 851 | 15.8 | 260 | conifer | AM | non-N-fixing | mineral | NA | NA | NA | NA | 75.58 | NA | NA | NA | NA | NA | NA | NA | NA | NA |
| Perez-Bejarano et al. 2010 | 38.38333 | -0.98333 | 851 | 15.8 | 260 | boradleaf | AM | non-N-fixing | mineral | NA | NA | NA | NA | 64.42 | NA | NA | NA | NA | NA | NA | NA | NA | NA |
| Perez-Bejarano et al. 2010 | 38.38333 | -0.98333 | 851 | 15.8 | 260 | boradleaf | ECM | non-N-fixing | mineral | NA | NA | NA | NA | 87.524 | NA | NA | NA | NA | NA | NA | NA | NA | NA |
| Perez-Bejarano et al. 2010 | 38.38333 | -0.98333 | 851 | 15.8 | 260 | conifer | ECM | non-N-fixing | mineral | NA | NA | NA | NA | 91.441 | NA | NA | NA | NA | NA | NA | NA | NA | NA |
| Ramesh et al. 2013 | 25.68917 | 91.92361 | 980 | 18.7 | 2208.5 | boradleaf | AM | non-N-fixing | mineral | 5.1245 | NA | NA | NA | 22.125 | NA | NA | NA | NA | NA | NA | NA | NA | 4.6 |
| Ramesh et al. 2013 | 25.68917 | 91.92361 | 980 | 18.7 | 2208.5 | boradleaf | AM | N-fixing | mineral | 3.5392 | NA | NA | NA | 23.336 | NA | NA | NA | NA | NA | NA | NA | NA | 4.59 |
| Ramesh et al. 2013 | 25.68917 | 91.92361 | 980 | 18.7 | 2208.5 | boradleaf | AM | non-N-fixing | mineral | 4.7675 | NA | NA | NA | 31.242 | NA | NA | NA | NA | NA | NA | NA | NA | 4.38 |
| Ramesh et al. 2013 | 25.68917 | 91.92361 | 980 | 18.7 | 2208.5 | conifer | ECM | non-N-fixing | mineral | 6.253 | NA | NA | NA | 22.706 | NA | NA | NA | NA | NA | NA | NA | NA | 4.36 |
| Turner et al. 1993 | 44.3333 | -123.9167 | 349 | 11.5 | 1210 | conifer | ECM | non-N-fixing | mineral | NA | NA | NA | 17 | 217 | 12.5 | NA | NA | NA | NA | NA | NA | NA | 3.8 |
| Turner et al. 1993 | 44.3333 | -123.9167 | 349 | 11.5 | 1210 | conifer | AM | non-N-fixing | mineral | NA | NA | NA | 19 | 217 | 11.4 | NA | NA | NA | NA | NA | NA | NA | 3.8 |
| Turner et al. 1993 | 44.3333 | -123.9167 | 349 | 11.5 | 1210 | conifer | ECM | non-N-fixing | mineral | NA | NA | NA | 20 | 316 | 15.4 | NA | NA | NA | NA | NA | NA | NA | 3.6 |
| Wang et al. 2013 | 22.1 | 106.7667 | 350 | 19.6 | 1400 | boradleaf | AM | non-N-fixing | mineral | NA | NA | NA | 12.28 | 35.11 | 2.86 | NA | NA | NA | NA | NA | NA | NA | 3.93 |
| Wang et al. 2013 | 22.1 | 106.7667 | 350 | 19.6 | 1400 | boradleaf | ECM | non-N-fixing | mineral | NA | NA | NA | 15.63 | 32.55 | 2.11 | NA | NA | NA | NA | NA | NA | NA | 4.26 |
| Wang et al. 2013 | 22.1 | 106.7667 | 350 | 19.6 | 1400 | conifer | ECM | non-N-fixing | mineral | NA | NA | NA | 16.69 | 26.46 | 1.59 | NA | NA | NA | NA | NA | NA | NA | 5.18 |
| Wang et al. 2013 | 22.1 | 106.7667 | 350 | 19.6 | 1400 | boradleaf | AM | non-N-fixing | mineral | NA | NA | NA | 12.07 | 30.25 | 2.5 | NA | NA | NA | NA | NA | NA | NA | 3.86 |
| Wang et al. 2013 | 22.1 | 106.7667 | 350 | 19.6 | 1400 | boradleaf | ECM | non-N-fixing | mineral | NA | NA | NA | 14.33 | 28.58 | 2.03 | NA | NA | NA | NA | NA | NA | NA | 4.15 |
| Wang et al. 2013 | 22.1 | 106.7667 | 350 | 19.6 | 1400 | conifer | ECM | non-N-fixing | mineral | NA | NA | NA | 14.88 | 23.94 | 1.6 | NA | NA | NA | NA | NA | NA | NA | 4.96 |
| Li et al. 2012 | 31.65 | 119.2833 | 300 | 17.5 | 1149 | conifer | AM | non-N-fixing | mineral | NA | NA | NA | 13.36 | 24.71 | 1.87 | NA | NA | NA | NA | NA | NA | NA | 4.4 |
| Li et al. 2012 | 31.65 | 119.2833 | 300 | 17.5 | 1149 | boradleaf | ECM | non-N-fixing | mineral | NA | NA | NA | 13.21 | 36.84 | 2.8 | NA | NA | NA | NA | NA | NA | NA | 4.36 |
| Li et al. 2012 | 31.65 | 119.2833 | 300 | 17.5 | 1149 | conifer | ECM | non-N-fixing | mineral | NA | NA | NA | 13.52 | 16.2 | 1.21 | NA | NA | NA | NA | NA | NA | NA | 4.5 |

Litter: annual litter production (); Cs: C stock; CN: C :N ratio; Ns: forest floor N stock; Nc: mineral soil N concentration; PAP: plant available phosphorus; MBC: microbial biomass C; MBN: microbial biomass nitrogen; MBP: microbial biomass phosphorus; respiration: soil base respiration.

**Table S2** Estimates of the linear mixed models assessing the effects of tree species type, mycorrhizal association, and N-fixing ability on mineral soil chemical and biotic properties. Values in bold indicate statistically significant difference between broadleaf and conifer, AM and ECM, or N-fixing and non-N-fixing. The corresponding statistical results are shown in Table 1.

| Soil properties | Tree species effect on soil properties | | | | | |
|---|---|---|---|---|---|---|
| | Species type | | Mycorrhizal association | | N-fixing ability | |
| | Broadleaf | Conifer | AM | ECM | N-fixing | non-N-fixing |
| $NH_4^+$ (mg kg$^{-1}$) | 214.8 | 575.9 | 309.1 | 411.2 | 327.5 | 377.5 |
| $NO_3^-$ (mg kg$^{-1}$) | 14.7 | 13.2 | **17.2** | **11.9** | 17.2 | 14.0 |
| PAP (mg kg$^{-1}$) | 312.1 | 463.8 | 293.9 | 447.8 | 246.3 | 391.4 |
| MBC (mg kg$^{-1}$) | 599.8 | 577.7 | **684.8** | **535.9** | **560.7** | **594.8** |
| MBN (mg kg$^{-1}$) | **85.7** | **57.2** | 76.4 | 73.5 | 70.8 | 75.2 |
| MBP (mg kg$^{-1}$) | 26.4 | 15.4 | 19.8 | 24.9 | 23.3 | 23.4 |
| Basal respiration (mg $CO_2$ g$^{-1}$ day$^{-1}$) | **27.6** | **29.4** | **24.2** | **29.7** | **27.1** | **28.8** |

**Appendix 1:** List of the 143 primary articles from which the data were extracted for this study